\documentclass[iop,apj,tighten]{emulateapj}

\shorttitle{INTERSTELLAR ABUNDANCE OF BORON}
\shortauthors{RITCHEY ET AL.}

\begin{document}
\title{The Abundance of Boron in Diffuse Interstellar Clouds\altaffilmark{1}}
\author{A. M. Ritchey\altaffilmark{2}$^,$\altaffilmark{3}, S. R. 
Federman\altaffilmark{2}$^,$\altaffilmark{3}, Y. 
Sheffer\altaffilmark{2}$^,$\altaffilmark{4}, and D. L. Lambert\altaffilmark{5}}
\altaffiltext{1}{Based in part on observations made with the NASA/ESA 
\emph{Hubble Space Telescope}, obtained from the Multimission Archive at the 
Space Telescope Science Institute (MAST). STScI is operated by the Association 
of Universities for Research in Astronomy, Inc., under NASA contract 
NAS5-26555.}
\altaffiltext{2}{Department of Physics and Astronomy, University of Toledo, 
Toledo, OH 43606; adam.ritchey@utoledo.edu; steven.federman@utoledo.edu.}
\altaffiltext{3}{Guest Observer, McDonald Observatory, University of Texas at 
Austin, Austin, TX 78712.}
\altaffiltext{4}{Department of Astronomy, University of Maryland, College Park, 
MD 20742; ysheffer@astro.umd.edu.}
\altaffiltext{5}{W.J. McDonald Observatory, University of Texas at Austin, 
Austin, TX 78712; dll@astro.as.utexas.edu.}

\begin{abstract}
We present a comprehensive survey of boron abundances in diffuse interstellar 
clouds from observations made with the Space Telescope Imaging Spectrograph 
(STIS) of the \emph{Hubble Space Telescope}. Our sample of 56 Galactic sight 
lines is the result of a complete search of archival STIS data for the 
B~{\small II} $\lambda$1362 resonance line, with each detection confirmed by 
the presence of absorption from O~{\small I}~$\lambda$1355, 
Cu~{\small II}~$\lambda$1358, and Ga~{\small II}~$\lambda$1414 (when available) 
at the same velocity. Five previous measurements of interstellar B~{\small II} 
from Goddard High Resolution Spectrograph observations are incorporated in our 
analysis, yielding a combined sample that more than quadruples the number of 
sight lines with significant boron detections. Our survey also constitutes the 
first extensive analysis of interstellar gallium from STIS spectra and expands 
on previously published results for oxygen and copper. The observations probe 
both high and low-density diffuse environments, allowing the density-dependent 
effects of interstellar depletion to be clearly identified in the gas-phase 
abundance data for each element. In the case of boron, the increase in relative 
depletion with line-of-sight density amounts to an abundance difference of 
0.8~dex between the warm and cold phases of the diffuse interstellar medium. 
The abundance of boron in warm, low-density gas is found to be 
B/H~=~$(2.4\pm0.6)\times10^{-10}$, which represents a depletion of 60\% 
relative to the meteoritic boron abundance. Beyond the effects of depletion, 
our survey reveals sight lines with enhanced boron abundances that potentially 
trace the recent production of $^{11}$B, resulting from spallation reactions 
involving either cosmic rays or neutrinos. Future observations will help to 
disentangle the relative contributions from the two spallation channels for 
$^{11}$B synthesis.
\end{abstract}


\section{INTRODUCTION}

The origins of the two stable isotopes of boron, $^{10}$B and $^{11}$B, remain 
uncertain despite numerous theoretical and observational advances over the past 
four decades (see the reviews by Reeves 1994; Vangioni-Flam et al. 2000; 
Prantzos 2007). As one of the rare light elements (a group that also includes 
lithium and beryllium), boron cannot be synthesized through quiescent burning 
in stellar interiors, where it is destroyed at temperatures that exceed 
$5\times10^6$ K, nor is it produced in significant quantities by the standard 
model of the Big Bang. The spallation of interstellar nuclei by relativistic 
Galactic cosmic rays (GCR), as originally proposed by Reeves et al. (1970), is 
known to be an effective means of producing the light elements. However, 
detailed models of GCR spallation (e.g., Meneguzzi et al. 1971, hereafter MAR; 
Ramaty et al. 1997), while successful at reproducing the solar system abundance 
of $^{10}$B (as well as $^9$Be and, by including the $\alpha$-$\alpha$ fusion 
reactions, $^6$Li), fail to adequately account for $^{11}$B (and $^7$Li). 
Standard GCR nucleosynthesis predicts a boron isotopic ratio of 
$^{11}$B/$^{10}$B~=~2.4 (MAR), in conflict with the value measured for solar 
system material (4.0; Lodders 2003). Still, since the threshold energies for 
spallation reactions leading to $^{11}$B are generally lower than those for 
reactions that produce $^{10}$B, it is possible to account for the discrepancy 
by adopting a cosmic-ray spectrum enhanced at low energies (particularly in the 
range 5$-$40~MeV nucleon$^{-1}$; see MAR; Meneguzzi \& Reeves 1975). The GCR 
energy spectrum below 1 GeV nucleon$^{-1}$ is highly uncertain due to the 
effects of solar modulation, meaning that there are no direct observational 
constraints against a large flux of low-energy cosmic rays.

In the present-day interstellar medium (ISM), the dominant contribution to 
cosmic-ray nucleosynthesis comes from forward spallation reactions (i.e., 
energetic protons and $\alpha$-particles impinging on interstellar CNO nuclei). 
Forward spallation is a secondary production mechanism because it depends on 
the metallicity of the ISM (i.e., the CNO abundances) and on the rate of 
supernovae occurring in the Galaxy (supernovae being the presumed sources of 
GCR acceleration). Over Galactic evolutionary timescales, the abundances of 
light elements synthesized via forward spallation should scale quadratically 
with the abundances of the metals that serve as interstellar targets. However, 
this fact is contrary to the well-known primary behavior exhibited by beryllium 
and boron in the halo. Studies of metal-poor halo stars (e.g., Gilmore et al. 
1992; Duncan et al. 1992, 1997; Garc\'ia L\'opez et al. 1998; Boesgaard et al. 
1999) have consistently shown that Be and B abundances increase approximately 
linearly with metallicity. Thus, many investigations (e.g., Cass\'e et al. 
1995; Ramaty et al. 1996; Vangioni-Flam et al. 1996) have focused on scenarios 
involving reverse spallation reactions (i.e., accelerated CNO nuclei being 
spalled from ambient interstellar H and He). In these models, low-energy, 
metal-enriched cosmic rays are accelerated in superbubbles (Parizot \& Drury 
1999, 2000; Parizot 2000) by shocks associated with Type II supernovae (SNe II) 
or by the winds of massive Wolf-Rayet stars, and then interact with nearby 
interstellar material. Since light elements produced in superbubbles result 
from the breakup of freshly synthesized CNO nuclei, the superbubble model 
represents a primary mechanism that can operate throughout the lifetime of the 
Galaxy and may be particularly important at early times.

Neutrino-induced spallation in SNe II (the $\nu$-process; Woosley et al. 1990) 
offers an alternative explanation for primary boron production and could 
ameliorate the problem in standard GCR nucleosynthesis of a low predicted 
$^{11}$B/$^{10}$B ratio. A significant amount of $^{11}$B (though virtually no 
$^{10}$B) is expected to be synthesized during the collapse of a massive star's 
core as the immense flux of neutrinos interacts with $^{12}$C in the 
carbon-rich shell. Models of core-collapse supernovae that incorporate the 
$\nu$-process (see Woosley \& Weaver 1995, hereafter WW95) also predict 
substantial yields for $^7$Li (produced in the helium shell) and $^{19}$F (in 
the neon shell). Timmes et al. (1995), adopting the WW95 yields for their model 
of Galactic chemical evolution (GCE), conclude that a major portion of the 
cosmic abundances of $^{11}$B and $^{19}$F, and about half of the $^7$Li 
abundance, can be attributed to neutrino nucleosynthesis. However, these 
prescriptions may be difficult to accommodate if other sources [e.g., 
cosmic-ray spallation for $^{11}$B and $^7$Li, Big Bang nucleosynthesis for 
$^7$Li, stellar processing in asymptotic giant branch (AGB) stars for $^7$Li 
and $^{19}$F] must also contribute. In the spallation models of Fields et al. 
(2000), which include a GCR, superbubble, and neutrino component, the 
$\nu$-process yields of WW95 had to be reduced by 60\% in order to avoid the 
overproduction of $^{11}$B. The $\nu$-process interactions chiefly involve the 
$\mu$- and $\tau$-neutrinos because these have higher temperatures, yet the 
temperatures are uncertain and the light element yields are strongly dependent 
on these values. Yoshida et al. (2005, 2008) constrained the $\mu$- and 
$\tau$-neutrino temperatures by requiring that $\nu$-process synthesis not 
result in $^{11}$B overproduction, in accordance with GCE models that include 
cosmic-ray spallation (e.g., Ramaty et al. 2000; Fields et al. 2000). The 
constraints place the temperatures between 60\% and 80\% of the value adopted 
in WW95, though even lower temperatures are inferred if neutrino oscillation 
effects are taken into account (see Yoshida et al. 2006, 2008).

The extent to which the various nucleosynthetic processes contribute to the 
enhancement in the solar system abundance of $^{11}$B over that predicted by 
standard GCR spallation is still unclear. However, since all of the boron 
production mechanisms occur in, or are closely associated with, the ISM, 
additional clues could potentially be gleaned from a careful study of 
interstellar boron abundances and isotopic ratios. Boron was first detected in 
the ISM by Meneguzzi \& York (1980), who used the \emph{Copernicus} satellite 
to measure absorption from the B~{\small II}~$\lambda$1362 resonance line 
toward $\kappa$~Ori. They obtained an interstellar abundance (B/H = 
$1.5\times10^{-10}$) in good agreement with the then-current stellar value 
($2\times10^{-10}$; Boesgaard \& Heacox 1978), assumed to be the Galactic 
value. Later investigations with the Goddard High Resolution Spectrograph 
(GHRS) on board the \emph{Hubble Space Telescope} (\emph{HST}; e.g., Federman 
et al. 1993; Jura et al. 1996) added new sight lines to the list of detections 
(all toward nearby stars) and found that the interstellar abundances were 
substantially lower than both the solar and stellar values. This led Jura et 
al. (1996) to propose a scenario involving either the recent infall of 
metal-poor material in the vicinity of the Sun or the depletion of boron onto 
interstellar grains. Federman et al. (1996a), examining GHRS data for the line 
of sight to $\delta$~Sco, provided the first measurement of the 
$^{11}$B/$^{10}$B ratio outside the solar system. Their result ($^{11}$B/$^{10}$B 
= 3.4$^{+1.3}_{-0.6}$), along with the subsequent work of Lambert et al. (1998), 
showed that the solar system ratio is not anomalous but probably representative 
of the local Galactic neighborhood. The B~{\small II} survey by Howk et al. 
(2000) expanded the sample of interstellar boron abundances to include some of 
the more extended sight lines accessible to the Space Telescope Imaging 
Spectrograph (STIS) of \emph{HST}. These authors found that the gas-phase B/O 
ratio decreases with increasing average hydrogen density and interpreted the 
trend as an indication of interstellar depletion. From the least depleted sight 
line, they derived a lower limit to the present-day total interstellar boron 
abundance of B/H $\gtrsim$ $(2.5\pm0.9)\times10^{-10}$.

The discovery by Knauth et al. (2000) of newly synthesized lithium toward $o$ 
Per, a member of the Per OB2 association, was an important new development in 
the study of light element nucleosynthesis. At least one of the clouds in the 
$o$ Per direction has a low $^7$Li/$^6$Li ratio ($\sim2$; Knauth et al. 2000, 
2003b), consistent with the ratio predicted by standard GCR spallation (1.5; 
MAR) and much lower than the solar system value (12.2; Lodders 2003). The line 
of sight to $o$ Per passes very near the massive star-forming region IC~348 and 
measurements of interstellar OH suggest that the sight line possesses an 
order-of-magnitude higher cosmic-ray flux compared to other sight lines in Per 
OB2 (Federman et al. 1996b). The implication is that accelerated particles 
supplied by the star-forming region are interacting with ambient interstellar 
material in the direction of $o$ Per, leading to an enhancement in $^6$Li 
relative to $^7$Li. This unexpected result prompted our STIS program (GO 8622) 
aimed at measuring $^{11}$B/$^{10}$B ratios in Per OB2 from observations of the 
B~{\small II} line toward four stars: 40 Per, $o$ Per, $\zeta$ Per, and X Per. 
Ultimately, the acquired STIS spectra lacked the signal-to-noise (S/N) ratio 
necessary to yield meaningful results on $^{11}$B/$^{10}$B, though the data did 
provide accurate B~{\small II} column densities. Therefore, we redirected our 
efforts to obtaining elemental boron abundances for a larger, more 
statistically significant sample of Galactic sight lines. Such a sample is 
needed to determine conclusively the level of boron depletion in interstellar 
gas. This information would then enable a more robust interpretation of the 
observed abundances and possibly allow the detection of intrinsic abundance 
variations, which, if discovered, could offer vital clues to the 
nucleosynthetic origin of boron and other light elements. 

In this investigation, we more than quadruple the number of interstellar sight 
lines with significant detections of the B~{\small II} line by taking advantage 
of the wealth of UV data provided by the \emph{HST}/STIS archive. The remainder 
of this paper is organized as follows. We describe the observations and our 
reduction of the data in \S{}~2. The methods used to obtain column densities 
(i.e., the integration of apparent column density profiles and the method of 
profile synthesis) are detailed in \S{}~3, where we also provide a comparison 
with previous studies. The profile synthesis results are utilized in \S{}~4 to 
determine elemental abundances (\S{}~4.1) and depletions (\S{}~4.2) and trends 
are sought with various measures of gas density. In \S{}~4.3, the boron 
abundances are examined in detail to search for intrinsic variations 
superimposed on the general trend due to depletion. The results of our analyses 
are discussed in \S{}~5 in the context of light element nucleosynthesis and the 
summary and conclusions, along with suggestions for future studies, are given 
in \S{}~6.

\section{OBSERVATIONS AND DATA REDUCTION}

The B~{\small II} resonance line at 1362.46~\AA{} probes the dominant 
ionization stage of boron in diffuse clouds because the ionization potential of 
neutral boron (8.30~eV) is below the Lyman limit. Currently, the Space 
Telescope Imaging Spectrograph is the only UV-sensitive instrument capable of 
observing B~{\small II} at high enough spectral resolution to study 
interstellar absorption profiles in detail. Our sample of 56 Galactic sight 
lines (see Table~1) is the result of a comprehensive search of all archival 
STIS datasets\footnote{Our initial search of the STIS archive was performed in 
2006, at which time we examined all datasets acquired before the failure of the 
Side 2 electronics in 2004. Our survey does not include any observations made 
since the successful completion of servicing mission SM4 in 2009.} with the 
necessary wavelength coverage for absorption from B~{\small II}~$\lambda$1362. 
The STIS data for five of these sight lines (HD~104705, HD~121968, HD~177989, 
HD~218915, and HDE~303308) provided the basis for the previous analysis of 
interstellar boron by Howk et al. (2000). These data are reexamined here for 
consistency, enabling an important comparison between the results of two 
independent investigations.

Along with the B~{\small II} line, the UV spectra yield information on 
O~{\small I}~$\lambda$1355, Cu~{\small II}~$\lambda$1358, and 
Ga~{\small II}~$\lambda$1414. The interstellar absorption profiles resulting 
from these transitions are useful for determining the cloud component structure 
to be applied to the weaker B~{\small II} profile. Since O$^0$, Cu$^+$, and 
Ga$^+$ are each the dominant ion of their element in neutral diffuse clouds, 
these species should coexist with B$^+$ along the diffuse and translucent sight 
lines studied here. For many directions, high resolution ground-based data on 
Ca~{\small II}~$\lambda$3933 and K~{\small I}~$\lambda$7698 acquired at 
McDonald Observatory or obtained from the literature serve to confirm the 
component structure derived from the somewhat lower resolution UV spectra. 
Interstellar lines from K~{\small I}~$\lambda$7698 trace the same neutral 
atomic gas probed by the UV measurements and yet are typically quite narrow, 
allowing precise determinations of component velocities. Absorption profiles of 
Ca~{\small II} are generally more optically thick and arise from gas more 
widely distributed in velocity than the other atomic species (Pan et al. 2005), 
yet the line widths of individual Ca~{\small II} components track those of the 
dominant ions in diffuse clouds (Welty et al. 1996). By incorporating the 
ground-based data into our analysis, we are able to evaluate the effect that 
unresolved component structure in the UV lines has on the derived interstellar 
abundances.

\begin{figure}[!t]
\centering
\includegraphics[width=0.45\textwidth]{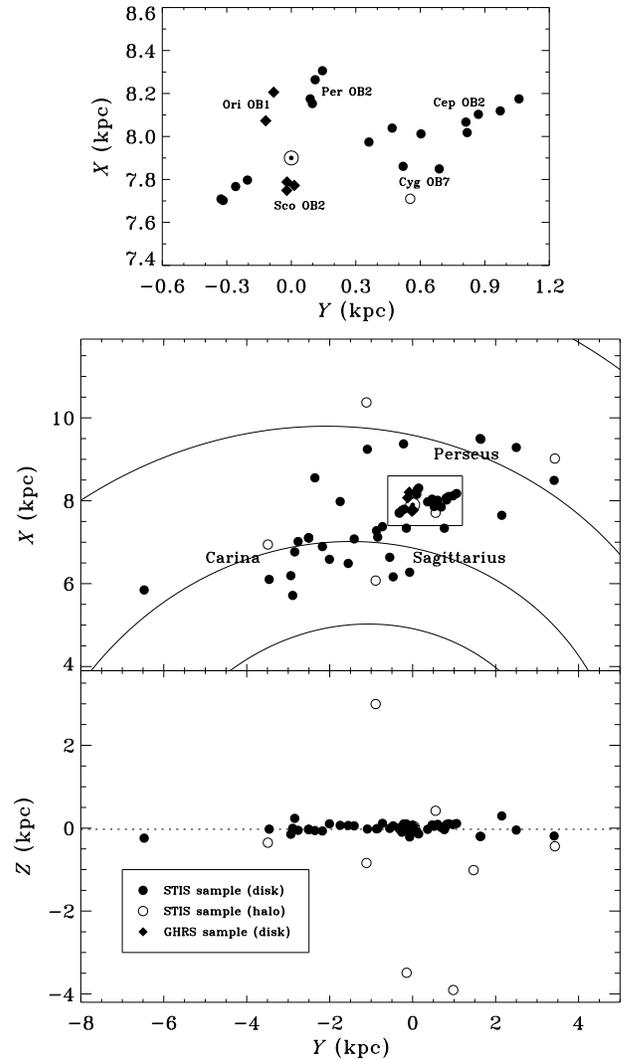}
\caption[Galactic distribution of stars in the boron sample]{Distribution of 
stars in the boron sample in Galactic cartesian coordinates with the origin at 
the Galactic center. The position of the Sun at ($X=R_0=7.9$, $Y=0$, $Z=0$) is 
indicated. Solid symbols denote stars that reside in the disk, while open 
symbols are used for halo stars (see legend). The upper panel shows the local 
solar neighborhood with prominent OB associations identified. The curves in the 
middle panel indicate the positions of the spiral arms based on the logarithmic 
model of Vall\'ee (2005). The nearest arms to the Sun are labeled. The dotted 
line in the lower panel designates the Galactic plane, which is offset from the 
$Z$-coordinate of the Sun by $-26$ pc (Majaess et al. 2009).}
\end{figure}

\subsection{Description of the Sample}

Table~1 provides the relevant stellar data for the 56 O and B-type stars that 
comprise the STIS sample, along with the five stars that have published 
interstellar boron abundances from GHRS (Jura et al. 1996; Lambert et al. 1998; 
Howk et al. 2000). The GHRS sight lines are included in our abundance analysis 
(\S{} 4), though we do not rederive B~{\small II} column densities in these 
directions. The spectral types, $V$ magnitudes, and Galactic coordinates in 
Table~1 are from the SIMBAD database, while, for most sight lines, the values 
for interstellar reddening and distance are taken from the same references that 
provided column densities of H~{\small I} or, if data on atomic hydrogen are 
lacking, H$_2$ (see \S{}~4.1). The distance derived from the \emph{Hipparcos} 
parallax (Perryman et al. 1997) is adopted instead if the measurement is 
significant at the 3-$\sigma$ level or greater. In some cases, we determined 
the distance from the method of spectroscopic parallax, using absolute visual 
magnitudes from Schmidt-Kaler (1982). The typical uncertainty for distances 
derived spectroscopically or obtained from the literature is 30\%. Table~1 also 
lists any known associations for the program stars.

The full boron sample probes a diverse array of interstellar environments, 
spanning at least two orders of magnitude in average sight-line density and 
more than four orders of magnitude in molecular hydrogen fraction. Over 50\% of 
the sample stars reside in OB associations or young stellar clusters, another 
three stars are associated with H~{\small II} regions, and eight are located in 
the Galactic halo. The rest are presumably field stars of the Galactic disk or 
possibly unidentified members of OB associations. Figure~1 presents the 
distribution of the program stars in Galactic cartesian coordinates with the 
origin at the Galactic center (assuming $R_0$ = 7.9~kpc; Vall\'ee 2005). While 
there is a significant range in heliocentric distance in the sample, it is 
likely that much of the interstellar material traced by the STIS and GHRS 
observations of these stars resides in the local ISM. This can be inferred from 
the velocities of the dominant absorption components, which are typically near 
the velocity of the local standard of rest. Still, many of the extended sight 
lines observed with STIS exhibit double-peaked absorption profiles, indicating 
that the sight lines pass through both local gas and gas associated with either 
the Sagittarius-Carina or Perseus spiral arm. Figure~1 confirms that stars 
displaying such features in their interstellar spectra indeed lie either behind 
or within one of the two nearest spiral arms from the vantage point of the 
Sun\footnote{For two stars (HD~108639 and HD~114886), we initially adopted 
distances obtained from the literature (Andersson et al. 2002; Cartledge et al. 
2008), which were based on measurements of trigonometric parallax. Yet these 
are significantly smaller than the distances derived spectroscopically. 
Ultimately, we chose to use the larger spectroscopic distances, which place the 
stars within the Sagittarius-Carina spiral arm, since the interstellar profiles 
are clearly double-peaked.}. These directions offer the unique opportunity to 
study variations in elemental abundances in clouds at different distances along 
the line of sight. The overall diversity of sight-line properties in our sample 
allows us to thoroughly investigate the role played by physical environment in 
the observed abundances of boron in diffuse clouds.

\subsection{\emph{HST}/STIS Data}

STIS observations employing the FUV Multi-Anode Microchannel Array (MAMA) 
detector and either the high-resolution (E140H) or medium-resolution (E140M) 
grating in both prime and secondary modes were examined to identify sight lines 
with unambiguous interstellar absorption from O~{\small I}~$\lambda$1355, 
Cu~{\small II}~$\lambda$1358, and Ga~{\small II}~$\lambda$1414. Subsequent 
searches for absorption from B~{\small II}~$\lambda$1362 at similar velocities 
yielded 67 preliminary detections. Throughout the data selection process, 
spectra acquired using the secondary E140H mode at 1271~\AA{} required special 
consideration. The B~{\small II} line is located in order 309 of the E140H 
grating. When the central wavelength is set to 1271 \AA{}, this order typically 
falls close to the edge of the FUV MAMA detector. Because the spectral format 
of STIS is periodically shifted by the Mode Select Mechanism (MSM) to avoid 
charge depletion in localized regions of the MAMA, a given exposure with a 
central wavelength of 1271~\AA{} may or may not include order 309 (see Kim 
Quijano et al. 2003). For each promising sight line with data at this setting, 
raw echelle images were inspected to determine whether or not the order of 
interest was present in the observations. This procedure contributed 22 of the 
initial 67 detections of interstellar B~{\small II}. From the preliminary 
sample, only those sight lines with estimated B~{\small II} line strengths of 
3~$\sigma$ or greater were retained for further analysis. The final sample 
contains 37 such detections from E140H data and another 19 exclusively from 
E140M. Two sight lines (HD~88115 and CPD$-$59 2603) have data at both settings 
covering the B~{\small II} transition, allowing a direct comparison between the 
column densities derived from high- and medium-resolution absorption profiles 
(see \S{} 3.2.3).

All STIS datasets for the final sample of interstellar sight lines were 
obtained from the Multimission Archive at STScI (MAST) after the close-out 
re-calibration of STIS archival data, completed in 2007. The re-calibrated 
datasets benefit from an improved flux calibration and blaze-shift correction 
for all echelle modes. However, because the observations used to calibrate the 
echelle sensitivities employed non-zero offsets of the MSM, some edge orders 
present in previous calibration observations are now missing (see Aloisi et al. 
2007). These orders are not included in the updated photometric conversion 
table (PHOTTAB), which gives the throughput as a function of spectral order, 
and were not extracted by the CALSTIS pipeline during re-calibration. In 
particular, orders 309 and 310, the latter of which contains the O~{\small I} 
and Cu~{\small II} absorption lines, are not present in re-calibrated data with 
a central wavelength of 1271 \AA. These orders had to be extracted manually 
from the flat-fielded science files obtained from the archive. This was 
accomplished by re-running X1D, the one-dimensional spectral extraction portion 
of CALSTIS, within the Space Telescope Science Data Analysis System (STSDAS) 
using an older version of the PHOTTAB reference table. The older sensitivities 
have higher systematic uncertainties and there is no blaze-shift correction in 
the table for secondary modes. As a result, the flux calibration based on the 
older sensitivities can differ from the new re-calibration by as much as 
5$-$8\% (Aloisi et al. 2007). However, the analysis of interstellar absorption 
lines depends only on the relative intensity between the line and the continuum 
and should be unaffected by errors in absolute flux calibration. Order 311, 
which also includes the O~{\small I} line due to overlap in the echelle format, 
was extracted in the same way as were orders 309 and 310 for consistency, 
allowing us to verify that there was no significant difference between the 
manual and pipeline extraction procedures.

Table~2 lists the STIS datasets used in this investigation and gives exposure 
times and details of the optical setup for each observation. The resolving 
power $R$ adopted for a given combination of grating and aperture was derived 
from the CO spectral synthesis fits of Sheffer et al. (2007). Multiple 
exposures of a target acquired with the same echelle grating were coadded to 
improve the S/N ratio in the final spectrum. Typical values of S/N for this 
sample are in the range 40$-$80 (per pixel), with values as low as 20 for the 
faintest stars or those with the shortest exposure times. For the bright 
Per~OB2 stars, the significantly longer exposure times result in S/N ratios in 
the range 100$-$300. When a feature appeared in adjacent orders, these orders 
were combined as long as the overlapping portion of the spectrum had sufficient 
continua on both sides of the line. This was the case for most of the 
O~{\small I} and Ga~{\small II} lines in data obtained with the E140H grating 
(resulting in an increase in S/N of almost 50\% over Cu~{\small II} and 
B~{\small II} from these data) and for all of the B~{\small II} lines with 
E140M (yielding a 35\% increase in S/N). It should be noted that one of the 
most commonly used STIS spectroscopic modes with E140H is the secondary mode at 
1271 \AA. Hence, many of the observations yielding O~{\small I}, 
Cu~{\small II}, and B~{\small II} spectra were acquired at this setting. 
However, since the spectral format at 1271 \AA{} does not include the orders 
containing the Ga~{\small II} line (orders 297 and 298), obtaining data on this 
species requires observations using a different spectroscopic mode, which may 
be lacking. Of the 56 sight lines in the STIS sample, 11 do not have 
Ga~{\small II} data for this reason.

Final coadded spectra were normalized to the continuum by fitting low-order 
polynomials to regions free of interstellar absorption within a spectral 
window, usually 2~\AA{} wide, centered on the line of interest. Occasionally, a 
smaller portion of the spectrum surrounding the interstellar line was used in 
order to avoid a strong stellar absorption feature. For one star (HD~108610), 
the task of distinguishing between stellar and interstellar absorption for the 
purpose of normalizing the continuum was deemed unfeasible. The UV spectrum of 
this B3 subgiant/dwarf has many narrow absorption lines uncharacteristic of an 
early-type star. Because HD~108610 is the primary component in a double system, 
the observed features are likely due to contamination by the faint, late-type 
secondary. As a result of the large uncertainties in continuum placement, this 
sight line was removed from the analysis of boron abundances. For all other 
sight lines, the UV spectra were normalized concurrently, taking care to 
separate weak interstellar components, often seen at large negative velocities, 
from noise in the continuum. To this end, we compared the various UV absorption 
profiles for a given sight line in velocity space to check for consistency. 
When ground-based data were available (see \S{} 2.3), they provided an 
additional means of determining the likelihood that a weak feature seen in the 
UV was a real interstellar absorption component. If a questionable feature is 
not seen in Ca~{\small II} or K~{\small I} absorption, it is likely to be an 
artifact of noise or of instrumental origin.

\subsection{McDonald Observatory Data}

Seventeen stars were observed with the Harlan J. Smith 2.7~m telescope at 
McDonald Observatory using the high-resolution mode (cs21) of the Tull 
(2dcoud\'e) spectrograph (Tull et al. 1995) during two observing runs in 2007 
October and 2008 June\footnote{Additional spectra were acquired during an 
observing run in 2007 December/2008 January designed to measure 
$^{12}$C/$^{13}$C ratios in diffuse molecular clouds (see Ritchey et al. 
2010).}. Two configurations of the cross-dispersed echelle spectrometer were 
needed to obtain data on Ca~{\small II}~K~$\lambda$3933 and 
K~{\small I}~$\lambda$7698. Both settings employed the 79~gr~mm$^{-1}$ grating 
(E1), the 145~$\mu$m slit (Slit 2), and a 2048~$\times$~2048 CCD (TK3). For the 
Ca~{\small II} observations, order 56 of the E1 grating spectrum was centered 
at 4080~\AA. This setting simultaneously provides data on CN~$\lambda$3874, 
Ca~{\small I}~$\lambda$4226, CH$^+$~$\lambda$4232, and CH~$\lambda$4300, in 
addition to Ca~{\small II}~K. Observations of the K~{\small I} line were 
acquired by centering order 31 of the grating spectrum at 7210~\AA. The grating 
tilt was moved slightly each night to lessen the effect of fixed pattern noise 
and prevent any instrumental glitches at specific locations on the CCD from 
interfering with the observations. Stellar exposures were limited to 30 minutes 
per frame to minimize the number of cosmic ray hits being recorded by the 
detector during a single integration. Dark frames were obtained on the first 
night of a run, while exposures for bias correction and flat fielding were 
taken each night. Comparison spectra from a Th-Ar lamp were recorded throughout 
the night at intervals of 2$-$3 hr. To correct for the presence of telluric 
absorption near the K~{\small I} line, an unreddened, early-type star should 
ideally be observed each night that the red setup is used. For the observing 
run in 2007 October, one of the program stars (40~Per) served as the telluric 
standard due to its simple, well understood interstellar absorption profile. 
For the 2008 June run, the bright, unreddened star $\alpha$~Lyr (Vega; A0 V; 
$V=0.03$) was observed for this purpose. The 40~Per exposures were acquired at 
an airmass of 1.4, while that of $\alpha$~Lyr at an airmass of 1.0.

The list of McDonald targets was compiled from the STIS sample after an 
extensive literature search for reliable determinations of interstellar 
Ca~{\small II} and K~{\small I} component structure from spectra with 
sufficiently high resolution ($\Delta$$v\leq2$ km s$^{-1}$; e.g., Welty et al. 
1996; Welty \& Hobbs 2001; Pan et al. 2004). Only nine stars had existing data 
on both species with the required spectral resolution. One star (55~Cyg) had 
very high resolution (0.6~km~s$^{-1}$) K~{\small I} measurements from Welty 
\& Hobbs (2001), but lacked similar data on Ca~{\small II}. Six others had 
already been observed by us at McDonald, but only with the blue setting. These 
observations were part of a program designed to yield cloud component 
structures for CO and H$_2$ (see Sheffer et al. 2008). Of the 47 stars in the 
sample without high-resolution measurements of both lines, only the 17 selected 
for observation had declinations accessible to McDonald Observatory. Table~3 
gives details of the McDonald observations, including the total exposure times 
and resulting S/N ratios per resolution element for each of the 17 targets. Due 
to adverse weather conditions during the 2008 June run, we were unable to 
acquire Ca~{\small II} profiles for two of the planned targets (HD~156110 and 
HD~177989). In the end, however, ground-based data on both species at high 
resolution were obtained from McDonald observations or from the literature for 
a statistically significant fraction (40\%) of the STIS sample.

The McDonald data were reduced using standard Image Reduction and Analysis 
Facility (IRAF) routines. Low-order polynomial fits to the overscan region were 
subtracted from the raw images before removing the average bias from the darks, 
flats, stellar exposures, and comparison lamp frames. None of the observations 
were dark corrected because the level of dark current was always found to be 
insignificant after subtracting off the bias. The thresholds for cosmic-ray 
rejection were set to 50 and 25 times the mean of the surrounding pixels for 
the stellar and comparison lamp exposures, respectively. Any cosmic rays 
present in individual flat frames were effectively removed by taking the median 
of all flats for a given night. Scattered light was modeled in the dispersion 
and cross-dispersion directions and subtracted from the stellar exposures and 
from the median flat. The flat was then normalized to unity and divided into 
the stellar and comparison lamp frames to correct for variations in pixel 
sensitivity across the detector. One-dimensional spectra were extracted from 
the processed images and the stellar exposures were calibrated in wavelength 
after identifying emission lines in the Th-Ar comparison spectra, typically 5 
per order. The wavelength solution applied to a given stellar exposure was 
interpolated from the solutions in the two nearest Th-Ar frames, the one 
preceding and the one following the stellar exposure. Finally, the calibrated 
spectra were shifted to the reference frame of the local standard of rest 
(LSR).

Before Doppler correcting data obtained with the red setting, a template for 
telluric absorption was divided into each of the orders containing 
K~{\small I}. The template was derived by fitting the telluric lines in the 
spectrum of the standard star and then scaling the fit to the airmass of the 
observation being corrected. Since each star was observed close to the 
meridian, the airmass varied only slightly from one observation to the next. 
After Doppler correction, all stellar exposures of the same target acquired 
with the same spectrograph configuration were coadded and the resulting spectra 
continuum normalized in the same way as were the UV data. From the widths of 
thorium emission lines in the Th-Ar lamp spectra, we determined that a 
resolving power of $R=185,000$ ($\Delta$$v$~=~1.6~km~s$^{-1}$) was achieved 
with the blue setting. The red configuration resulted in a resolving power of 
approximately 165,000 (1.8~km~s$^{-1}$). An extensive collection of UV and 
visible spectra for sight lines in the STIS boron sample is presented in 
Ritchey (2009).

\begin{figure}[!t]
\centering
\includegraphics[width=0.45\textwidth]{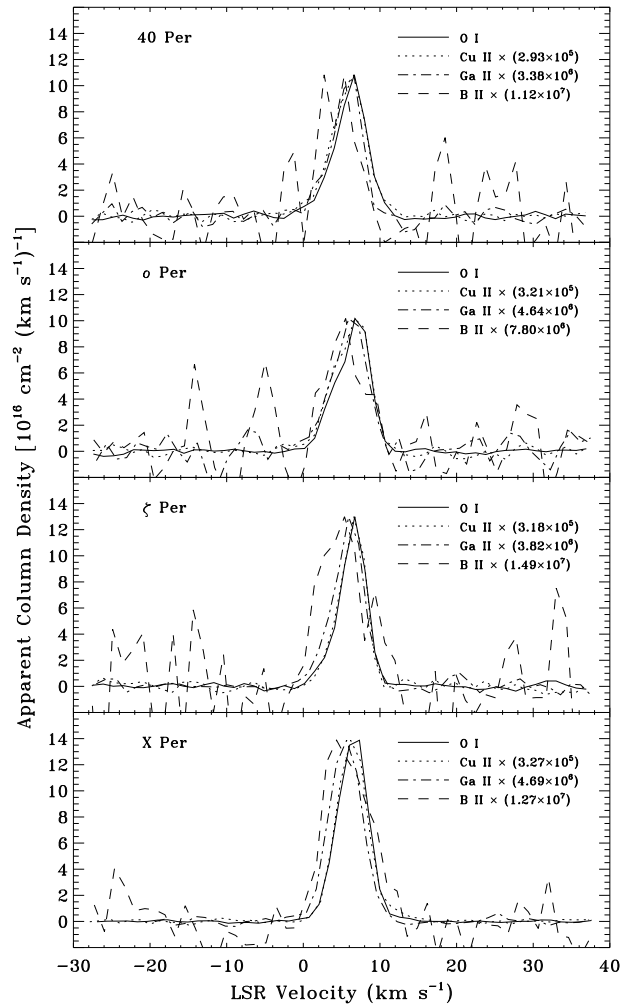}
\caption[Apparent column density profiles toward 40~Per, $o$~Per, $\zeta$~Per, 
and X~Per]{Apparent column density profiles of O~{\scriptsize I}~$\lambda$1355, 
Cu~{\scriptsize II}~$\lambda$1358, Ga~{\scriptsize II}~$\lambda$1414, and 
B~{\scriptsize II}~$\lambda$1362 toward 40~Per, $o$~Per, $\zeta$~Per, and 
X~Per, all members of the Per OB2 association. The line profiles in these 
directions are among the simplest in the STIS sample and these data have the 
highest signal to noise. In each panel, the Cu~{\scriptsize II}, 
Ga~{\scriptsize II}, and B~{\scriptsize II} profiles have been scaled (by the 
factors indicated) to the O~{\scriptsize I} profile. This has the effect of 
amplifying the noise in the B~{\scriptsize II} spectrum in particular. 
Considering the uncertainties, the profile shapes for the four species are very 
similar.}
\end{figure}

\begin{figure}[!t]
\centering
\includegraphics[width=0.45\textwidth]{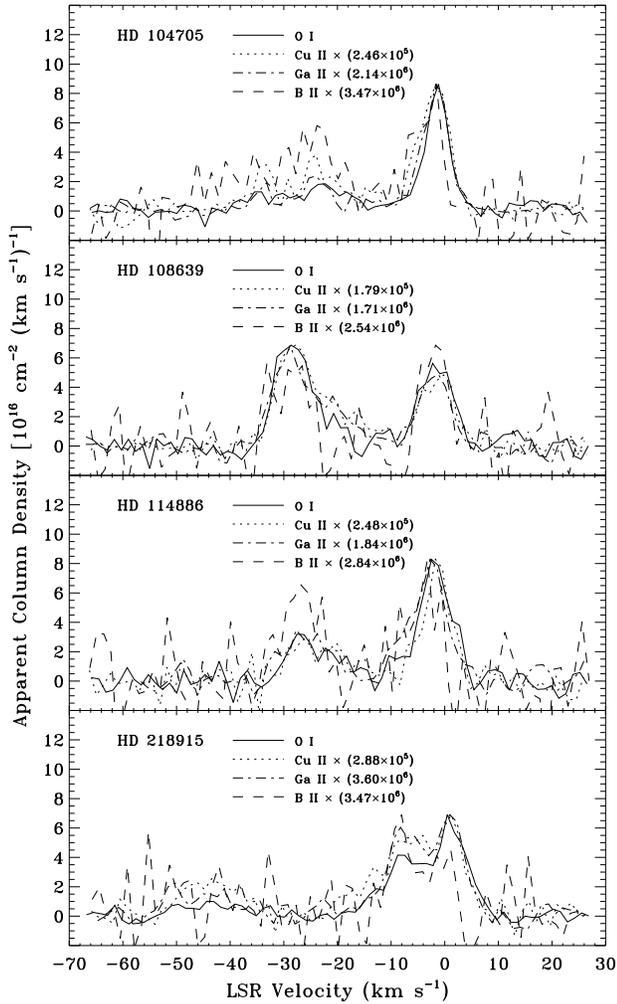}
\caption[Apparent column density profiles toward HD~104705, HD~108639, 
HD~114886, and HD~218915]{Same as Figure~2 except toward HD~104705, HD~108639, 
HD~114886, and HD~218915. All of these sight lines trace both local gas (near 
$v_{\mathrm{LSR}}$~=~0~km~s$^{-1}$) and gas associated with one of the two 
nearest spiral arms (at more negative velocities). The sight lines to 
HD~104705, HD~108639, and HD~114886 probe the Sagittarius-Carina spiral arm, 
while HD~218915 exhibits absorption from the Perseus arm (near 
$v_{\mathrm{LSR}}$~=~$-$44~km~s$^{-1}$). Note the greater relative strength of 
the B~{\scriptsize II} components compared to those of the other species in the 
Sagittarius-Carina arm toward HD~104705 and HD~114886.}
\end{figure}

\begin{figure}[!t]
\centering
\includegraphics[width=0.45\textwidth]{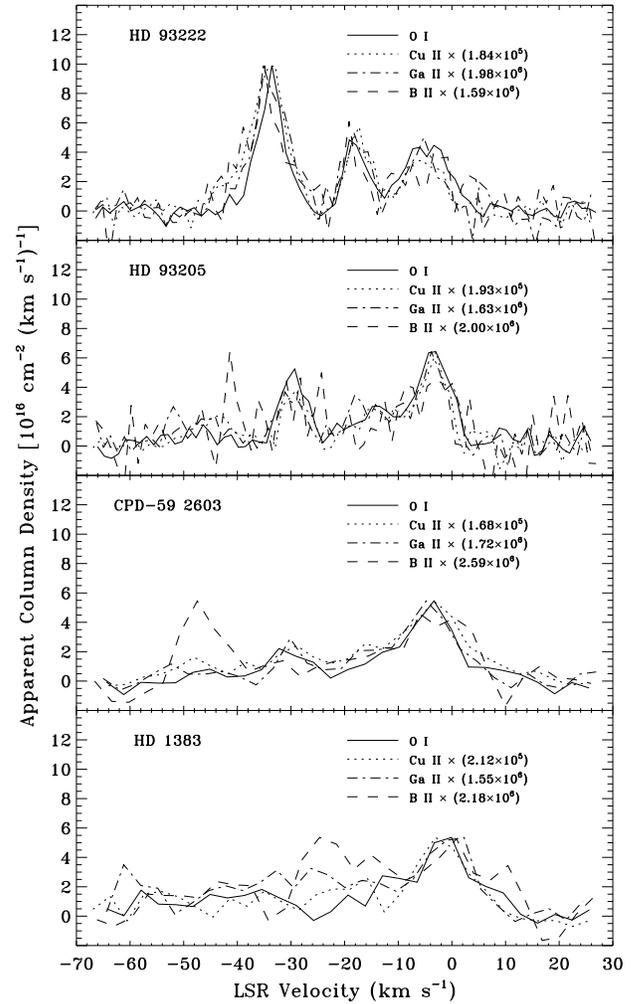}
\caption[Apparent column density profiles toward HD~93222, HD~93205, 
CPD$-$59~2603, and HD~1383]{Same as Figure~2 except toward HD~93222, HD~93205, 
CPD$-$59~2603, and HD~1383. The data shown for CPD$-$59~2603 and HD~1383 were 
acquired at medium resolution. Each of these sight lines exhibits absorption 
from three distinct cloud complexes. The stars HD~93222, HD~93205, and 
CPD$-$59~2603 reside in the Carina Nebula, while HD~1383 lies behind the 
Perseus spiral arm. The strong B~{\scriptsize II} component at 
$v_{\mathrm{LSR}}$~=~$-$46~km~s$^{-1}$ toward CPD$-$59~2603 is associated with 
the near side of the globally expanding H~{\scriptsize II} region in Carina, as 
is the weaker component at this velocity toward HD~93205 (see Walborn et al. 
2002). The $-$23~km~s$^{-1}$ component toward HD~1383, which is strongest in 
B~{\scriptsize II}, somewhat weaker in Ga~{\scriptsize II} and 
Cu~{\scriptsize II}, and essentially undetectable in O~{\scriptsize I}, 
probably also arises in an H~{\scriptsize II} region along the line of sight.}
\end{figure}

\section{RESULTS ON COLUMN DENSITIES}

\subsection{Apparent Optical Depth Calculations}

In an effort to directly evaluate the presumed similarity among the different 
UV absorption profiles for a given line of sight, the apparent column density 
as a function of velocity, $N_a(v)$, was computed for each species (see Savage 
\& Sembach 1991). Briefly, when an absorption line is not fully resolved by the 
spectrograph, the true intensity profile is blurred by the instrumental line 
spread function and may differ significantly from the observed intensity, 
depending on the resolution. The apparent optical depth per unit wavelength 
interval, $\tau_a(\lambda)$, is defined in terms of the observed intensity, 
$I_{\mathrm{obs}}(\lambda)$, according to 
$\tau_a(\lambda)=\mathrm{ln}[I_0(\lambda)/I_{\mathrm{obs}}(\lambda)]$, where 
$I_0(\lambda)$ is the intensity in the continuum in the absence of absorption 
and is equal to unity, by definition, for normalized spectra. Following Savage 
\& Sembach (1991), we can write the apparent optical depth in terms of the 
apparent column density, expressing both as functions of velocity,

\begin{equation}
\tau_a(v)= \frac{\pi e^2}{m_e c}f\lambda N_a(v)=2.654\times10^{-15}f\lambda 
N_a(v),
\end{equation}

\noindent
where $f$ is the oscillator strength and $\lambda$ the rest wavelength of the 
transition. The constant in Equation (1) is valid when $\lambda$ is in \AA{} 
and $N_a(v)$ is in units of cm$^{-2}$ (km s$^{-1})^{-1}$. The total apparent 
column density can then be found by solving for $N_a(v)$ and integrating over 
the entire line profile,

\begin{eqnarray}
N_a & = & \int N_a(v)dv=\frac{1}{2.654\times10^{-15}f\lambda}\int \tau_a(v)dv 
\nonumber \\
 & = & \frac{1}{2.654\times10^{-15}f\lambda}\int 
\mathrm{ln}\frac{I_0(v)}{I_{\mathrm{obs}}(v)}dv.
\end{eqnarray}

\noindent
Equation (2) yields a lower limit to the true column density of the material 
when the absorption profile contains unresolved, optically thick components. 
However, if the absorption is optically thin, then $\tau_a\approx\tau$ and 
$N_a$ can be taken as a good approximation of the true column density.

In our analysis, each of the normalized UV absorption profiles for a particular 
line of sight was integrated to obtain the apparent column density as well as 
the integrated equivalent width. The integration was performed with the 
Interactive Data Language (IDL) procedure INT\underline{ }TABULATED, which 
integrates tabulated data using a five-point Newton-Cotes formula. The limits 
of integration, where the optical depth goes to zero, were fixed by the 
O~{\small I} profile due to the strength of this line and its typically higher 
S/N ratio. The apparent optical depth (hereafter AOD) calculations provided 
useful consistency checks on the equivalent widths and column densities derived 
by other means (see \S{} 3.2.3 for a comparison between apparent column 
densities and column densities derived through profile synthesis). However, the 
primary motivation for performing these calculations was simply to compare the 
resulting apparent column density profiles to test for similarity among 
species. In general, the shapes of the O~{\small I}, Cu~{\small II}, 
Ga~{\small II}, and B~{\small II} profiles are quite similar for a given line 
of sight (see Figues~2$-$4), validating the assumption that these species 
coexist. However, it is also evident that the relative strengths of certain 
groups, or complexes, of components are not always preserved from species to 
species (see, e.g., the apparent column density profiles toward HD~104705 and 
HD~114886 in Figure~3). Integrated (i.e., line-of-sight) column densities are 
inadequate for the purpose of studying changes in elemental abundance ratios 
from one absorption complex to another. For such an analysis, it is necessary 
to synthesize the observed component structure.

\begin{figure*}[!t]
\centering
\includegraphics[width=0.95\textwidth]{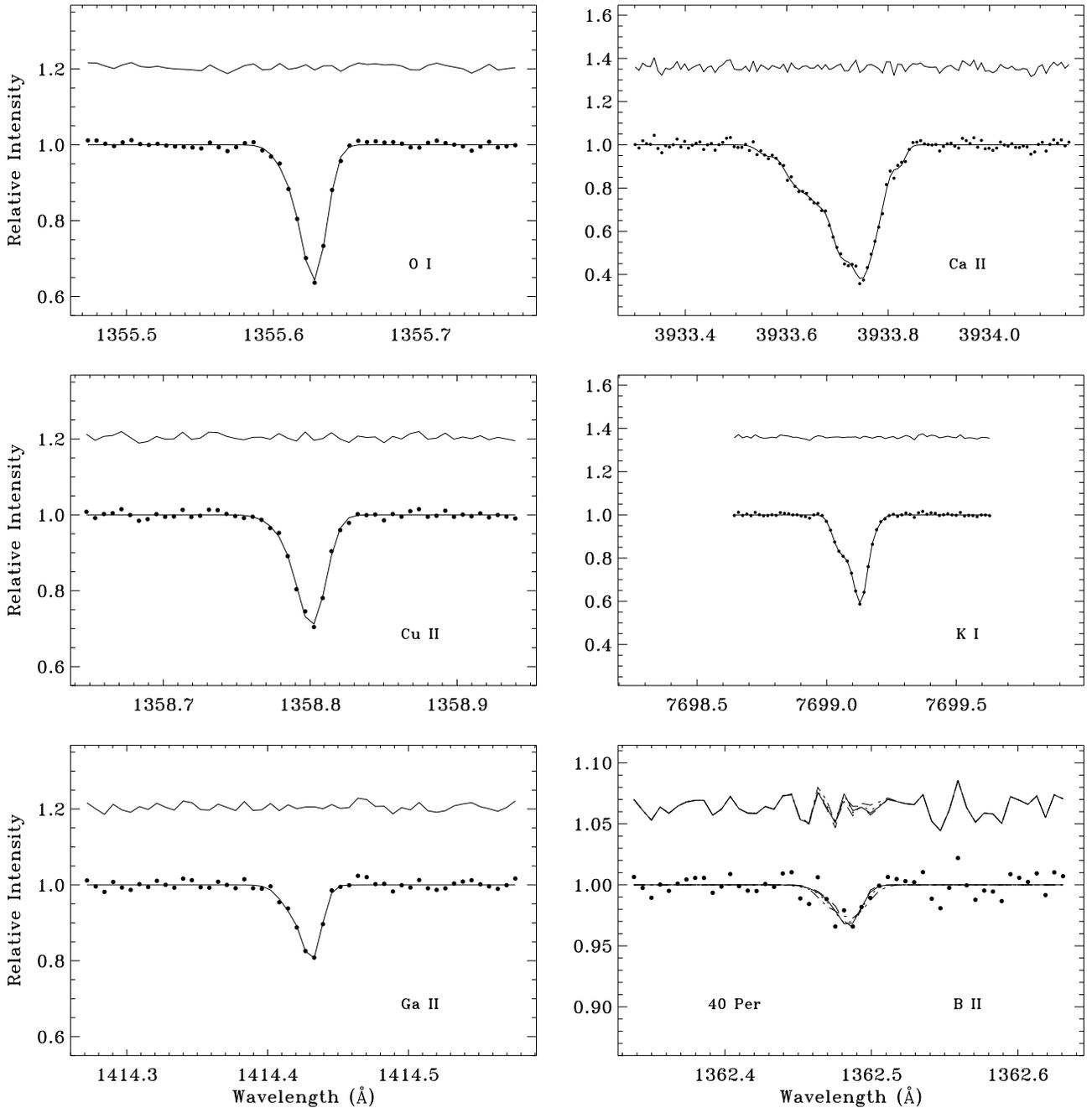}
\caption[Profile synthesis fits for O~{\scriptsize I}, Cu~{\scriptsize II}, 
Ga~{\scriptsize II}, Ca~{\scriptsize II}, K~{\scriptsize I}, and 
B~{\scriptsize II} toward 40~Per]{Profile synthesis fits to the 
O~{\scriptsize I}~$\lambda$1355, Cu~{\scriptsize II}~$\lambda$1358, 
Ga~{\scriptsize II}~$\lambda$1414, Ca~{\scriptsize II}~$\lambda$3933, 
K~{\scriptsize I}~$\lambda$7698, and B~{\scriptsize II}~$\lambda$1362 lines 
toward 40~Per. The same range in velocity is displayed in each panel. The 
synthetic profiles are shown as solid lines (in most cases) passing through 
data points that represent the observed spectra. For the B~{\scriptsize II} 
line, four separate profile templates were fitted to the observations: three 
templates based on the component structure found for O~{\scriptsize I} 
(\emph{solid line}), Cu~{\scriptsize II} (\emph{dotted line}), and 
Ga~{\scriptsize II} (\emph{dashed line}), and a high-resolution template 
derived from the results for Ca~{\scriptsize II} and K~{\scriptsize I} 
(\emph{dash-dotted line}). Residuals for each fit are plotted above the 
observed spectrum.}
\end{figure*}

\begin{figure*}[!t]
\centering
\includegraphics[width=0.95\textwidth]{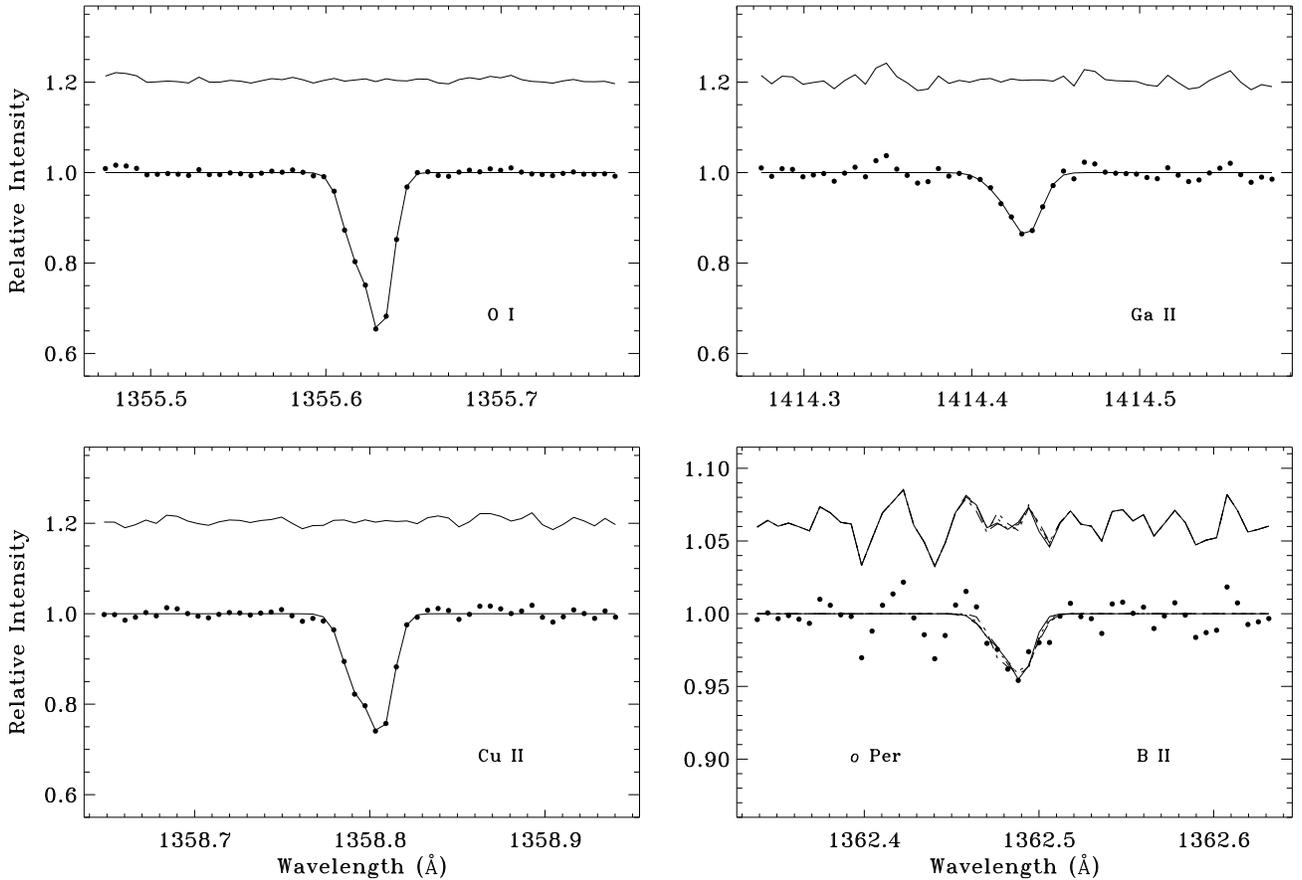}
\caption[Profile synthesis fits for O~{\scriptsize I}, Cu~{\scriptsize II}, 
Ga~{\scriptsize II}, and B~{\scriptsize II} toward $o$~Per]{Same as Figure~5 
except for the O~{\scriptsize I}, Cu~{\scriptsize II}, Ga~{\scriptsize II}, and 
B~{\scriptsize II} lines toward $o$~Per. High-resolution data on 
Ca~{\scriptsize II} and K~{\scriptsize I} are available for this sight line 
from the literature (Welty et al. 1996; Welty \& Hobbs 2001).}
\end{figure*}

\begin{figure*}[!t]
\centering
\includegraphics[width=0.95\textwidth]{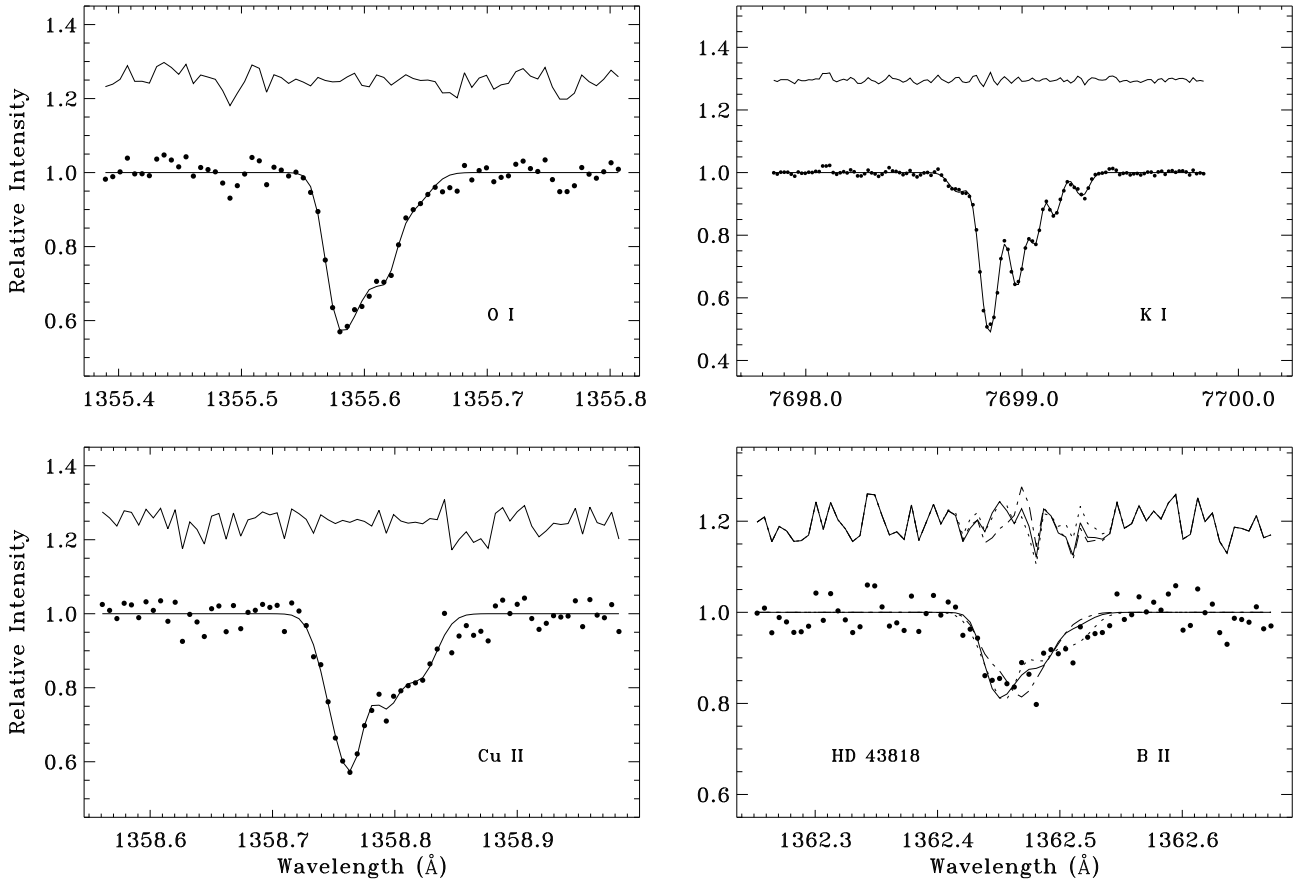}
\caption[Profile synthesis fits for O~{\scriptsize I}, Cu~{\scriptsize II}, 
K~{\scriptsize I}, and B~{\scriptsize II} toward HD~43818]{Same as Figure~5 
except for the O~{\scriptsize I}, Cu~{\scriptsize II}, K~{\scriptsize I}, and 
B~{\scriptsize II} lines toward HD~43818. No Ga~{\scriptsize II} data are 
available for this sight line. High-resolution data on Ca~{\scriptsize II} are 
available from previous McDonald observations.}
\end{figure*}

\begin{figure*}[!t]
\centering
\includegraphics[width=0.95\textwidth]{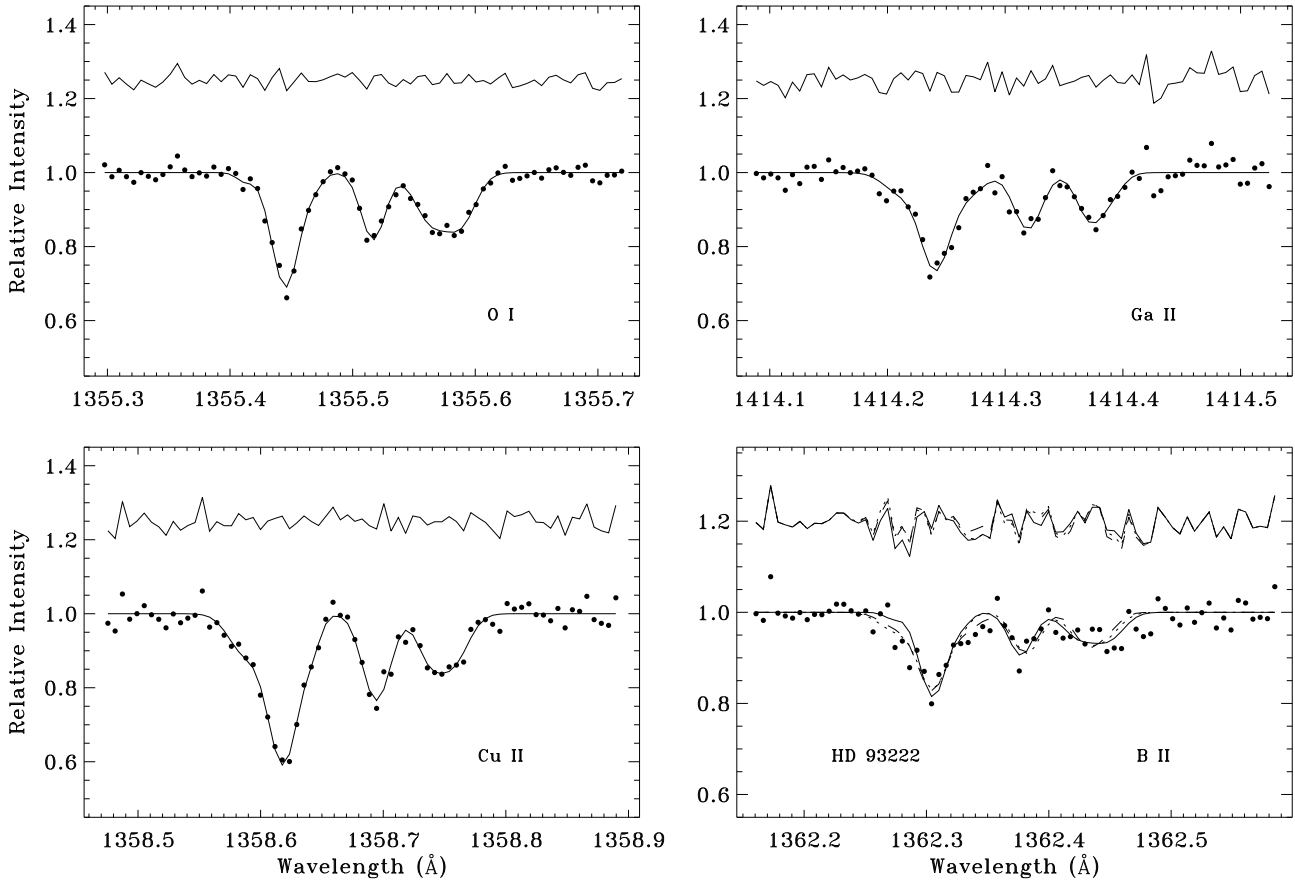}
\caption[Profile synthesis fits for O~{\scriptsize I}, Cu~{\scriptsize II}, 
Ga~{\scriptsize II}, and B~{\scriptsize II} toward HD~93222]{Same as Figure~5 
except for the O~{\scriptsize I}, Cu~{\scriptsize II}, Ga~{\scriptsize II}, and 
B~{\scriptsize II} lines toward HD~93222. No high-resolution data on 
Ca~{\scriptsize II} or K~{\scriptsize I} are available for this sight line. In 
fitting the B~{\scriptsize II} profile, the various templates were applied to 
the three absorption complexes, independently.}
\end{figure*}

\begin{figure*}[!t]
\centering
\includegraphics[width=0.95\textwidth]{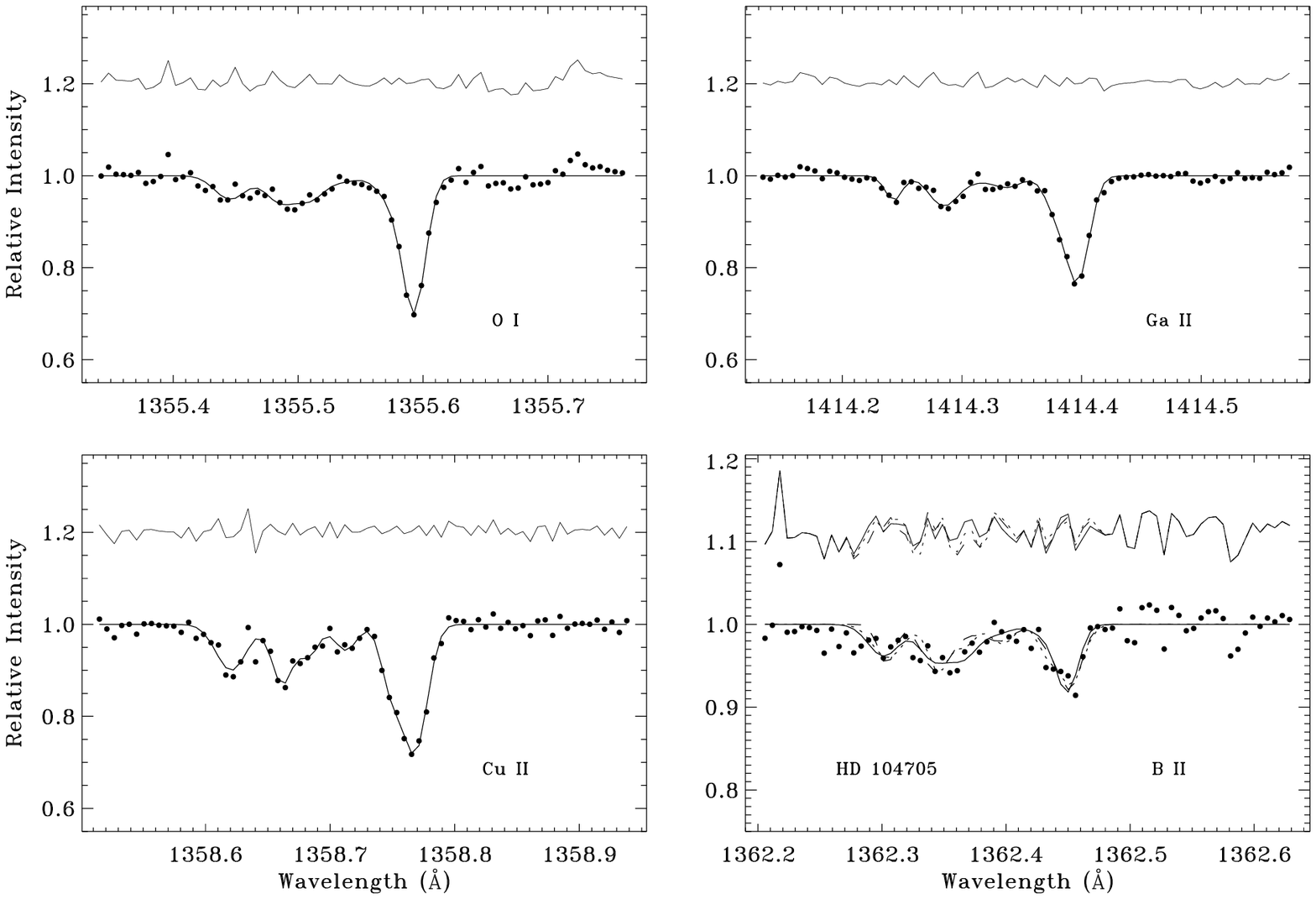}
\caption[Profile synthesis fits for O~{\scriptsize I}, Cu~{\scriptsize II}, 
Ga~{\scriptsize II}, and B~{\scriptsize II} toward HD~104705]{Same as Figure~5 
except for the O~{\scriptsize I}, Cu~{\scriptsize II}, Ga~{\scriptsize II}, and 
B~{\scriptsize II} lines toward HD~104705. No high-resolution data on 
Ca~{\scriptsize II} or K~{\scriptsize I} are available for this sight line. The 
fit to the B~{\scriptsize II} profile assumes two independent absorption 
complexes.}
\end{figure*}

\begin{figure*}[!t]
\centering
\includegraphics[width=0.95\textwidth]{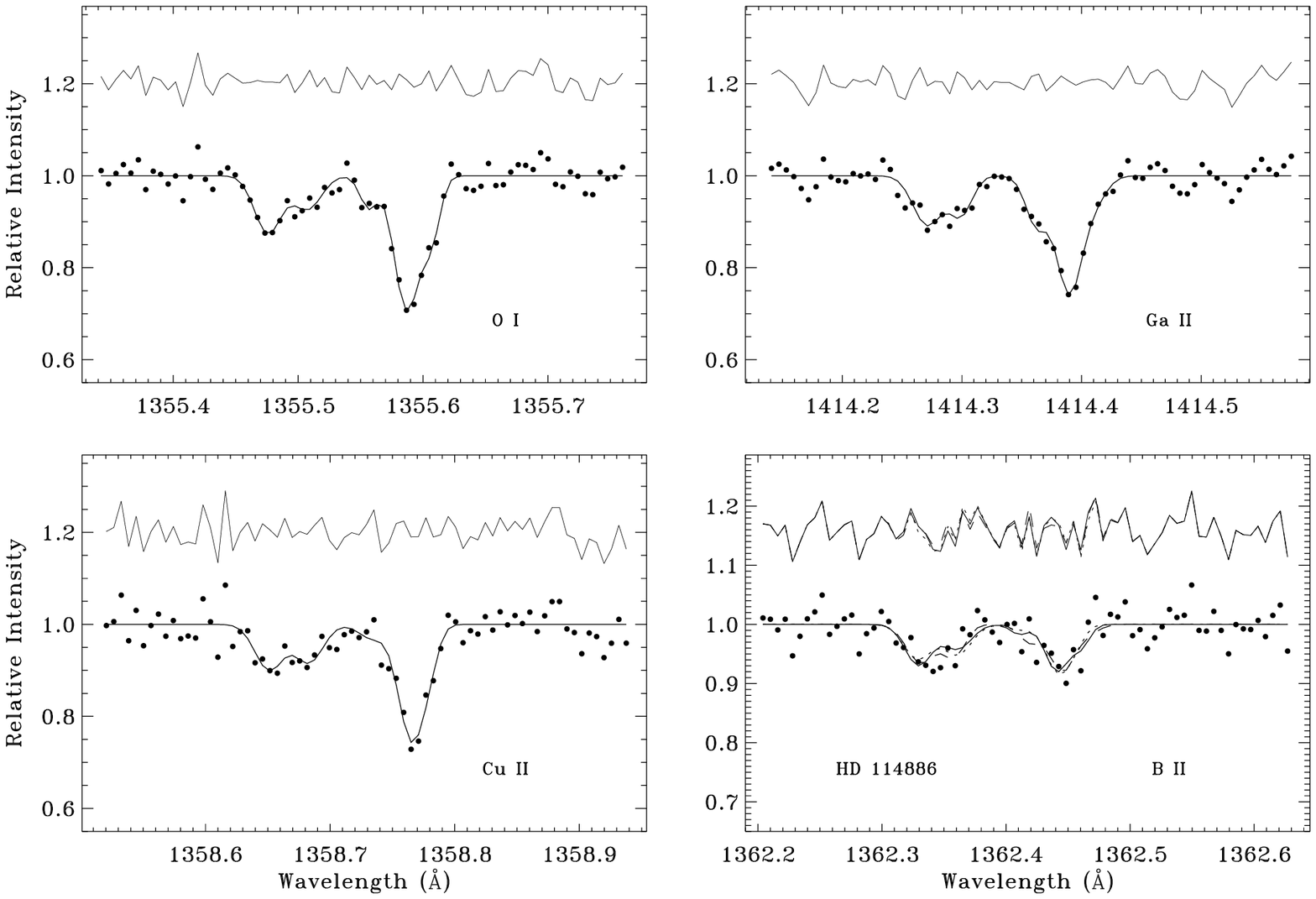}
\caption[Profile synthesis fits for O~{\scriptsize I}, Cu~{\scriptsize II}, 
Ga~{\scriptsize II}, and B~{\scriptsize II} toward HD~114886]{Same as Figure~5 
except for the O~{\scriptsize I}, Cu~{\scriptsize II}, Ga~{\scriptsize II}, and 
B~{\scriptsize II} lines toward HD~114886. No high-resolution data on 
Ca~{\scriptsize II} or K~{\scriptsize I} are available for this sight line. The 
fit to the B~{\scriptsize II} profile assumes two independent absorption 
complexes.}
\end{figure*}

\begin{figure*}[!t]
\centering
\includegraphics[width=0.95\textwidth]{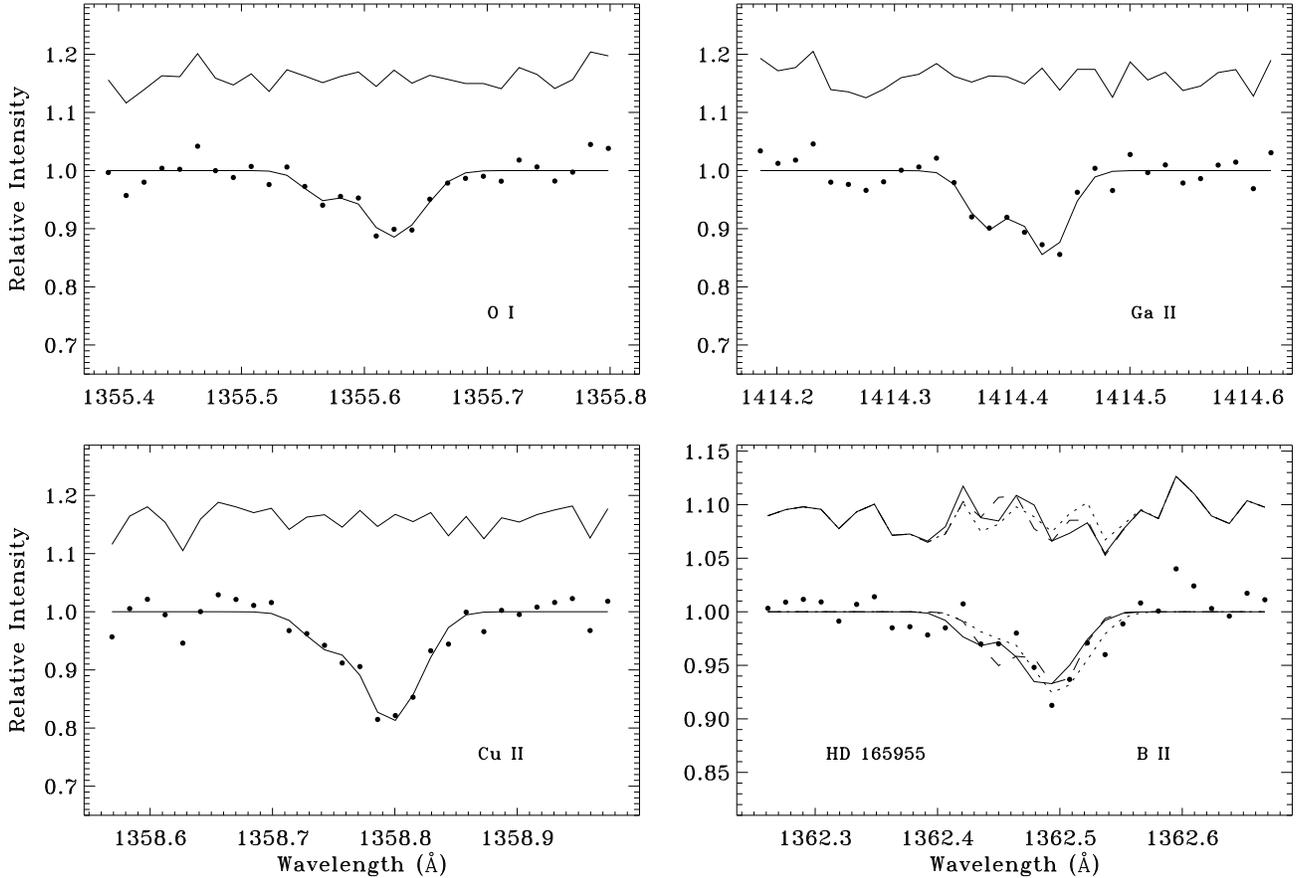}
\caption[Profile synthesis fits for O~{\scriptsize I}, Cu~{\scriptsize II}, 
Ga~{\scriptsize II}, and B~{\scriptsize II} toward HD~165955]{Same as Figure~5 
except for the O~{\scriptsize I}, Cu~{\scriptsize II}, Ga~{\scriptsize II}, and 
B~{\scriptsize II} lines toward HD~165955. These data were acquired at medium 
resolution. No high-resolution data on Ca~{\scriptsize II} or K~{\scriptsize I} 
are available for this sight line.}
\end{figure*}

\begin{figure*}[!t]
\centering
\includegraphics[width=0.95\textwidth]{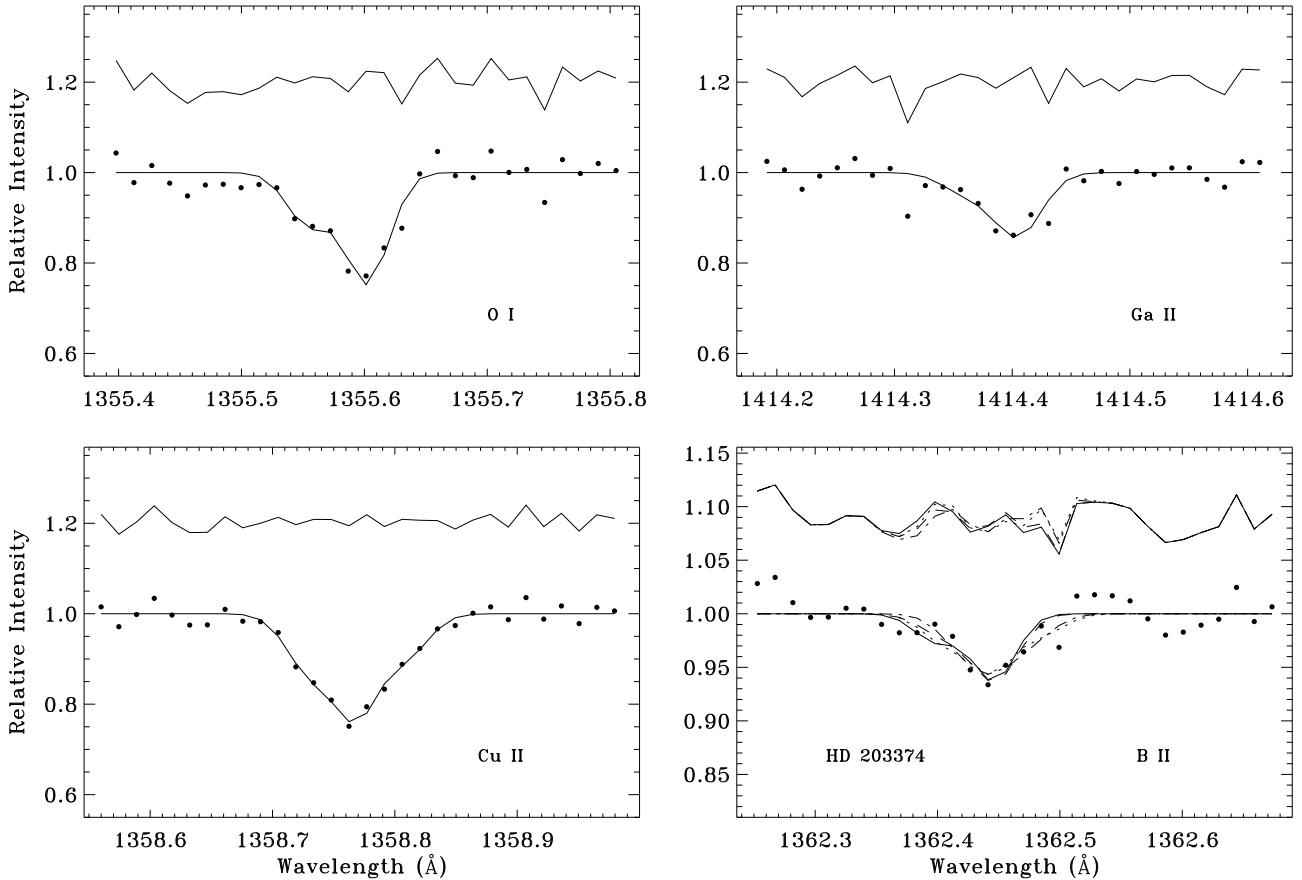}
\caption[Profile synthesis fits for O~{\scriptsize I}, Cu~{\scriptsize II}, 
Ga~{\scriptsize II}, and B~{\scriptsize II} toward HD~203374]{Same as Figure~5 
except for the O~{\scriptsize I}, Cu~{\scriptsize II}, Ga~{\scriptsize II}, and 
B~{\scriptsize II} lines toward HD~203374. These data were acquired at medium 
resolution. High-resolution data on Ca~{\scriptsize II} and K~{\scriptsize I} 
are available for this sight line from the literature (Pan et al. 2004).}
\end{figure*}

\begin{figure*}[!t]
\centering
\includegraphics[width=0.95\textwidth]{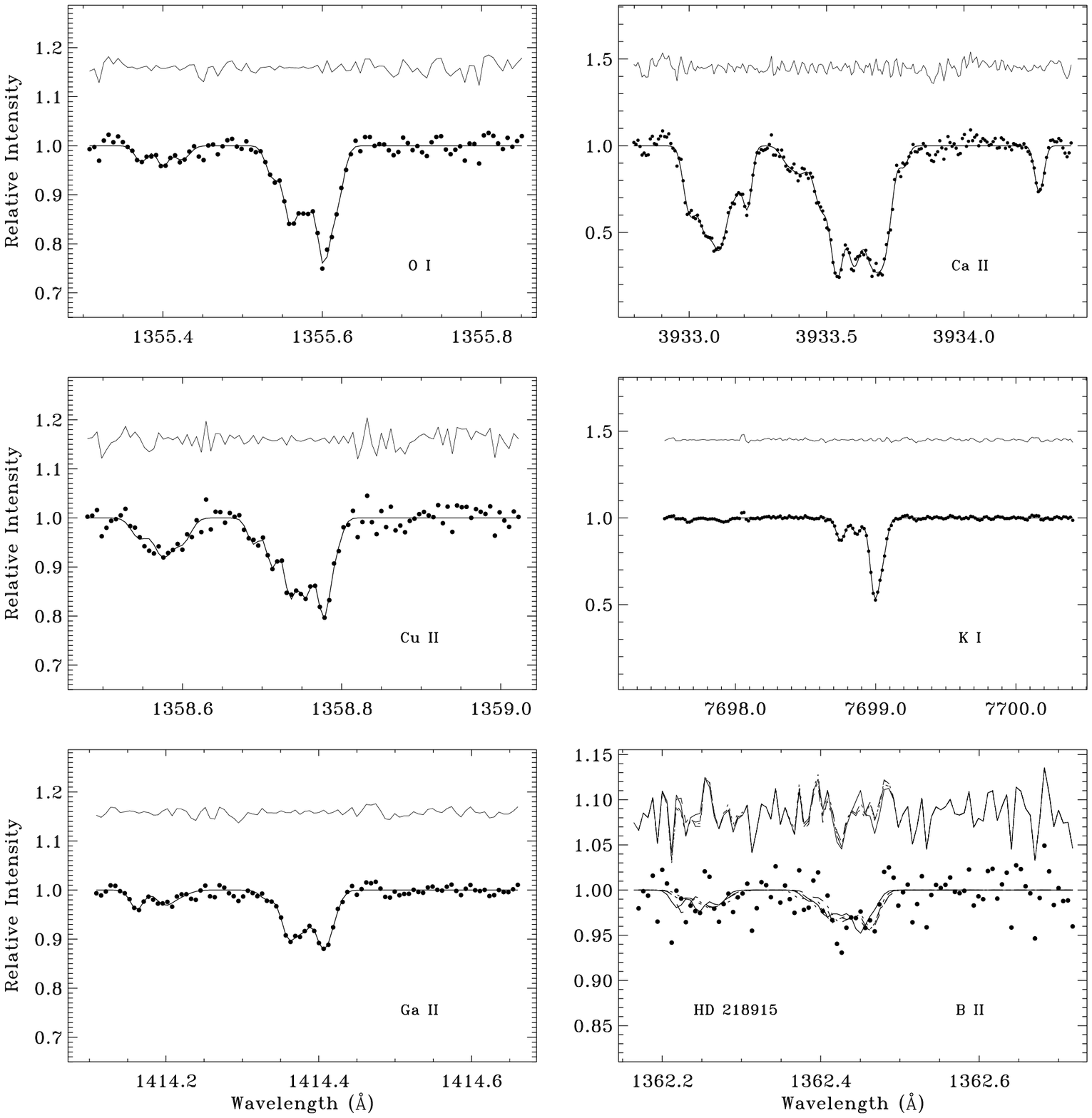}
\caption[Profile synthesis fits for O~{\scriptsize I}, Cu~{\scriptsize II}, 
Ga~{\scriptsize II}, Ca~{\scriptsize II}, K~{\scriptsize I}, and 
B~{\scriptsize II} toward HD~218915]{Same as Figure~5 except for the 
O~{\scriptsize I}, Cu~{\scriptsize II}, Ga~{\scriptsize II}, 
Ca~{\scriptsize II}, K~{\scriptsize I}, and B~{\scriptsize II} lines toward 
HD~218915. The fit to the B~{\scriptsize II} profile assumes two independent 
absorption complexes. Note the very weak K~{\scriptsize I} components tracing 
the Perseus spiral arm.}
\end{figure*}

\subsection{Profile Synthesis}

Synthetic absorption profiles were constructed and then fitted to the observed 
profiles using the rms-minimizing code ISMOD (Y. Sheffer, unpublished), which 
assumes a Voigt profile function for each absorption component comprising the 
fit. The Voigt components are essentially Gaussians in practice, because the 
absorption lines studied here are too weak for the radiatively damped wings to 
contribute significantly to the shape of the profile. In order to duplicate the 
broadening that results from the finite resolving power of the spectrograph, 
the synthetic spectra were convolved with an instrumental profile, represented 
by a Gaussian with a width corresponding to the resolution of the data, before 
being fit to the observations. The profile synthesis routine treats the 
velocity $v$, the Doppler parameter $b$, and the column density $N$ of each 
component as a free parameter, though the number of free parameters can be 
reduced in cases where the data are of lesser quality. The velocities and 
column densities were computed using atomic data (i.e., wavelengths and 
oscillator strengths) from Morton (2003), which we list in Table~4 for 
reference.

Preliminary examination of the observed profiles within IRAF provided the 
component structure that served as the initial input to the profile fitting 
routine. All of the UV spectra for a particular sight line were analyzed 
together to derive a consistent component decomposition across species. 
Component results for different species were deemed to be consistent with each 
other if the derived velocities agreed to within approximately 1/3 of a 
resolution element. The measured line widths were also expected to be similar 
from one component to the next and from one species to another (at least for 
the dominant ions). Before accepting a particular decomposition, the residual 
intensity was examined to ensure that it was indistinguishable from the noise 
in the continuum. We stress the importance of obtaining an accurate solution 
for the component structure along each line of sight. The overall weakness of 
the B~{\small II} absorption line means that reliable B~{\small II} column 
densities cannot be obtained through direct synthesis of the absorption 
profiles due to the effects of noise in the spectra. In this investigation, as 
in our previous work (Federman et al. 1996a, Lambert et al. 1998), column 
densities of B~{\small II} are derived by fitting the absorption profiles with 
fixed templates of interstellar component structure constructed from the other 
observed species. Thus, obtaining consistent component results is a crucial 
step in the analysis.

\subsubsection{Templates from UV Data}

Once a preliminary component decomposition was obtained for a particular line 
of sight, the absorption profiles of O~{\small I}, Cu~{\small II}, and 
Ga~{\small II} were synthesized, allowing the initial input parameters to vary 
freely. The same initial component structure, usually that obtained from the 
O~{\small I} profile, was applied to all three lines. Typically, multiple 
iterations were required to bring the component velocities for the different 
species into satisfactory agreement. In some cases, it was necessary to add or 
remove components from the fit or to hold the relative velocities, or, in rarer 
circumstances, the absolute velocities, fixed. Additionally, the derived 
$b$-values were expected to fall within certain limits characteristic of 
dominant ion species (i.e., $0.5\lesssim b \lesssim3.5$; see Pan et al. 2004). 
The majority of fits, especially of the higher resolution data, naturally 
yielded acceptable $b$-values. When the $b$-value for a particular component 
fell outside the expected range, a maximum or minimum value, either estimated 
directly from the data or determined from high-resolution Ca~{\small II} 
spectra, was strictly enforced by the fit. Since restrictions on line width can 
affect the velocity at line center, both parameters must be considered with 
each iteration until a satisfactory solution is found yielding realistic 
$b$-values and consistent component velocities in each species. The general 
philosophy was to adopt the smallest number of components producing an 
acceptable fit.

Every effort was made to retain the maximum number of free parameters in the 
profile synthesis fits. In cases where this was not possible, the restrictions 
placed on the fits were justified by the quality of the data. As an example, 
consider spectra obtained with the E140M grating at a resolution of 
7.9~km~s$^{-1}$. Essentially all Galactic sight lines have multiple absorption 
components that can be separated by as little as a few km~s$^{-1}$ or less 
(e.g., Welty \& Hobbs 2001). In lower resolution data, the presence of such 
components is often revealed by asymmetries in the absorption profiles. If 
accurate column densities are to be obtained, it is necessary to include each 
component. However, if an absorption component is defined by three parameters 
and there are three or four components that make up the profile, it is possible 
in E140M data for there to be more free parameters in the fit than there are 
pixels sampling the profile. When this type of situation was encountered, the 
techniques described above effectively reduced the number of free parameters to 
an appropriate level so as to avoid overfitting the data. Still, the $b$-values 
obtained from medium resolution spectra are not particularly well constrained, 
though this likely has minimal impact on the derived column densities because 
the lines are relatively weak and tend to lie on the linear portion of the 
curve of growth.

After the O~{\small I}, Cu~{\small II}, and Ga~{\small II} syntheses were 
complete, the derived values of $v_{\mathrm{LSR}}$, $N$, and $b$ for individual 
components (see Table~5) were used to create three profile templates, one from 
each species. The B~{\small II} absorption profile was then fitted with each 
template, holding the relative velocities, the relative component strengths, 
and the $b$-values constant. The only free parameters in the B~{\small II} fits 
were the absolute velocity of the template, which was allowed to shift to 
account for any small velocity offsets due to misaligned spectra, and the total 
column density. In some instances, when either a low S/N ratio or coarse 
resolution prevented a reliable determination of velocity, only the total 
column density was permitted to vary. If a sight line displayed multiple 
complexes of absorption components well separated in velocity, the various 
templates were created for each complex and fitted to that portion of the 
B~{\small II} profile, independently. For most sight lines in this category, 
two groups of templates were sufficient to fit the B~{\small II} profile. 
However, four lines of sight were found to exhibit distinct absorption from 
three separate cloud complexes. Three of these (HD~93205, HD~93222, and 
CPD$-$59 2603) are toward stars in the Carina Nebula, and the fourth (HD~1383) 
lies behind the Perseus spiral arm. The absorption profiles toward HD~93222 are 
perhaps the most striking, with three clearly-defined groups of components 
almost completely resolved from one another (see Figure~4). For this sight 
line, B~{\small II} column densities can confidently be derived for each 
distinct group, in contrast to the situation with the other sight lines, where 
not all of the individual absorption complexes are strongly detected in 
B~{\small II}.

\subsubsection{Templates from Visible Data}

The Ca~{\small II}~K and K~{\small I} profiles acquired at McDonald Observatory 
were synthesized with ISMOD in a manner similar to that in which the 
O~{\small I}, Cu~{\small II}, and Ga~{\small II} data were analyzed. The 
component structure found from direct examination of the spectra within IRAF 
again served as the initial input to the routine. However, the same structure 
was not applied to both profiles as it is recognized that Ca~{\small II} will 
have many more components than K~{\small I} and the K~{\small I} line widths 
will be, on average, somewhat narrower (e.g., Welty et al. 1996; Welty \& Hobbs 
2001; Pan et al. 2004, 2005). Multiple iterations were performed until the 
syntheses met all of the usual criteria. That is, the fits were expected to 
yield not only consistent velocities for components common to both 
Ca~{\small II} and K~{\small I}, but also realistic $b$-values appropriate for 
the particular species. In practice, the K~{\small I} profiles were synthesized 
first because the S/N was higher and the narrow components were well resolved 
allowing accurate velocities to be determined. These velocities were then 
incorporated into fitting the more blended and more optically thick 
Ca~{\small II} profiles. Since the exposures containing the Ca~{\small II}~K 
line also provided data on Ca~{\small I}, CH$^+$, CH, and CN, these species 
were analyzed along with K~{\small I} and Ca~{\small II}, in an effort to 
further constrain the line-of-sight component structure. The weaker species can 
provide valuable information on the strongest K~{\small I} and Ca~{\small II} 
components (see Pan et al. 2005 for a detailed discussion), though the 
usefulness of these lines in the present data is diminished due to the low S/N 
in exposures taken with the blue setup. Ultimately, the K~{\small I} profiles 
themselves provided the most robust constraints on the cloud component 
structure.

For each of the 24 sight lines that possessed the necessary measurements on 
Ca~{\small II} and K~{\small I}, a ``high-resolution'' template was created to 
be applied to the B~{\small II} line. Individual values of $v_{\mathrm{LSR}}$, 
$N$, and $b$ came either from profile synthesis fits to McDonald observations 
(see Table~6), or from published values in literature studies of comparable 
quality (i.e., Welty et al. 1996; Welty \& Hobbs 2001; Sonnentrucker et al. 
2003; Pan et al. 2004; Sonnentrucker et al. 2007). In general, velocities from 
K~{\small I} were chosen as the template velocities due to the precision with 
which these values can be determined. The $b$-values and relative column 
densities of the template components, however, were taken from the 
corresponding components in Ca~{\small II}, as this species is expected to 
track more closely the (relative) strengths and widths of absorption lines from 
dominant ions (Welty et al. 1996). The number of components in each template 
depended on the range of velocities spanned by the UV profiles. Typically, all 
components seen in K~{\small I} were included, though, in rare cases, a minor 
component on the periphery of the K~{\small I} profile was removed. More 
commonly, it was necessary to add Ca~{\small II} components not detected in 
K~{\small I} in order to fill in extra absorption at the edges of the UV 
profiles.

The high-resolution templates from visible data were fitted to the 
B~{\small II} lines exactly as were the UV templates, applying separate 
template fits for profiles with multiple complexes of absorption components. In 
addition, the O~{\small I}, Cu~{\small II}, and Ga~{\small II} lines were 
fitted with the same high-resolution component structure to check the validity 
of the method. These fits showed some variation between the shape of the 
templates derived from visible data and the actual profiles of the stronger UV 
species, but the resulting total column densities were very similar to those 
based on direct profile synthesis. More aesthetically pleasing fits were 
obtained by allowing the $b$-values and column densities of the high-resolution 
components to vary while holding the relative velocities from the template 
fixed. Here, the total column densities were virtually indistinguishable from 
those based on direct synthesis (see Table~7), demonstrating that unresolved 
component structure has very little influence on the column densities derived 
from the lower resolution UV spectra. A sample of profile synthesis fits to the 
UV and visible data for nine sight lines is presented in Figures~5$-$13.

\subsubsection{Final Column Densities}

Table~7 lists the total (line-of-sight) column densities of O~{\small I}, 
Cu~{\small II}, and Ga~{\small II} obtained by direct profile synthesis 
(\S{}~3.2.1) and by applying fixed velocity templates from visible data as 
described above. For these species, the column densities resulting from direct 
profile synthesis are adopted for further analysis since there are a larger 
number of sight lines lacking visible data. Uncertainties in $N$ for individual 
components were derived from the fitted line widths and the rms variations in 
the continuum. The individual uncertainties for all of the components in a 
given absorption profile were then summed in quadrature to yield the error in 
total column density. Sight lines with two row entries in Table~7 have data at 
both grating setups (i.e., E140M and E140H). For each of these sight lines, the 
total column densities obtained by synthesizing the medium-resolution and 
high-resolution profiles are found to agree within their mutual 1-$\sigma$ 
uncertainties. We adopt the medium-resolution results here, because the E140H 
data either do not include the B~{\small II} line or have lower S/N in the 
vicinity of B~{\small II} compared to E140M. Adopting the E140M results for all 
species in these instances ensures consistency when comparing different 
elemental abundances along a particular line of sight.

The total B~{\small II} column densities resulting from the various profile 
template fits are also given in Table~7. Final values of $N$(B~{\small II}) 
were determined by taking the unweighted mean of the available results for a 
given sight line, since the uncertainties associated with each result are based 
simply on the noise characteristics in the spectrum (along with the line widths 
and number of components in the profile template) and are not related to any 
goodness-of-fit parameter. In every case, the dispersion in the various 
determinations of $N$(B~{\small II}) is found to be less than the uncertainties 
in the individual results, with the average dispersion equal to 0.02~dex. None 
of the templates yield systematically higher or lower column densities compared 
to the mean value of $N$(B~{\small II}) for a specific line of sight, giving us 
confidence in our overall methodology. In the final analysis, the B~{\small II} 
column densities that we obtain toward eight stars [HD~108002, HD~111934, 
HD~124314, HD~147683, HD~195965, HD~198478 (55~Cyg), HD~210809, and HD~212791] 
each have a statistical significance of less than 2~$\sigma$ (i.e., 
$N$/$\sigma_N<2$). When discussing elemental abundances in \S{} 4.1, the boron 
column densities in these directions are given as 3-$\sigma$ upper limits. 
Another ten sight lines have only tentative detections of the B~{\small II} 
line, which result in column densities that are between 2 and 2.5 times the 
associated uncertainties. However, 37 sight lines from the STIS sample exhibit 
B~{\small II} absorption at the 2.5-$\sigma$ level or better, and along eight 
of these [HD~1383, HD~23180 ($o$~Per), HD~24532 (X~Per), HD~43818 (11~Gem), 
HD~93222, HD~165955, HD~177989, and CPD$-$59~2603] the B~{\small II} line is 
detected with a significance of greater than 4.5~$\sigma$.

\begin{figure*}[!t]
\centering
\includegraphics[width=0.8\textwidth]{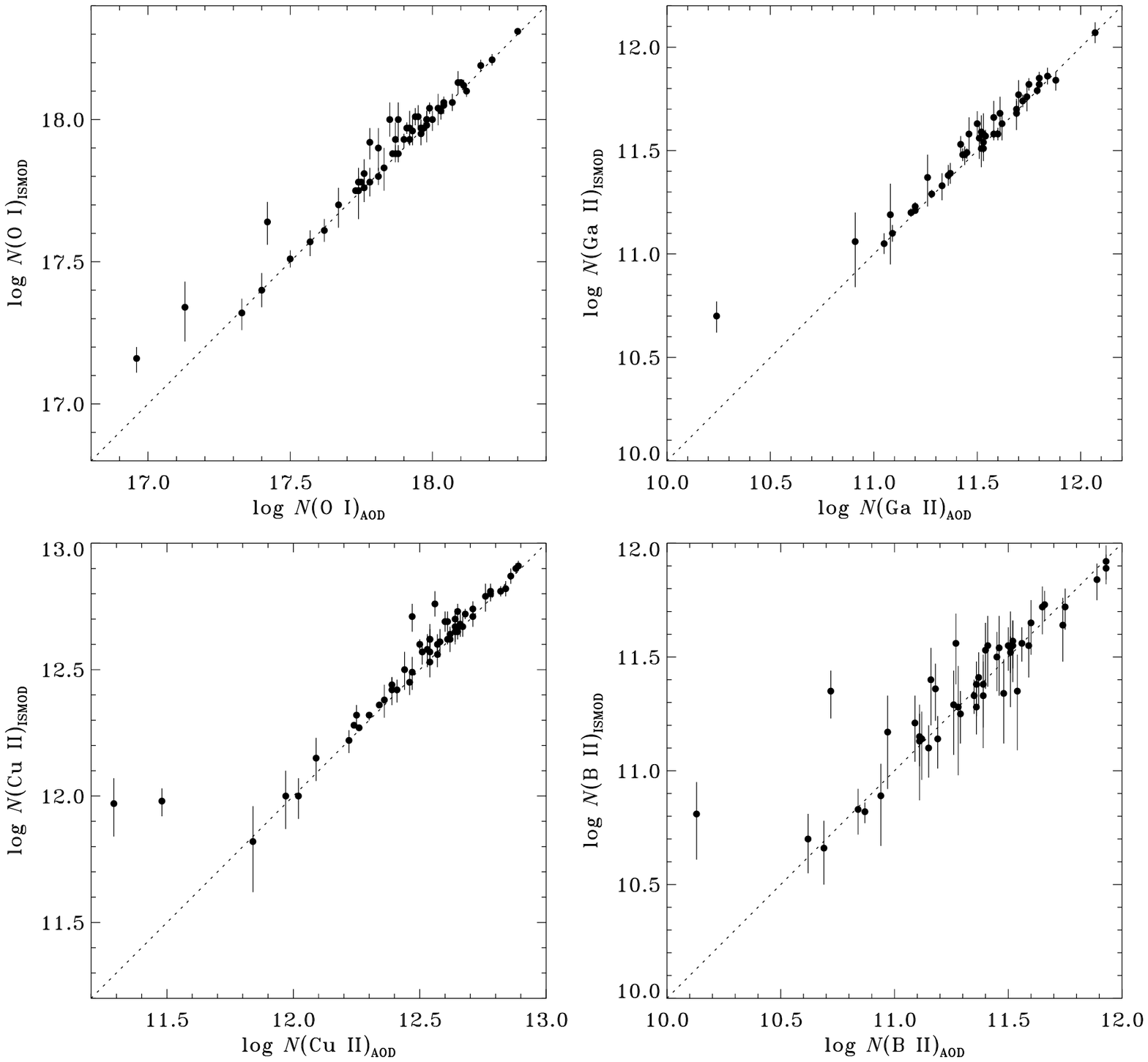}
\caption[Comparison between profile synthesis and AOD results]{Comparison 
between column densities of O~{\scriptsize I}, Cu~{\scriptsize II}, 
Ga~{\scriptsize II}, and B~{\scriptsize II} derived through profile synthesis 
(labeled ``ISMOD'') and those obtained by integrating the apparent column 
density profiles (labeled ``AOD''). The dotted line represents an exact 
correspondence. The data points are expected to lie on or above this line since 
the AOD calculations may yield only a lower limit to the true column density of 
the material. The discrepant points at low column density correspond to 
HD~36841, HD~93237, and HD~187311, all of which show narrow lines poorly 
sampled by the observations, leading to errors in the integrated column 
densities.}
\end{figure*}

In Figure~14, we give a comparison between the final column densities derived 
through profile synthesis and those obtained by integrating the apparent column 
density profiles (\S{} 3.1). The close correspondence between the results of 
the two methods is evident, even though the apparent column densities have not 
been corrected for any optical depth effects, confirming that the majority of 
lines are optically thin. For O~{\small I}, Cu~{\small II}, and Ga~{\small II}, 
virtually no data points lie below the ``line of equality'', as expected if the 
apparent column density represents a lower limit to the true column density of 
the material. For B~{\small II}, however, the points are more randomly 
distributed since the B~{\small II} profiles are optically thin yet dominated 
by noise to a larger extent than are the other UV lines. The discrepant points 
at low column density correspond to the same group of sight lines in each panel 
(i.e., HD~36841, HD~93237, and HD~187311). These are the only sight lines in 
the STIS sample where the fits employed single absorption components for all 
species. In each discrepant case, it is the integration of the profile that 
fails because there are too few pixels sampling the line (and the pixels sample 
only the core of the line and not the wings). The reliability of the profile 
synthesis fits should not be affected by these difficulties.

\subsubsection{Sight Lines with Multiple Absorption Complexes}

Table~8 presents the mean B~{\small II} results for sight lines where separate 
templates were fit to distinct cloud complexes seen in the absorption profiles. 
The sight lines are divided into those that probe the Perseus spiral arm and 
those that trace the Sagittarius-Carina spiral feature. Unless otherwise noted, 
the velocity given for a particular absorption complex is the column density 
weighted mean LSR velocity in O~{\small I} for that group of components. We 
list 3-$\sigma$ upper limits for any complexes with derived column densities 
below the 2-$\sigma$ level. For HD~13268 and HD~13745, the complexes at 
$v_{\mathrm{LSR}}$~=~$-$36 and $-$42~km~s$^{-1}$, respectively, which are seen in 
the other observed species and are presumably associated with gas in the 
Perseus arm, could not be discerned in the B~{\small II} profiles. Thus, no 
fits were attempted for these features. Similarly, the nominal fits to the 
absorption complexes at $v_{\mathrm{LSR}}$~=~$-$24 and $-$38 km~s$^{-1}$ toward 
HD~116781 and HDE~303308, respectively, yielded column densities that were 
smaller than the uncertainties. In these latter two cases, it was clear that 
the fits did not go deep enough to match the observed flux level, a result of 
excessive noise in the spectra. In order to correct the total B~{\small II} 
column densities along each of these sight lines, a constant 
$N$(B~{\small II})/$N$(O~{\small I}) ratio was assumed for the entire profile, 
consistent with the noise at the expected positions of the ``missing'' 
components. Two sight lines in Table~8 (HD~1383 and CPD$-$59~2603) have strong 
B~{\small II} components without significant detections of O~{\small I} at 
similar velocities. The component at $v_{\mathrm{LSR}}$~=~$-$46 km~s$^{-1}$ 
toward CPD$-$59~2603 traces the near side of the globally expanding 
H~{\small II} region associated with the Carina Nebula (see Walborn et al. 
1998, 2002). The $-$23 km~s$^{-1}$ component toward HD~1383 probably also 
arises in an H~{\small II} region along the line of sight, considering that 
the strength of the component in the UV absorption profiles examined here 
decreases with the ionization potential of the species. 

In the context of Galactic differential rotation, interstellar absorption 
detected near $v_{\mathrm{LSR}}$~=~0~km~s$^{-1}$ can reliably be ascribed to the 
local ISM in most cases (excluding sight lines in the direction of the Galactic 
center or anticenter), whereas absorption components with higher absolute 
velocities typically originate in more distant gas. Each of the sight lines in 
Table~8 tracing the Sagittarius-Carina spiral arm has an absorption complex 
near $v_{\mathrm{LSR}}$~=~$-$30 km s$^{-1}$. The velocity of the feature is seen 
to increase roughly monotonically with increasing Galactic longitude, from 
$v_{\mathrm{LSR}}\approx-33$ km~s$^{-1}$ at $l=288^{\circ}$ to $-$18 km~s$^{-1}$ at 
$l=336^{\circ}$. Such systematic variation confirms that we are observing a 
coherent structure and is consistent with parcels of gas moving on circular 
orbits within a spiral arm. Since each sight line also exhibits absorption near 
0~km~s$^{-1}$, it is possible to compare the elemental B/O ratios\footnote{For 
the remainder of this paper, we assume that an elemental abundance ratio is 
equivalent to the column density ratio of the two elements in their dominant 
ionization stages. For example, we define the logarithmic boron-to-oxygen ratio 
as 
log~(B/O)~$\equiv$~log~$N$(B~{\scriptsize II})~$-$~log~$N$(O~{\scriptsize I}).} 
in the Sagittarius-Carina arm with those in the local ISM for the same 
directions. The mean value of log (B/O) for the detected Sagittarius-Carina arm 
components is $-6.3\pm0.2$, while the average for local gas along the same 
lines of sight is $-6.5\pm0.1$. We therefore find suggestive evidence that the 
B/O ratio is higher (by $\sim0.2$ dex) in the inner Sagittarius-Carina spiral 
arm than it is in the vicinity of the Sun, though the scatter in the 
measurements, particularly for the spiral arm components, precludes a 
definitive interpretation of the results. If confirmed, however, an elevated 
B/O ratio in the inner Galaxy would seem to indicate that boron is currently 
produced as a secondary element. The abundances of secondary elements increase 
relative to those of primary ones with decreasing Galactocentric radius due to 
the enhanced rates of star formation and stellar nucleosynthesis in the inner 
disk. A secondary behavior for boron would then cast doubt on the efficiency of 
the $\nu$-process as this is a primary production mechanism.

Considering the implications of this finding, it would be useful to compare the 
mean B/O ratios given above with a similar mean for gas in the Perseus spiral 
arm, since this structure contains clouds at larger Galactocentric radii. 
Absorption from the Perseus arm is found near $v_{\mathrm{LSR}}$~=~$-$40 
km~s$^{-1}$ along each of the sight lines in Table~8 sampling this structure. 
Unfortunately, the data yield only upper limits on the B~{\small II} column 
densities for these absorption complexes. There are, of course, other effects 
that can lead to variations in B/O among different absorption complexes along a 
given line of sight, such as differential ionization and grain destruction. In 
fact, Howk et al. (2000) interpreted the UV absorption profiles toward 
HD~104705 as showing increased boron depletion in cold clouds (the narrow 
complex near $v_{\mathrm{LSR}}$~=~0~km~s$^{-1}$) relative to warmer diffuse gas 
(the complex centered at $v_{\mathrm{LSR}}$~=~$-$25~km~s$^{-1}$). However, it is 
unclear how such an effect would lead to a systematic increase in the B/O ratio 
for components associated with the Sagittarius-Carina spiral arm. Additionally, 
there are no similar enhancements in the Cu/O or Ga/O ratios for 
Sagittarius-Carina arm components as would be expected if differences in 
depletion were the cause.

\subsection{Comparison with Previous Studies}

Of the 56 sight lines in the STIS sample, 32 have previously been analyzed for 
the purpose of deriving O~{\small I} column densities, and 16 of these have 
also been studied for Cu~{\small II}. An extensive comparison of the 
O~{\small I} and Cu~{\small II} column densities derived along these lines of 
sight is presented in Table~9. Since most of the previous investigations (e.g., 
Howk et al. 2000; Andr\'e et al. 2003; Cartledge et al. 2004, 2006; Jensen et 
al. 2005) used the same STIS datasets examined here, the generally good 
agreement that can be seen between our results and those from the literature is 
perhaps not surprising. Significant discrepancies (i.e., greater than 
3~$\sigma$) are found for only six sight lines, and each of the discrepancies 
is in the O~{\small I} column density. The value given by Snow et al. (1998) 
for $N$(O~{\small I}) toward X~Per, obtained from GHRS data, is 0.24 dex higher 
than our more precise result from STIS spectra. Our value agrees with the more 
recent determinations by Knauth et al. (2003a) and Jensen et al. (2005), which 
were also made using STIS. In their analysis, Snow et al. (1998) assume a 
$b$-value of 1.0 km s$^{-1}$ for the O~{\small I} line, whereas we model the 
profile using three components with $b$-values of 1.7, 1.1, and 1.5 
km~s$^{-1}$. Since the total equivalent widths in the two cases are virtually 
identical, an overcorrection for optical depth on the part of Snow et al. 
(1998) suffices to explain the discrepancy. Jensen et al. (2005) derive an 
O~{\small I} column density toward HD~207538, from a curve-of-growth analysis 
of three O~{\small I} lines including the weak line at 1355~\AA, that is 0.19 
dex higher than our profile synthesis result. While their effective $b$-value 
of 5.4 km s$^{-1}$ seems reasonable for the overall width of the 1355 \AA{} 
line, the column density they obtain implies a large optical depth correction 
for this line, which seems unlikely. Furthermore, our result for 
$N$(O~{\small I}) toward HD~207538 is more consistent with the O~{\small I} 
column densities that we derive along nearby sight lines in Cep OB2.

Cartledge et al. (2004), using a profile synthesis approach, obtain a result 
for $N$(O~{\small I}) toward HD~122879 that is larger than ours by 0.21 dex. In 
this case, a difference in continuum normalization led Cartledge et al. (2004) 
to include components in their fit at intermediate velocities between the two 
main absorption complexes (HD~122879 being one of the sight lines probing the 
Sagittarius-Carina spiral arm; see Table~8). We did not include components at 
these velocities because the features are indistinguishable from the noise 
based on our placement of the continuum and similar components are not seen in 
the absorption profile of either Cu~{\small II} or Ga~{\small II}. For our 
investigation, it was essential to perform the continuum normalization and 
profile synthesis procedures on multiple species, simultaneously, since our 
main objective was to derive a consistent component structure in the species 
used as templates for the B~{\small II} line. Still, Cartledge et al. (2006) 
find a larger value for $N$(Cu~{\small II}) as well along this line of sight 
(though it is just barely within 3~$\sigma$ of our result). These authors 
generally tend to include more components in their profile synthesis fits than 
we do here, leading them to derive slightly larger column densities in both 
O~{\small I} and Cu~{\small II} (see the results for HD~111934 in Table~9 as a 
further example). Again, it was our policy to adopt the fewest components that 
produced consistent results, so as not to overfit the spectra, which, in many 
cases, have only moderate S/N.

The other three significant discrepancies in $N$(O~{\small I}) arise in 
comparison with the results of Andr\'e et al. (2003), who determined final 
column densities through a combination of the AOD and profile fitting methods. 
The O~{\small I} column densities that these authors derive toward HD~93205, 
HD~210809, and HD~218915 are each smaller than the values we obtain (by 0.11, 
0.16, and 0.15~dex, respectively). For the latter two sight lines, Andr\'e et 
al. (2003) also find smaller O~{\small I} equivalent widths because they omit 
certain portions of the absorption profiles that are included in our syntheses. 
Our result for HD~210809 agrees with that of Cartledge et al. (2004) since 
these authors adopt a similar profile synthesis fit in this case. The 
components missing from the Andr\'e et al. (2003) analysis for HD~218915 are 
the large negative velocity components associated with the Perseus spiral arm. 
Howk et al. (2000) also overlook these components when integrating the apparent 
column density profiles of both O~{\small I} and B~{\small II}, although they 
derive an O~{\small I} column density that is identical to ours because they 
apply a saturation correction to their measurement. We report detections of the 
Perseus arm components toward HD~218915 at about the 2.5-$\sigma$ level in 
O~{\small I}, and at even greater significance levels in Ga~{\small II} and 
Cu~{\small II}. The components are also found in weak K~{\small I} absorption 
and are very prominent in the Ca~{\small II} profile (see Figure~13). For 
HD~93205, our O~{\small I} equivalent width ($W_{\lambda}=20.2\pm1.4$ m\AA) 
essentially agrees with the Andr\'e et al. (2003) value ($19.5\pm1.1$ m\AA), 
yet our column density [log~$N$(O~{\small I})~=~18.06] is approximately 30\% 
larger. It is unclear how Andr\'e et al. (2003) derive a column density as low 
as log~$N$(O~{\small I})~=~17.95 along this line of sight, since our value is 
only $\sim8$\% (0.03 dex) larger than the value given by the weak-line 
approximation. Our O~{\small I} column density for HD~93205 also leads to more 
consistent results among the four sight lines probing gas in the Carina Nebula.

As noted in \S{} 2, the B~{\small II} profiles for five of the STIS sight lines 
examined here were analyzed previously by Howk et al. (2000). In Table~10, we 
compare the B~{\small II} equivalent widths and column densities that we obtain 
for these sight lines with those reported in Howk et al. (2000). While there 
are some differences in the various determinations, the largest discrepancy 
(relative to the uncertainties) is only $\sim2$~$\sigma$. There are no 
systematic differences between the two studies and the variations in column 
density simply mirror the variations in equivalent width, as expected for very 
weak lines where corrections for optical depth are insignificant. Since Howk et 
al. (2000) used the AOD method to obtain final results, we also compared their 
values against our own integrated column densities and equivalent widths for 
both B~{\small II} and O~{\small I}. In this comparison, we find virtually the 
same minor discrepancies between the two studies, since our AOD results show 
little disagreement with our results from profile synthesis. The random 
variations in $N$(B~{\small II}) between our study and that of Howk et al. 
(2000) may reflect differences either in continuum placement or in the range of 
velocities over which the profiles were integrated (or fitted). As already 
mentioned, Howk et al. (2000) do not include absorption at large negative 
velocities in their integration for HD~218915, which results in a smaller 
B~{\small II} column density. For HDE~303308, they again exclude components 
with large negative velocities, but also they include additional absorption at 
more extreme positive velocities than are considered in our analysis. Overall, 
our results lead to reduced scatter in the B/O ratios for this small sample of 
five sight lines.

\section{ABUNDANCE ANALYSIS}

\subsection{Elemental Abundances}

The determination of an elemental abundance requires knowledge of the total 
column densities of atomic and molecular hydrogen along the line of sight. The 
majority of sight lines in the boron sample have H~{\small I} column densities 
provided by Diplas \& Savage (1994; hereafter DS94), who examined archival 
Ly$\alpha$ absorption data from the \emph{International Ultraviolet Explorer} 
(\emph{IUE}) satellite toward over 500 O and B-type stars out to 11~kpc. This 
study built upon earlier work by Bohlin et al. (1978; hereafter BSD78), who 
relied on \emph{Copernicus} observations of bright stars that were mostly 
within 1 kpc. A complementary survey with \emph{Copernicus} of molecular 
hydrogen in the $J$ = 0 and 1 rotational levels of the ground vibrational state 
was presented by Savage et al. (1977). Since these \emph{Copernicus} studies 
were restricted to bright targets, however, their applicability to the present 
investigation is limited. Fortunately, many of the sight lines in our sample 
not analyzed by DS94 have been examined more recently using \emph{HST}/STIS 
observations of H~{\small I} Ly$\alpha$ (e.g., Andr\'e et al. 2003; Cartledge 
et al. 2004). Results on molecular hydrogen are also readily obtainable from 
\emph{Far Ultraviolet Spectroscopic Explorer} (\emph{FUSE}) measurements of the 
Lyman ($B$$-$$X$) and Werner ($C$$-$$X$) bands (e.g., Rachford et al. 2002; Pan 
et al. 2005; Sheffer et al. 2008).

Table~11 compiles the atomic and molecular hydrogen column densities along all 
of the lines of sight in the boron sample for which data were available. The 
total hydrogen column density, 
$N_{\mathrm{tot}}$(H)~=~$N$(H~{\small I})~+~2$N$(H$_2$), is also given if the 
value could be reliably determined. Errors in $N_{\mathrm{tot}}$(H) were derived 
through standard error propagation using the reported uncertainties in 
$N$(H~{\small I}) and $N$(H$_2$). These errors are typically 20\% (0.08~dex). 
For one sight line (HD~99872), the adopted value of $N_{\mathrm{tot}}$(H) was 
obtained from the column density of Kr~{\small I}, assuming a constant 
interstellar krypton abundance equal to 
log~(Kr/H)$_{\mathrm{ISM}}$~=~$-9.02\pm0.02$ (Cartledge et al. 2008). For four 
other directions (HD~36841, HD~43818, HD~52266, and HD~208947), we based the 
values of $N$(H$_2$) on the amount of CH along the line of sight, according to 
the correlation between CH and H$_2$ given in Sheffer et al. (2008), which can 
be expressed as 
log~$N$(CH)~=~($0.97\pm0.07$)~log~$N$(H$_2$)~$-$~($6.80\pm1.50$). The quoted 
errors in $N$(H$_2$) for these sight lines reflect the scatter in the relation. 
Table~11 also lists two common sight-line parameters derived from the hydrogen 
data that characterize the density of the absorbing gas. These are the fraction 
of hydrogen in molecular form, defined as 
$f$(H$_2$)~=~2$N$(H$_2$)/$N_{\mathrm{tot}}$(H), and the average line-of-sight 
hydrogen density, given by 
$\langle n_{\mathrm{H}} \rangle$~=~$N_{\mathrm{tot}}$(H)/$d$, where $d$ is the 
distance to the background star (from Table~1).

Essentially all boron in diffuse clouds is in the singly-ionized state. Thus, 
the final column densities of B~{\small II} obtained through profile synthesis 
(\S{}~3.2.3) can be taken as the total boron column densities in the gas phase. 
These values are given in Table~12 together with the adopted oxygen column 
densities, the B/O ratios, and the elemental boron abundances\footnote{An 
elemental abundance is defined in logarithmic terms as, for example, 
log~(B/H)~$\equiv$~log~$N$(B~{\scriptsize II})~$-$~log~$N_{\mathrm{tot}}$(H).}. 
Because oxygen is only lightly depleted in diffuse gas (see \S{}~4.2), the B/O 
ratio can serve as a proxy for B/H for those sight lines that lack 
determinations of $N_{\mathrm{tot}}$(H). Quoted uncertainties in B/O and B/H 
include the propagated errors resulting from the uncertainties in both 
$N$(B~{\small II}) and $N$(O~{\small I}) or $N_{\mathrm{tot}}$(H). The five sight 
lines with published B~{\small II} column densities from GHRS spectra (Jura et 
al. 1996; Lambert et al. 1998; Howk et al. 2000) are included in Tables~11 and 
12 since they are incorporated into our abundance analysis. The oxygen 
measurements in these directions are based on GHRS data, except for that toward 
$\delta$ Sco, which is derived from \emph{Copernicus} observations (see Meyer 
et al. 1998). The elemental abundances of oxygen, copper, and gallium toward 
stars in the boron sample are also given in Table~12. For these elements, the 
abundance uncertainties are dominated by the errors in total hydrogen column 
density. We note that these are the first extensive results to be published on 
interstellar gallium abundances from STIS spectra\footnote{In a future paper, 
the gallium abundances discussed here will be examined along with interstellar 
measurements for other neutron-capture elements, such as arsenic, cadmium, tin, 
and lead, which have been identified in archival STIS spectra. The combined 
results will help to illuminate the role played by neutron capture, both in AGB 
stars through the main $s$-process and in massive stars through the weak 
$s$-process and the $r$-process, in Galactic chemical evolution.}.

\subsection{Interstellar Depletion}

In this section, we explore trends in the elemental abundance data by plotting 
the values against various measures of gas density. This analysis will allow us 
to determine the overall level of depletion in each element and to identify the 
density-dependent effects that are anticipated based on prior interstellar 
studies of oxygen and copper, specifically, but also of elements such as 
germanium, manganese, iron, and nickel (e.g., Jenkins et al. 1986; Cartledge et 
al. 2006). Ultimately, what is sought is evidence concerning the 
nucleosynthetic origin of boron. Again, it is unclear which spallation channel 
is responsible for the increase in the $^{11}$B abundance over that predicted 
in standard GCR nucleosynthesis. If any directions yield enhanced boron 
abundances relative to lines of sight with similar densities, then recent 
production of (mostly) $^{11}$B, by either cosmic-ray or neutrino-induced 
spallation, may be the cause. Before such a conclusion is reached, however, the 
variations in line-of-sight abundance resulting from differences in depletion 
must be thoroughly considered. Since physical densities in the gas are 
difficult to obtain directly, other (albeit imperfect) measures such as 
$\langle n_{\mathrm{H}} \rangle$ and $f$(H$_2$) will serve as indicators of the 
physical conditions averaged over the line of sight. The use of these two 
parameters, in particular, enables us to compare our results for O, Cu, B, and 
Ga with those published previously for these and other elements.

\subsubsection{Average Line-of-Sight Hydrogen Density}

Many investigators have employed $\langle n_{\mathrm{H}} \rangle$ as a proxy for 
gas density and the parameter has often been shown to correlate well with 
interstellar gas-phase depletion (e.g., Savage \& Bohlin 1979; Jenkins et al. 
1986; Crinklaw et al. 1994). The connection between 
$\langle n_{\mathrm{H}} \rangle$ and depletion was interpreted by Jenkins et al. 
(1986) within the framework of the Spitzer (1985) model of an idealized neutral 
ISM. According to this model, three different types of neutral atomic gas 
(i.e., warm low-density gas, cold diffuse clouds, and cold dense clouds) may 
contribute to the depletion seen along a given interstellar sight line. The 
warm and cold phases will each exhibit a characteristic level of depletion, 
where the specific levels will vary for different elements. For most elements, 
the cold phase will be associated with a stronger depletion signature. Thus, 
the overall level of depletion will depend on the exact mixture of warm and 
cold neutral material along the line of sight, which, in this framework, is 
parameterized by the average sight-line density of hydrogen. The typical 
behavior can be seen in Figure~15, where we plot the gas-phase elemental 
abundances of O, B, Ga, and Cu against $\langle n_{\mathrm{H}} \rangle$. The 
panels for the different elements are shown from top to bottom in order of 
increasing condensation temperature, $T_{\mathrm{cond}}$, defined as the 
temperature at which 50\% of the element has condensed from the gas phase (see 
Table~13 for the adopted values from Lodders 2003). Two effects are readily 
discernible in the figure. First, a pattern of density-dependent depletion is 
evident in the case of each element. The abundances are seen to vary between 
the warm-gas value (represented by the solid line in each panel) at low 
$\langle n_{\mathrm{H}} \rangle$ and the cold-cloud value (the dashed line) at 
the highest $\langle n_{\mathrm{H}} \rangle$. The trend is more clearly 
delineated when the observational data are binned by average density 
(Figure~16). Second, the depletion in the lowest density gas is found to 
increase with $T_{\mathrm{cond}}$, consistent with the idea that the atoms are 
condensing onto interstellar grains.

\begin{figure}[!t]
\centering
\includegraphics[width=0.45\textwidth]{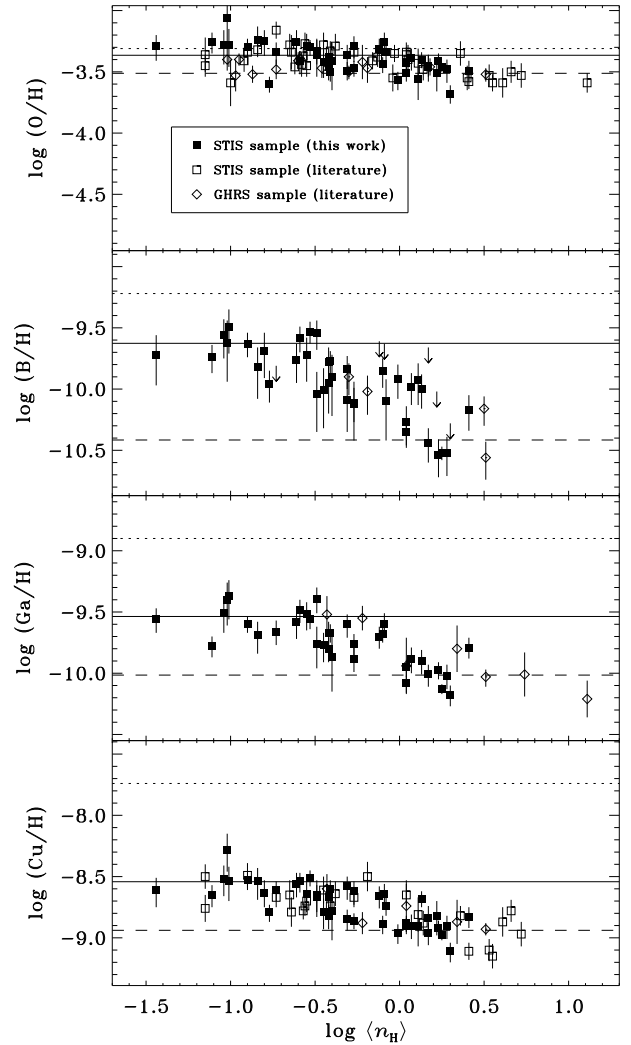}
\caption[Gas-phase O, B, Ga, and Cu abundances versus average hydrogen 
density]{Gas-phase elemental abundances of O, B, Ga, and Cu versus average 
line-of-sight hydrogen density. The panels for different elements are arranged 
from top to bottom in order of increasing condensation temperature. The dotted 
line in each panel represents the solar system abundance (Lodders 2003), while 
the solid and dashed lines correspond to the mean abundances in the warm and 
cold phases, respectively, of the diffuse ISM (see text). Solid symbols (and 
upper limits) represent the STIS boron sample (this work). Open symbols are 
used for abundances taken from the literature (see legend for description of 
symbols; see text for references).}
\end{figure}

\begin{figure}[!t]
\centering
\includegraphics[width=0.45\textwidth]{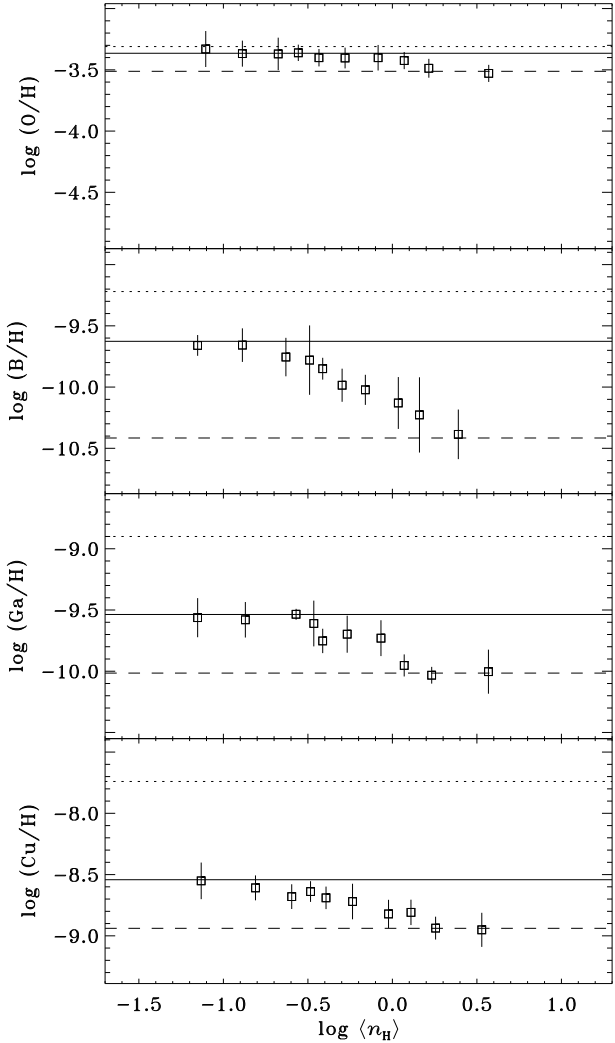}
\caption[Mean O, B, Ga, and Cu abundances versus average hydrogen density]{Mean 
gas-phase elemental abundances of O, B, Ga, and Cu in successive density bins. 
The overplotted lines have the same meaning as in Figure~15. An equal number of 
density bins is used in each panel to provide uniform sampling of the 
observational data. The error bars on each plotted point represent the 
dispersion in the elemental abundances at that density. The mean dispersions in 
the O, B, Ga, and Cu abundances are 0.09, 0.17, 0.13, and 0.11~dex, 
respectively.}
\end{figure}

Our analysis of these effects differs from previous investigations (i.e., 
Jenkins et al. 1986; Cartledge et al. 2004, 2006) in that we do not assume a 
functional form for the dependence of sight-line depletion on 
$\langle n_{\mathrm{H}} \rangle$. The mean abundances, and hence depletions, in 
the warm and cold phases (Table~13) were determined for each element by 
averaging all of the line-of-sight abundances within the appropriate density 
regimes. No weights were applied in averaging, since the errors in individual 
abundance measurements are based on the S/N values in the spectra and are not 
related to how closely a given measurement represents the ``true'' abundance 
for that density regime. This allows for the possibility that certain sight 
lines will exhibit enhanced abundances. The warm-gas values include all sight 
lines with $\langle n_{\mathrm{H}} \rangle$~$<$~0.14~cm$^{-3}$ 
(log~$\langle n_{\mathrm{H}} \rangle$~$<$~$-$0.85), while the cold-cloud 
averages are obtained from sight lines with 
$\langle n_{\mathrm{H}} \rangle$~$>$~1.4~cm$^{-3}$ 
(log~$\langle n_{\mathrm{H}} \rangle$~$>$~0.15). Although these threshold 
densities were somewhat subjectively determined, we note that our value for the 
warm gas threshold is identical to the ``smoothed'' density of warm gas adopted 
by Spitzer (1985). The high density threshold was essentially set so as to 
include the Per~OB2 stars in the averages for cold clouds (with the exception 
of $o$~Per, which has a slightly lower value of 
$\langle n_{\mathrm{H}} \rangle$). These stars have the lowest interstellar 
boron abundances in the STIS sample, and sight lines toward Per~OB2 are known 
to harbor relatively dense gas. Altering these critical densities within 
acceptable limits would modify the derived mean abundances by less than 0.03 
dex. In calculating the various mean values, we used all available sight lines 
with STIS or GHRS abundance measurements for the elements of interest. Thus, 
the means for oxygen, gallium, and copper include abundances taken from the 
literature (Hobbs et al. 1993; Meyer et al. 1998; Welty et al. 1999; Cartledge 
et al. 2001, 2004, 2006; Federman et al. 2003; Andr\'e et al. 2003; Knauth et 
al. 2003a; Jensen et al. 2005) toward stars not in the boron sample. Where 
necessary, these literature values have been corrected to be consistent with 
the set of $f$-values used in our analysis (i.e., Morton 2003). The depletion 
levels are based on the recommended solar abundances from Lodders (2003). 
Quoted errors in mean abundance are 1-$\sigma$ standard deviations for each 
group.

Our results on oxygen and copper depletions can be directly compared with those 
obtained by Cartledge et al. (2004, 2006). These authors fit trends of 
elemental abundance versus $\langle n_{\mathrm{H}} \rangle$ with a 
four-parameter function, modified from the form used by Jenkins et al. (1986), 
finding respective warm and cold-phase depletions of $-$0.10 and $-$0.24 dex 
for oxygen and $-$0.97 and $-$1.27 dex for copper\footnote{The copper depletion 
levels reported by Cartledge et al. (2006) have been adjusted here by +0.16 dex 
to account for the newer $f$-value adopted in this investigation (i.e., that of 
Morton 2003). Brown et al. (2009) have recently conducted beam-foil experiments 
to measure lifetimes and derive oscillator strengths for several ultraviolet 
transitions in Cu~{\scriptsize II}. The oscillator strength they obtain for the 
1358.773~\AA{} line ($f=0.273\pm0.028$) is in very good agreement with the 
recommended value given by Morton (2003), strengthening the abundance and 
depletion results for copper presented here.}. They report the scatter in their 
O and Cu abundances (with respect to the function they fit to the data) as 
being 0.09 and 0.13 dex, respectively. Thus, while Cartledge et al. find 
slightly more depletion in each case, their results are consistent, considering 
the uncertainties, with the results of our more simplified approach (Table~13). 
In Figure~16, we demonstrate that the general trend of decreasing gas-phase 
elemental abundance with increasing average sight-line density can be 
elucidated without adopting an arbitrary fitting function. This figure was 
created by calculating mean abundances in successive density bins, thereby 
removing the bulk of the scatter due to observational uncertainties (as well as 
any intrinsic abundance variations) and revealing the underlying trend. In each 
panel of Figure~16, the binned data are found to exhibit a smooth transition 
between the warm-gas and cold-cloud levels, except, perhaps, in the case of 
gallium. The mean gallium abundances appear to flatten out at intermediate 
densities, although the difference here may simply reflect the small sample 
size and the fact that the data are somewhat irregularly distributed with 
respect to $\langle n_{\mathrm{H}} \rangle$. In contrast, the trend for the mean 
boron abundance shows very little scatter, despite the sample size being the 
smallest in this case. This particular plot will be useful for determining the 
degree to which any individual sight line possesses an elevated abundance of 
boron, as the individual abundance will need to be compared to the mean 
appropriate for the specific value of $\langle n_{\mathrm{H}} \rangle$ 
associated with the line of sight.

\begin{figure*}[!t]
\centering
\includegraphics[width=0.95\textwidth]{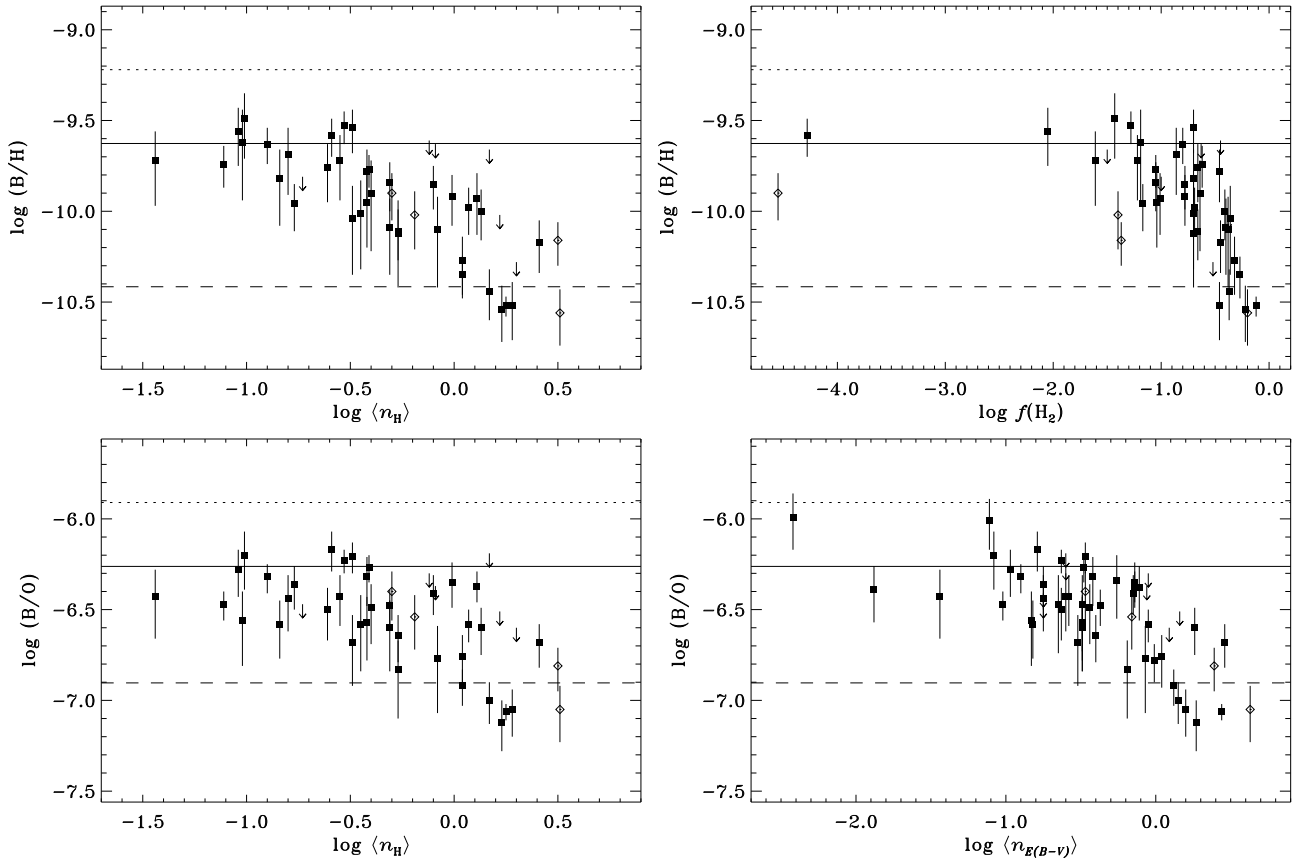}
\caption[Gas-phase B abundances and B/O ratios versus measures of gas 
density]{Gas-phase elemental boron abundances and boron-to-oxygen ratios 
plotted against various measures of gas density and dust content. The 
overplotted lines have the same meaning as in Figure~15, except that in the 
lower panels the lines indicate abundance ratios rather than abundances. Solid 
symbols (and upper limits) represent the STIS boron sample (this work). Open 
symbols denote GHRS measurements from the literature (see text for references). 
\emph{Upper left:} Elemental B abundances versus average line-of-sight hydrogen 
density. \emph{Lower left:} Elemental B/O ratios versus average line-of-sight 
hydrogen density. \emph{Upper right:} Elemental B abundances versus molecular 
hydrogen fraction. The sight line with the lowest value of $f$(H$_2$) in the 
STIS sample is HD~165955. The GHRS sight line at low $f$(H$_2$) is 
$\kappa$~Ori. \emph{Lower right:} Elemental B/O ratios versus interstellar 
reddening ``density'', defined as 
$\langle n_{E(B-V)} \rangle \equiv E(B-V) \times 
\langle N_{\mathrm{tot}}$(H)/$E(B-V) \rangle /d$ (see text). Elevated B/O ratios 
(consistent with the solar value) are found toward HD~187311 and HD~156110 (in 
order of increasing $\langle n_{E(B-V)} \rangle$).}
\end{figure*}

In Figure~17, we present a comparison of the trends seen in the B/H and B/O 
ratios when the data are plotted against various measures of gas density. On 
the left-hand side of the figure, in particular, the B/H and B/O ratios are 
both plotted against $\langle n_{\mathrm{H}} \rangle$. Again, since we are 
attempting to identify sight lines that show evidence of recent boron 
production, we will need to ensure that an elevated B/H (or B/O) ratio is not 
caused by a spurious hydrogen (or oxygen) measurement. There is minimal 
variation between the two trends, however, since the O/H ratios are relatively 
constant with line-of-sight density. Naturally, the distinction between the 
warm and cold-phase depletion levels is less pronounced in B/O than it is in 
B/H due to the slight enhancement in oxygen depletion at high densities. The 
scatter in the B/O versus $\langle n_{\mathrm{H}} \rangle$ relation is also 
marginally higher (by 0.02 dex) than the relation in B/H, although this likely 
results from the fact that B/H and $\langle n_{\mathrm{H}} \rangle$ are not 
completely independent parameters, given that both depend on 
$N_{\mathrm{tot}}$(H). When B/O is plotted instead of B/H, a small amount of 
additional scatter is contributed by the measurement uncertainties (and/or 
intrinsic variations) in the O~{\small I} column densities.

\subsubsection{Molecular Fraction and Reddening}

The other readily available measure of gas density is the molecular hydrogen 
fraction, $f$(H$_2$), which should reflect the local conditions in the 
absorbing clouds more directly than $\langle n_{\mathrm{H}} \rangle$, yet has 
typically been found to yield a poorer correlation with various gas-phase 
elemental abundances. This must partly be related to the fact that the 
fractional abundance of H$_2$ is not solely determined by the local density, 
but is also influenced by factors such as the intensity of the ambient UV 
radiation field, which is responsible for dissociating the H$_2$ molecule. By 
this reasoning, the molecular fraction will perform poorly as an indicator of 
density, and thus of depletion, if the interstellar material lies in close 
proximity to a strong source (or strong sources) of UV radiation, like the hot 
stars of an OB association, unless the H$_2$ column density is large enough for 
the molecule to be self shielded. The upper right panel of Figure~17 gives the 
plot of gas-phase boron abundance versus $f$(H$_2$), where the overall increase 
in scatter over the relation with $\langle n_{\mathrm{H}} \rangle$ is apparent. 
Still, the molecular fraction does reasonably well at predicting whether a 
particular sight line will exhibit enhanced depletion. For instance, the most 
molecule-poor sight line in the STIS sample (HD 165955) has a boron abundance 
that is remarkably consistent with the warm gas average, indicating that the 
B/H ratio remains fairly constant over three orders of magnitude in $f$(H$_2$). 
This is followed by a sharp decline in B/H for values of log $f$(H$_2$) greater 
than $-$1.0. As would be expected, the sight lines with the lowest gas-phase 
boron abundances are also the most molecule rich. These are the sight lines to 
stars in Per~OB2 (40~Per, $o$~Per, $\zeta$~Per, and X~Per), Cep~OB2 (HD~207308 
and HD~207538), and Sco~OB2 ($\zeta$~Oph), which are all regions shaped by the 
formation and evolution of massive stars. Detailed abundance measurements in 
such environments have the potential to offer unique insight into the 
mechanisms responsible for boron production. Unfortunately, any boron 
enhancements in these directions will be difficult to discern due to the heavy 
depletion in the relatively dense gas.

Since interstellar grains are presumed to provide the surface upon which the 
atoms from the gas are depleted, one would expect to see a connection between 
enhanced depletion and high dust content. To seek such a correlation, we 
converted each of the values of $E$($B-V$) in Table~1 into a reddening 
``density'' by multiplying by the mean ratio of total hydrogen column density 
to color excess, $\langle N_{\mathrm{tot}}$(H)/$E(B-V) \rangle = 
5.8~\times~10^{21}$~cm$^{-2}$~mag$^{-1}$ (Bohlin et al. 1978), and dividing by 
the distance $d$. The resulting parameter, which we denote as 
$\langle n_{E(B-V)} \rangle$, is essentially just the reddening normalized by 
the pathlength and scaled to have units similar to 
$\langle n_{\mathrm{H}} \rangle$. The trend of B/O versus reddening density 
(lower right panel of Figure~17) bears a marked resemblance to that of B/O 
versus line-of-sight hydrogen density (lower left panel), undoubtedly a result 
of the known close correspondence between $E$($B-V$) and $N_{\mathrm{tot}}$(H) 
(see Bohlin et al. 1978). Other investigations that have explored possible 
correlations between depletion and some normalized reddening parameter (e.g., 
Cartledge et al. 2006, Jensen \& Snow 2007a, 2007b) have yielded similar 
findings. The advantage in our case is that the plot of B/O versus 
$\langle n_{E(B-V)} \rangle$ includes all of the sight lines in the boron 
sample, including those that lack determinations of total hydrogen column 
density, with two exceptions. It is difficult to estimate a $B-V$ color excess 
for the stars HD~203338 and $\alpha$~Sco~B, since both have rather 
poorly-determined photometric parameters (the former has a composite spectral 
type and the latter is a faint companion to the red supergiant star Antares). 
For another star (HD~212791), the values of $E$($B-V$) and $N_{\mathrm{tot}}$(H) 
were found to lie significantly outside the mean relation. The reddening in 
this direction (0.05 mag) implies a total hydrogen column density of 
log~$N_{\mathrm{tot}}$(H) = 20.46, whereas the observed value is larger by 0.76 
dex (a factor of six). Since HD~212791 is a known Be-type star, the 
interstellar reddening value is almost certainly the cause of the discrepancy. 
A more appropriate value, based on the observed $N_{\mathrm{tot}}$(H), would be 
0.29 mag, although this revised value is not adopted in the derivation of 
$\langle n_{E(B-V)} \rangle$. The boron measurement in this direction is only an 
upper limit, so the discrepant hydrogen and reddening parameters will have 
little impact on the interpretation of the trends in Figure~17.

The B/O ratios for two sight lines in the plot versus reddening density are 
substantially elevated above the mean appropriate for warm neutral gas. The 
anomalous ratios are found in the directions of HD~156110 and HD~187311, both 
of which are high Galactic latitude halo stars, at distances of 420~pc above 
and 3500~pc below the Galactic plane, respectively. These two stars are also 
the least reddened in the boron sample. If low reddening implies low dust 
content, it might be tempting to assume that these sight lines probe 
essentially undepleted gas. Indeed, the gas-phase B/O ratios in these 
directions are consistent with the solar value. However, since halo sight lines 
often trace regions of high ionization (e.g., Spitzer \& Fitzpatrick 1992), the 
elevated B~{\small II}/O~{\small I} ratios may instead indicate that the 
absorption from dominant ions along these lines of sight actually arises in 
partially ionized regions of space. It is possible to test this idea by 
searching for ions that, at their next lower ionization stage, have an IP 
between that of O~{\small I} and B~{\small II}, such as Fe~{\small III} or 
Al~{\small III}. While no Al~{\small III} data exist for either of the sight 
lines in question, the Fe~{\small III}~$\lambda$1122 line is available from 
\emph{FUSE} spectra. Interestingly, the Fe~{\small III} line does appear to be 
stronger, relative to the nearby Fe~{\small II}~$\lambda$1121 line, toward 
HD~156110 and HD~187311 than it is toward HD~121968, another high Galactic 
latitude halo star with low reddening but without an usually high interstellar 
B~{\small II}/O~{\small I} ratio. Some combination of increased ionization and 
reduced depletion must be responsible for the elevated B/O ratios seen along 
the rarefied paths to HD~156110 and HD~187311, as these are not regions 
expected to have boron abundances enhanced by nucleosynthetic processes.

\subsection{Enhanced Boron Abundances}

An ideal approach to determining the relative contribution of each spallation 
channel to boron nucleosynthesis would be to obtain precise interstellar 
$^{11}$B/$^{10}$B ratios in a diverse selection of Galactic environments, 
representing both the ambient ISM as well as regions where Type II supernovae 
have occurred (e.g., in supernova remnants or in superbubbles). Since 
supernovae are presumably the sources responsible for the acceleration of 
Galactic cosmic rays and are also sites of the $\nu$-process, enhancements in 
one or both of the boron isotopes are anticipated in their environs. If 
neutrino spallation contributes significantly to $^{11}$B production, then the 
$^{11}$B/$^{10}$B ratio in the ejecta of a core-collapse supernova should vastly 
exceed the ambient value. The $\nu$-process yields from a 25~$M_\odot$ 
progenitor of solar metallicity, for example, result in a ratio of 
$^{11}$B/$^{10}$B = 587 (Model S25A; WW95). Of course, this $^{11}$B-rich 
material will be diluted somewhat when the ejecta mixes with the surrounding 
interstellar gas, although an enhancement should still be discernible. In the 
case of cosmic-ray spallation, the relative isotopic enhancement will depend on 
the energy spectrum of the accelerated particles. A $^{11}$B/$^{10}$B ratio near 
the meteoritic value (or slightly higher) is predicted if supernovae can 
efficiently accelerate particles at low energies (e.g., Meneguzzi \& Reeves 
1975; Parizot \& Reeves 2001). Otherwise, the pure GCR spallation value (i.e., 
$^{11}$B/$^{10}$B = 2.4; MAR) would be expected. While the present sample of 
interstellar sight lines is diverse enough for such an investigation, the 
component structure toward most of the STIS targets is too complicated (and the 
STIS data have insufficient S/N) to allow any meaningful determinations of 
$^{11}$B/$^{10}$B. Nevertheless, since any recent boron production (in the 
vicinity of SNe II, for instance) should manifest itself as a local enhancement 
in the total boron abundance, the large database of abundances presented here 
can still potentially yield clues to boron nucleosynthesis. In this section, 
the interstellar boron abundances are examined in detail to isolate any 
intrinsic variations superimposed on the general trend due to depletion.

As a first cut in identifying sight lines with enhanced boron abundances, we 
consider the upper envelopes of the distributions of B/H and B/O with 
$\langle n_{\mathrm{H}} \rangle$. Requiring an enhancement (i.e., a significant 
increase over the mean) in both abundance parameters, in order to exclude cases 
where an anomalously low oxygen or hydrogen measurement appears as a boron 
enhancement, results in the following list of sight lines (in order of 
increasing $\langle n_{\mathrm{H}} \rangle$): HD~165955, HD~93222, HD~1383, 
HD~114886, HD~208947, HD~148937, HD~43818, HD~203374, and HD~99872. For both 
HD~1383 and HD~99872, however, the apparent enhancement in boron is more likely 
a result of differential ionization rather than an indication of recent 
nucleosynthesis. Considering that the profile synthesis for HD~1383 required 
fitting a B~{\small II} component not detected in O~{\small I}, this line of 
sight almost certainly crosses an H~{\small II} region. For HD~99872, the Ga/H 
and Cu/H ratios are elevated along with B/H, but the O/H ratio is normal for a 
sight line at high $\langle n_{\mathrm{H}} \rangle$. This strongly suggests that 
the line of sight to HD~99872 traces partially-ionized gas, in which both 
neutral hydrogen and neutral oxygen have been preferentially destroyed, but 
Cu$^+$, Ga$^+$, and B$^+$ remain due to their larger ionization potentials. It 
might also be appropriate to reject HD~165955 as exhibiting a boron 
enhancement, since the very low value of $f$(H$_2$) along this line of sight 
would group it among the low-depletion sight lines and both B/H and B/O are 
consistent with the warm-gas averages. (The discrepancy is that the value of 
$\langle n_{\mathrm{H}} \rangle$ for HD~165955 implies a higher degree of 
depletion, making the B/H and B/O ratios appear to be elevated.)

The remaining six sight lines with suggestively enhanced interstellar boron 
abundances constitute a sample that warrants further investigation. Five of 
these probe gas either in an OB association (i.e., HD~114886 in Cen OB1, 
HD~148937 in Ara OB1a, HD~43818 in Gem~OB1, and HD~203374 in Cep OB2) or in a 
young stellar cluster (i.e., HD~93222 in Collinder 228 in the Carina Nebula), 
where massive star formation is ongoing. Notably, the boron abundance toward 
HD~203374 is 0.27 and 0.44 dex larger than the respective abundances toward 
HD~207308 and HD~207538, other sight lines in Cep OB2 with comparable average 
densities and molecular fractions. A similar enhancement, after accounting for 
the difference in average density, is seen toward HD~208947, a star that lies 
somewhat in the foreground on the northern edge of the Cep OB2 association, 
though is not a member. Because a misplaced continuum could lead to a 
spuriously high abundance, the continuum fits for all of these sight lines were 
examined. In most cases, no apparent problems were discovered, although the 
continua surrounding the B~{\small II} lines toward HD~114886 and HD~148937 are 
somewhat poorly-defined. By lowering the level of the continuum to the extent 
tolerated by the spectrum, the reduction in equivalent width is less than 15\% 
for HD~148937, well within the quoted uncertainty. For HD~114886, however, the 
equivalent width can be reduced by almost 40\% with an alternate continuum fit, 
somewhat outside the 1-$\sigma$ error. If such reductions were adopted, an 
enhanced boron abundance would still be inferred for HD~148937, though an 
enhancement toward HD~114886 would no longer seem likely.

Any claim of enhanced abundances must of course be subject to certain scrutiny 
regarding the level of scatter expected in the data. The average uncertainty in 
log(B/H) for the boron sample is 0.13 dex, whereas the average dispersion in 
these abundances for a particular value of $\langle n_{\mathrm{H}} \rangle$ is 
0.17 dex (see Figure 16), implying that some degree of intrinsic variability is 
present. A portion of the excess scatter may be related to the 
$\langle n_{\mathrm{H}} \rangle$ parameter itself, since a variety of actual 
physical conditions can result in a single value of 
$\langle n_{\mathrm{H}} \rangle$, owing to differences in how the gas is 
distributed along the line of sight. Moreover, the average uncertainty in 
$\langle n_{\mathrm{H}} \rangle$ is 0.13~dex (assuming 30\% errors in distance 
and 20\% errors in total hydrogen column density). Nevertheless, all of the 
sight lines identified above have boron abundances that are more than 0.17 dex 
larger than the means appropriate for their respective values of 
$\langle n_{\mathrm{H}} \rangle$. Four of the sight lines (HD~114886, HD~208947, 
HD~148937, and HD~203374) show only mild boron enhancements, ranging from 0.18 
to 0.21~dex, while in two directions (HD~93222 and HD~43818) the enhancements 
are significantly larger (0.27 and 0.26~dex, respectively). An extensive 
analysis of all of the processes that could potentially result in the boron 
enhancements observed along these six lines of sight (an analysis which should 
consider in more detail the effects of ionization, depletion, and recent 
nucleosynthesis, for example) would be a worthy subject for future 
investigation. Here, we restrict our comments to describing a few of the more 
interesting cases.

\emph{HD~93222}---The boron abundance we derive in the direction of HD~93222 
[log~(B/H)~=~$-$9.53$^{+0.08}_{-0.10}$] is not only enhanced relative to sight 
lines with similar average densities, but is also significantly elevated 
compared to the abundances found toward three other stars in the Carina Nebula. 
The stars HD 93205, CPD$-$59 2603, and HDE 303308, all members of Trumpler 16, 
have interstellar boron abundances of $-$9.72$^{+0.14}_{-0.22}$, 
$-$9.77$^{+0.08}_{-0.10}$, and $-$9.95$^{+0.16}_{-0.25}$, respectively, and lie 
between 21$^{\prime}$ and 26$^{\prime}$ to the north of HD 93222 (i.e., within 
8$^{\prime}$ of the massive $\eta$~Car system). Our analysis of distinct 
absorption complexes toward HD~93222 (\S{} 3.2.4 and Table~8) revealed higher 
B/O ratios in the two more distant complexes (at $v_{\mathrm{LSR}}$~=~$-$34 and 
$-$18 km s$^{-1}$) compared to the local complex (at $-$5 km s$^{-1}$), 
suggesting that the enhancement is associated with gas in the Carina Nebula 
itself or with foreground material in the Sagittarius-Carina spiral arm. Since 
an enhancement in B~{\small II}, relative to O~{\small I} (and H~{\small I}), 
could be an ionization effect, the components near $-$34 and $-$18 km s$^{-1}$ 
may trace gas on the near side of the globally-expanding H~{\small II} region. 
However, these velocities are somewhat different from the more negative 
velocities typically associated with the H~{\small II} region in Carina (e.g., 
Walborn et al. 2002). In weak lines from dominant ions (such as in our 
Cu~{\small II} and Ga~{\small II} data), H~{\small II} region gas is detected 
at $v_{\mathrm{LSR}}$~=~$-$47~km~s$^{-1}$ toward both HD 93205 and CPD$-$59 2603 
(see the discussion of the B~{\small II} component at this velocity toward 
CPD$-$59 2603 in \S{} 3.2.4). Moreover, the strong absorption from O~{\small I} 
in the complexes at $-$34 and $-$18 km s$^{-1}$ toward HD~93222 seem to 
preclude a scenario in which much of this gas is ionized. Walborn et al. (2007) 
discuss high-velocity expanding structures seen in the interstellar absorption 
profiles toward a number of stars in the Carina Nebula in the context of a 
possible supernova remnant (SNR) in this direction. They note that the highest 
known interstellar velocities in the region occur in the spectrum of HD 93222.

\emph{HD~43818}---The line of sight to HD~43818, a member of the Gem OB1 
association, is characterized by a factor of four higher average density 
compared to sight lines in Carina. Yet, while the oxygen and copper abundances 
in this direction show enhanced depletion, the boron abundance 
[log~(B/H)~=~$-$9.93$^{+0.14}_{-0.20}$] is substantially elevated over those 
found for similarly dense sight lines. The proximity of this line of sight to 
IC 443, a young or intermediate-age (presumably core-collapse) SNR known to be 
interacting with nearby atomic and molecular gas (e.g., Snell et al. 2005; Lee 
et al. 2008, and references therein), suggests a nucleosynthetic origin for the 
enhancement. Recent observations of high and very-high energy $\gamma$-ray 
emission from IC 443 (e.g., Acciari et al. 2009; Tavani et al. 2010; Abdo et 
al. 2010) provide direct evidence for hadronic cosmic-ray acceleration by the 
remnant. In the emerging scenario, the $\gamma$-ray emission results from the 
decay of $\pi^0$ mesons produced by the interaction of accelerated protons (and 
other ions) with the ambient molecular cloud. Such interactions will also 
produce significant quantities of light nuclides through spallation, and should 
lead to enhanced light-element abundances in the immediate vicinity of the SNR. 
The sight line to HD 43818 passes 51$^{\prime}$ to the northeast of IC~443 (22 
pc at $d=1.5$ kpc), outside the region of active $\gamma$-ray emission. Thus, 
an enhancement here would have implications for the diffusion characteristics 
of the accelerated cosmic rays. We are currently pursuing $^7$Li/$^6$Li ratios 
and Li (and Rb) abundances toward stars closer to IC~443 with the Hobby-Eberly 
Telescope at McDonald Observatory in an effort to constrain the contribution 
from Type II supernovae to the synthesis of these elements.

\emph{$o$~Per}---While not identified above as having an enhanced abundance of 
boron for its particular value of $\langle n_{\mathrm{H}} \rangle$, the sight 
line to $o$ Per deserves special comment. The boron abundance we derive in this 
direction is 0.18~dex larger than the mean of the abundances found for the 
other three sight lines in Per OB2. Although the enhancement is only modest 
(50\%), it is significant considering that the other sight lines show very 
little scatter in their B/H ratios. For 40~Per, $\zeta$~Per, and X~Per, we 
derive values of log~(B/H) equal to $-$10.52$^{+0.13}_{-0.19}$, 
$-$10.54$^{+0.13}_{-0.18}$, and $-$10.52$^{+0.05}_{-0.06}$, respectively, while for 
$o$~Per we find log~(B/H)~=~$-$10.35$^{+0.10}_{-0.13}$. The higher abundance is 
not likely a result of reduced depletion toward $o$ Per, despite the slightly 
lower value of $\langle n_{\mathrm{H}} \rangle$ for this sight line (which is 
due to the somewhat larger adopted distance). Chemical models predict similar, 
if not larger, densities (as well as similar, if not lower, temperatures) for 
the gas in this direction compared to others in Per OB2 (e.g., Federman et al. 
1994; Federman et al. 1996b). The line of sight to $o$ Per is located just 
8$^{\prime}$ to the north of the young cluster IC 348 and contains at least one 
cloud with a low $^7$Li/$^6$Li ratio (Knauth et al. 2000, 2003b), implying that 
recent production of lithium, by cosmic rays accelerated by the star-forming 
region, has occurred. An enhanced cosmic-ray flux for the gas toward $o$ Per 
(relative to the sight lines to $\zeta$ Per and $\xi$ Per) had already been 
inferred from measurements of interstellar OH (Federman et al. 
1996b)\footnote{It is important to note that determinations of the primary 
cosmic-ray ionization rate ($\zeta_p$) toward $\zeta$ Per from measurements of 
H$_3^+$ (McCall et al. 2003; Indriolo et al. 2007) are an order-of-magnitude 
larger than those from OH and HD (Hartquist et al. 1978; Federman et al. 
1996b). Indriolo et al. (2007) do not definitively detect H$_3^+$ in the 
direction of $o$ Per. However, the upper limit on $\zeta_p$ that they report is 
still marginally consistent with a higher cosmic-ray flux for the gas in this 
direction (for a more detailed discussion, see Indriolo et al. 2007).}. 
Considering our new result that the boron abundance is significantly enhanced 
as well, evidence seems to be mounting of the effect of cosmic-ray spallation 
reactions on the synthesis of light elements near IC 348.

\section{DISCUSSION}

Boron is primarily synthesized in the ISM. Thus, its present-day interstellar 
abundance provides a direct constraint for models that employ various schemes 
to account for the chemical evolution of the light elements. Unfortunately, 
measurements of gas-phase interstellar boron abundances can yield only a lower 
limit to the total interstellar abundance since some depletion is expected even 
in the lowest-density warm gas. We find a gas-phase boron abundance for the 
warm diffuse ISM (from the mean abundance in the six lowest-density sight 
lines) of B/H = $(2.4\pm0.6)\times10^{-10}$. This result is consistent with the 
value of $(2.5\pm0.9)\times10^{-10}$ obtained by Howk et al. (2000), which was 
based on the analysis for only one line of sight (i.e., HD~121968, which has 
the lowest $\langle n_{\mathrm{H}} \rangle$ in the boron sample). Both results 
indicate a depletion of 60\% relative to the solar system (meteoritic) 
abundance of $(6.0\pm0.6)\times10^{-10}$ (Lodders 2003). The evidence for 
depletion is twofold. First, a density-dependent effect is seen in the 
gas-phase abundance data for each element examined here. Second, the difference 
between the solar system abundance and the mean abundance in the least depleted 
sight lines for a particular element increases with the condensation 
temperature of the element (see Figure~15). Note that this does not necessarily 
mean that the set of solar system abundances is the most appropriate cosmic 
abundance standard against which to measure interstellar depletion. Indeed, the 
abundances in F and G disk dwarfs of solar metallicity or hot B stars may be 
more suitable for such comparisons (e.g., Snow \& Witt 1996; Sofia \& Meyer 
2001; Przybilla et al. 2008).

In the case of boron, two studies in particular provide useful non-solar 
abundance standards: Cunha et al. (2000), who derived boron abundances for 14 
near-solar metallicity F and G dwarfs from GHRS spectra of the 
B~{\small I}~$\lambda$2497 line, and Venn et al. (2002), who compiled and 
updated a number of B-star abundances, notably the boron abundances obtained 
from the B~{\small III}~$\lambda$2066 line by Proffitt \& Quigley (2001). The 
average boron abundance in both of these samples, excluding any stars showing 
evidence of light element depletion (by stellar astration), is B/H = 
$3\times10^{-10}$. A similar value (i.e., B/H = $2.6\times10^{-10}$ at 
[Fe/H]~=~0.0) was obtained by Boesgaard et al. (2004) from their fit to the 
boron abundances in 20 solar-type dwarf stars of the Galactic disk, all with 
undepleted beryllium (implying that these stars have also retained their full 
initial abundances of boron). If such a value is representative of the 
present-day ISM, then boron would seem to be only lightly depleted along the 
lowest-density interstellar sight lines. However, many of the B-type stars in 
the Venn et al. (2002) sample may have had some of their initial boron 
destroyed through rotationally-induced mixing (though the most severe cases 
were not included in deriving the above mean value). The highest boron 
abundances in Venn et al. are consistent with the solar system value. 
Furthermore, the boron abundances in the Cunha et al. (2000) sample of F and G 
stars exhibit a well-defined positive correlation with the abundance of oxygen 
(Smith et al. 2001), indicating that the mean value of the sample may not 
reflect the present-day abundance in the ISM. Indeed, the most metal-rich star 
in the Cunha et al. sample (HD 19994) has a B/H ratio that is essentially 
identical to the meteoritic value. Since this star likely formed quite 
recently, the solar system abundance of boron may well represent the present 
interstellar value. The various boron abundances discussed here are listed in 
Table~14, where we also give the solar photospheric abundance [i.e., B/H = 
$(5.0^{+3.1}_{-1.2})\times10^{-10}$; Cunha \& Smith 1999], which is consistent 
with the more precise abundance derived from meteorites.

\begin{figure}[!t]
\centering
\includegraphics[width=0.45\textwidth]{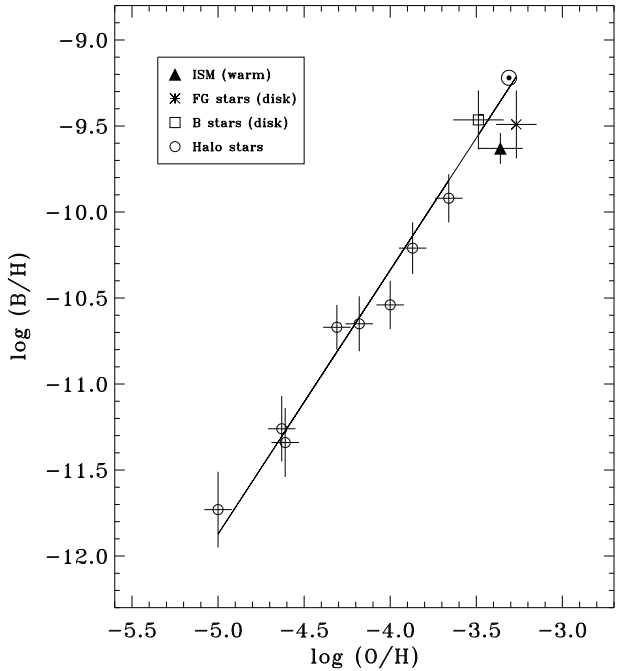}
\caption[Galactic evolution of boron]{Galactic evolutionary trend of boron 
versus oxygen. The halo boron and oxygen abundances are from Tan et al. (2010). 
Also shown are the mean abundances in the disk from the Cunha et al. (2000) 
sample of F and G stars (with oxygen from Smith et al. 2001) and the Venn et 
al. (2002) sample of B stars, along with the solar system and mean interstellar 
values. A linear fit to the solar and stellar data, which accounts for errors 
in both B/H and O/H, has a slope of $1.5\pm0.1$. The warm-gas interstellar 
values of B/H and O/H are not included in this fit since they represent lower 
limits to the total interstellar abundances.}
\end{figure}

It is important to establish the relationship between the abundances of boron 
and oxygen over both disk and halo metallicities if the origin and evolution of 
Galactic boron is to be understood (e.g., Duncan et al. 1997; Smith et al. 
2001; Tan et al. 2010). The oxygen abundance is a more appropriate metallicity 
indicator than [Fe/H] in this context because the evolutionary histories of 
boron and oxygen are closely linked. The spallative production of $^{10}$B and 
$^{11}$B results, at least in part, from interactions with $^{16}$O, either as 
an interstellar target for energetic protons and $\alpha$-particles or as an 
accelerated particle spalled from ambient interstellar H and He (MAR; Ramaty et 
al. 1997). Galactic oxygen is a product of helium burning in massive stars 
(WW95) and is released into the ISM by core-collapse supernovae, which may also 
produce boron during the $\nu$-process (Woosley et al. 1990). Critical insight 
into the mechanisms responsible for boron production may thus be gained by 
tracking the dependence of B/H on O/H over the lifetime of the Galaxy. The 
available data result in the trend displayed in Figure~18, where we have 
adopted the halo boron and oxygen abundances of Tan et al. (2010). These 
authors derived new non-local thermodynamic equilibrium (NLTE) 
corrections\footnote{Cunha et al. (2000) also applied NLTE corrections to their 
boron abundances based on the results presented in Kiselman \& Carlsson (1996), 
which may lead to a slight inconsistency in Figure~18 since the new corrections 
derived by Tan et al. (2010) are smaller than those of Kiselman \& Carlsson. 
However, the NLTE effects are small in general and less important for stars of 
solar metallicity.} for B~{\small I} and applied their results to existing 
observations of boron in metal-poor stars (i.e., Duncan et al. 1992, 1997; 
Garc\'ia L\'opez et al. 1998; Primas et al. 1999). Also included in the figure 
are the mean disk abundances of Cunha et al. (2000) and Venn et al. (2002) as 
well as the solar system and interstellar values discussed above. A simple 
linear fit to the data yields a slope of $1.5\pm0.1$, implying that a mixture 
of primary and secondary nucleosynthetic processes contributes to the Galactic 
evolution of boron.

The behavior of beryllium with O/H (or [Fe/H]) from measurements in both disk 
and halo stars is very similar to that of boron (e.g., Boesgaard et al. 1999; 
Tan et al. 2009), leading to an essentially constant B/Be ratio as a function 
of metallicity (Duncan et al. 1997; Boesgaard et al. 2004; Tan et al. 2010). In 
a recent analysis of beryllium abundances in stars spanning a range of 
metallicities (i.e., $-$3.5~$\lesssim$~[Fe/H]~$\lesssim$~$-$0.5), Rich \& 
Boesgaard (2009) find that a two-slope fit provides a good match to the Be-O 
trend. The measurements at low O/H are consistent with a slope of 
$0.74\pm0.11$, while a slope of $1.59\pm0.15$ fits the higher O/H points. 
Interestingly, the transition occurs at log~(O/H)~=~$-$4.9, which is at the 
metal-poor limit for stars with measured boron abundances (see Figure~18). Rich 
\& Boesgaard (2009) attribute the change in slope seen in their beryllium data 
to a change in the dominant mechanism of beryllium production (i.e., from a 
primary process to a secondary one). They suggest that the acceleration of CNO 
nuclei in the vicinity of SNe~II produced beryllium during the epoch of 
formation of the Galactic halo and that GCR bombardment of CNO in the ambient 
interstellar gas dominates its production today. Such a scenario is consistent 
with the higher B/O ratios we find for gas in the Sagittarius-Carina spiral 
arm, which seem to indicate that a secondary process also dominates the current 
production of boron in the Galaxy. Indeed, the slope determined for the trend 
of B/H versus O/H (Figure~18) agrees with the slope of Be/H versus O/H at the 
high metallicity end (Rich \& Boesgaard 2009). It remains to be seen whether 
observations of boron and oxygen in lower metallicity stars will confirm the 
flattening of the slope as observed in the beryllium data.

A limited number of sight lines analyzed in this work for interstellar boron 
also have measurements (or upper limits) on interstellar lithium and beryllium. 
While the sample size is small, a comparison of the abundances resulting from 
these measurements is important if the ultimate goal is a unified understanding 
of light element synthesis. Lithium elemental abundances have most recently 
been obtained by Knauth et al. (2003b), using ultra-high-resolution 
observations of the Li~{\small I}~$\lambda$6707 doublet toward stars in Per OB2 
and Sco OB2. The derived lithium abundances have large uncertainties, however, 
because Li~{\small I} represents a trace species in diffuse clouds, with most 
of the lithium being singly ionized (but unobservable). Beryllium has never 
been detected in interstellar space, though quite stringent limits on 
absorption from the Be~{\small II}~$\lambda$3130 doublet were provided by 
H\'ebrard et al. (1997). These authors obtained an upper limit on the 
equivalent width of Be~{\small II} toward $\zeta$~Per of 
$W_{\lambda}$~$\leq$~30~$\mu$\AA{}, which translates into an upper limit on the 
gas-phase interstellar beryllium abundance in this direction of 
Be/H~$\leq$~$7\times10^{-13}$. The current situation is summarized in Table~15, 
where we present the interstellar abundances of lithium, beryllium, and boron 
and the corresponding abundance ratios (or limits) along lines of sight that 
have results for at least two of these elements. The quoted abundances may 
differ from those in the original references because they are based on our 
adopted values for the total hydrogen column density.

Ideally, light element abundance ratios along individual sight lines, such as 
those presented in the last two columns of Table~15, can yield vital 
information on the nucleosynthetic history of the interstellar gas being 
probed. The present data is limited in its ability to provide such information, 
however, due to the uncertain corrections that must be made for the depletion 
in each element along each line of sight. Boron depletion was explored in 
detail in \S{}~4.2, where we found that a logarithmic depletion of $-$1.2~dex 
(see Table~13) is characteristic of sight lines through cold diffuse clouds 
(such as those in Table~15). Lithium depletion has been examined by Welty \& 
Hobbs (2001) and (along many of the lines of sight in Table~15, specifically) 
by Knauth et al. (2003b). The average depletion found by Knauth et al. (2003b) 
for the five sight lines they studied is $-$1.0~dex, while Welty \& Hobbs 
(2001) estimated a uniform amount of depletion for lithium equal to $-$0.6~dex. 
These values refer to cold clouds as well since detections of neutral lithium 
are restricted to such regions. Obviously, the amount of beryllium depletion 
along any line of sight is unknown, although, evidently, the depletion is at 
least $-$1.6 dex in the gas toward $\zeta$~Per. Lithium and beryllium have 
condensation temperatures of 1142 K and 1452 K, respectively (Lodders 2003). 
Thus, if their cold-phase depletion levels followed the general trend with 
$T_{\mathrm{cond}}$ that is observed for other elements (e.g., Ge, B, Ga, Cu, Mn, 
Fe, and Ni; Jenkins et al. 1986; Cartledge et al. 2006; this work), then 
depletions of $-$1.5 dex for lithium and $-$2.0 dex for beryllium would be 
expected, though the studies by Knauth et al. (2003b) and Welty \& Hobbs (2001) 
seem to indicate that lithium, at least, does not follow this trend.

The main problem with these types of analyses is the assumption that the solar 
system abundances represent the undepleted values for all sight lines. If each 
of the gas-phase interstellar abundances were corrected for depletion based on 
the difference between the abundance and the solar system value, then all of 
the elemental ratios would be altered to correspond with the solar system 
ratios and no new information would be revealed. Applying a uniform correction 
per element to all sight lines would leave abundance variations intact, but 
depletion should also vary with the local physical conditions of the gas in 
each direction. However, since all of the sight lines in Table~15 can be 
classified as ``high-depletion'' sight lines, a uniform correction may be 
appropriate. If we adopt depletion corrections of +1.0 dex for lithium, +2.0 
dex for beryllium, and +1.2 dex for boron, then the abundance ratios of Li/B 
would be altered to $3.9\pm2.0$ toward $o$~Per, $7.2\pm3.6$ toward $\zeta$~Per, 
$1.6\pm0.7$ toward X~Per, and $9.0\pm4.5$ toward $\zeta$~Oph, while the lower 
limits on the B/Be ratio would be 7 toward $\zeta$~Per, 1 toward $\delta$~Sco, 
and 0.6 toward $\zeta$~Oph. With these corrections, the B/Be lower limits are 
all consistent with the solar system ratio (23; Lodders 2003) as well as with 
the predictions of standard GCR spallation (15; MAR). The Li/B ratios are 
mostly consistent with the solar system value (3.2; Lodders 2003), although the 
ratios toward $\zeta$~Per and $\zeta$~Oph are still somewhat high. Again, the 
lithium abundances have an additional source of uncertainty related to the 
ionization balance in the gas since the observable resonance line originates 
from a trace neutral species.

Another line of investigation into light element nucleosynthesis involves the 
interstellar fluorine abundance (e.g., Federman et al. 2005; Snow et al. 2007), 
as $^{19}$F is expected to be produced by the $\nu$-process along with $^{7}$Li 
and $^{11}$B (Woosley et al. 1990; Timmes et al. 1995), yet is not made in 
substantial quantities by cosmic rays. The supernova yields from Model S25A in 
WW95, which include $\nu$-process contributions, result in an elemental 
abundance ratio of F/B~=~42, similar to the solar system ratio of 48 (Lodders 
2003). Since a much larger value would be anticipated without the 
$\nu$-process, the F/B ratio becomes an important diagnostic of 
neutrino-induced spallation. Fluorine abundances in diffuse clouds are obtained 
from the F~{\small I} line at 954~\AA, which is accessible from \emph{FUSE} 
spectra. However, absorption from H$_2$ is also present near 954~\AA{} and must 
be modeled and then removed if accurate fluorine abundances are to be 
determined. Federman et al. (2005) set an upper limit on the column density of 
H$_2$ that would allow a reliable measurement of the F~{\small I} line 
[log~$N$(H$_2$)~$\leq$~20.5]. Unfortunately, nearly half of the sight lines in 
the boron sample for which $N$(H$_2$) has been determined have H$_2$ column 
densities at or above this limit, making a general study of interstellar 
fluorine impossible for this sample. Nevertheless, literature results from 
secure detections of F~{\small I} are available for three of our sight lines. 
The F~{\small I} column densities toward HD~177989 
[log~$N$(F~{\small I})~=~13.49$^{+0.17}_{-0.23}$; Snow et al. 2007], HD~209339 
[log~$N$(F~{\small I})~=~13.46$^{+0.17}_{-0.29}$; Federman et al. 2005], and 
$\delta$~Sco [log~$N$(F~{\small I})~=~13.22$^{+0.15}_{-0.14}$; Snow \& York 1981; 
corrected using the Morton 2003 $f$-value] yield respective F/B ratios of 
$144\pm73$, $213\pm141$, and $196\pm93$.

All of these ratios are larger than that predicted by neutrino spallation, 
although the uncertainties are not inconsiderable and boron is expected to be 
more heavily depleted than fluorine (which has $T_{\mathrm{cond}}$~=~734~K; 
Lodders 2003). The depletion corrections will vary for these three sight lines 
because each lies in a different density regime: HD~177989 (in the low halo) is 
one of the low-depletion sight lines tracing mostly warm gas, the sight line to 
$\delta$~Sco presumably samples colder gas that should be characterized by 
relatively high depletion, and the gas toward HD~209339 should exhibit an 
intermediate degree of depletion between the warm-gas and cold-cloud levels. 
Still, the relative depletion corrections between fluorine and boron should be 
roughly the same for the different regimes. In the Snow et al. (2007) sample, 
fluorine is essentially undepleted along lines of sight with the lowest values 
of $\langle n_{\mathrm{H}} \rangle$, while boron exhibits a depletion of 
$-$0.4~dex in low-density, warm gas (see Table~13). If this difference in 
depletion persists for sight lines in the higher-density regimes, then each of 
the F/B ratios should be reduced by 60\% (i.e., F/B~=~$57\pm29$, $85\pm56$, and 
$78\pm37$ toward HD~177989, HD~209339, and $\delta$~Sco, respectively). Thus, 
even with a plausible correction for depletion, the ratios are higher than that 
expected from neutrino spallation. The result for the line of sight to 
HD~209339, which has the largest F/B ratio of the three, is particularly 
significant because this star, as a member of Cep OB2, is known to have formed 
in an expanding shell of gas created by SNe~II (Patel et al. 1998). Yet no 
evidence is found that the $\nu$-process has operated in the clouds in this 
direction. We do find evidence for an enhanced boron abundance toward 
HD~203374, a nearby member of the association. However, while this enhancement 
could be attributed to neutrino production, it might instead reflect the recent 
acceleration of cosmic rays from the surrounding star-forming regions 
associated with the Cepheus bubble. Unfortunately, the column density of H$_2$ 
toward HD~203374 [log~$N$(H$_2$)~=~20.70] will make any detection of 
F~{\small I} extremely difficult for this sight line, meaning that the F/B 
ratio will not be available to differentiate the two scenarios.

In the absence of convincing evidence for the $\nu$-process in Cep OB2 and with 
the possible detection of elevated B/O ratios in the Sagittarius-Carina spiral 
arm, a picture of light element nucleosynthesis begins to emerge. The stellar 
abundance results demonstrate that a mixture of primary and secondary processes 
are needed to explain the observed trends of boron and beryllium versus oxygen. 
For beryllium, the most metal-poor stars suggest a purely primary mechanism 
dominated light element production in the early Galaxy (Rich \& Boesgaard 
2009). This mechanism almost certainly was the acceleration of CNO nuclei by 
SNe~II because Be is not produced by the $\nu$-process. If this mechanism 
operated at early times, it likely is still at work today, though GCR 
spallation now dominates light element nucleosynthesis, giving rise to the 
observed slope of 1.5 in the logarithmic boron versus oxygen relation.

The remaining problem is the $^{11}$B/$^{10}$B ratio in the local solar 
neighborhood. The ratios measured in interstellar clouds (Federman et al. 
1996a; Lambert et al. 1998) and those determined for early B stars (see 
Proffitt et al. 1999) are all consistent with the meteoritic value of 4.0. 
While not much is known about the ratio beyond the solar neighborhood, because 
all of the existing measurements were made within 250~pc of the Sun, a 
satisfactory explanation of the meteoritic ratio has not yet emerged. If the 
$\nu$-process contribution to $^{11}$B synthesis is minimal, because a 
secondary process is required in the current epoch, the only remaining 
mechanism is an increased flux of low-energy Galactic cosmic rays, as 
originally proposed by MAR. Interestingly, some support for this mechanism 
comes from recent observations of H$_3^+$, which is a sensitive probe of the 
cosmic-ray ionization rate of molecular hydrogen (Indriolo et al. 2007, 2009). 
In order to match the ionization rate inferred from their observations of 
H$_3^+$, Indriolo et al. (2009) had to adopt a cosmic-ray energy spectrum 
enhanced at low energies, a natural consequence of which is the preferrential 
synthesis of $^{11}$B over $^{10}$B due to the lower thresholds for reactions 
leading to the former nuclide. The physical motivation for a low-energy 
cosmic-ray component comes from the recognition that some particles will be 
accelerated by weak shocks associated with star-forming regions, OB 
associations, and even the astrospheres of low mass stars (Scherer et al. 
2008), which are ubiquitous throughout the Galaxy.

\section{SUMMARY AND CONCLUSIONS}

Archival STIS spectra sampling diffuse interstellar gas along over 50 Galactic 
sight lines were analyzed to obtain column densities of B~{\small II}, and, in 
the process, O~{\small I}, Cu~{\small II}, and Ga~{\small II}. Before 
synthesizing the B~{\small II} absorption profile along a given line of sight, 
a consistent component decomposition was derived in the stronger UV lines. The 
component structure for many of these sight lines is confirmed through the 
analysis of high resolution Ca~{\small II} and K~{\small I} profiles. Both UV 
and visible data were used to construct profile templates that were fit to the 
B~{\small II} line, yielding the total column density along the line of sight. 
For sight lines with multiple complexes of absorption components, separate 
templates were fit to each complex, independently. This procedure allowed the 
detection of a slight increase in the boron-to-oxygen ratio in the inner 
Sagittarius-Carina spiral arm, a result which may indicate that boron 
production in the current epoch is dominated by a secondary process. Elemental 
abundances in the gas phase were determined for each of the UV species and were 
compared with line-of-sight measures of gas density to quantify the effects of 
depletion onto interstellar grains. The depletion in each element was found to 
increase with the average density of hydrogen, which characterizes the overall 
concentration of cold clouds relative to warm gas along the line of sight. Mean 
abundances were determined for the warm and cold phases of the diffuse ISM and 
the level of depletion in warm gas was shown to increase with the condensation 
temperature of the element. The gas-phase abundance of boron in the warm 
diffuse ISM was found to be B/H = $(2.4\pm0.6)\times10^{-10}$, which translates 
into a depletion of 60\% relative to the meteoritic boron abundance.

Knowledge of the trend of decreasing elemental abundance with increasing gas 
density allowed the identification of sight lines showing enhanced boron 
abundances. Many of these sight lines (e.g., HD~93222, HD~114886, HD~208947, 
HD~148937, HD~43818, and HD~203374) are near regions of massive star formation, 
and therefore (at least some of them) may be probing the recent production of 
$^{11}$B, resulting from spallation reactions induced by either cosmic rays or 
neutrinos. Further detailed analysis is needed to determine conclusively the 
degree of ionization and depletion in the gas in these directions. Observations 
of other light elements will also be required to disentagle the relative 
contributions from the two spallation channels, if recent nucleosynthesis can 
be confirmed. As with the majority of sight lines in our survey, the 
complicated component structure in most of the directions showing boron 
enhancements will not permit any meaningful determinations of the 
$^{11}$B/$^{10}$B ratio. Additional measurements on interstellar fluorine may 
help to clarify the situation if abundances can be obtained for a larger number 
of sight lines from the boron sample. The elemental F/B ratio is expected to be 
an important tracer of neutrino spallation, a process for which direct 
observational evidence is still lacking. From a limited sample, we find no 
indication that the $\nu$-process has operated in the Cepheus bubble (as probed 
by the line of sight to HD~209339), though this region is known to have been 
shaped by core-collapse supernovae in the recent past.

Future searches for observational signatures of light element production in the 
Galaxy should focus on specific regions where active nucleosynthesis is likely 
to be occurring. Among the more promising candidates are the interstellar 
clouds in the vicinity of the interacting supernova remnant IC~443 and those 
near the massive star-forming region IC~348. Observations of Li~{\small I} 
along lines of sight through the supernova remnant are currently being analyzed 
and similar data exist on Li~{\small I} toward members of IC 348. Complementary 
observations of B~{\small II} toward stars in both of these regions should now 
be acquired so that a more complete picture of light element synthesis can be 
obtained. Such observations have recently become feasible with the installation 
of the Cosmic Origins Spectrograph (COS) on \emph{HST}. In conjunction with 
future studies in the Galactic ISM, it will be important to expand the 
investigation of interstellar boron to the Large and Small Magellanic Clouds. 
Interstellar lithium has very recently been detected in the Small Magellanic 
Cloud (Howk 2010) and boron could likely be discovered there with COS or even 
STIS, now that it has been refurbished. The Magellanic Clouds present a unique 
opportunity to study light element nucleosynthesis in metal-poor environments 
with regions of active star formation, providing suitable analogs of our own 
Galaxy at earlier times. A similarly important discovery will be the eventual 
detection of beryllium in interstellar space, as the sole stable isotope of 
beryllium ($^9$Be) can be produced only through cosmic-ray spallation. While 
existing searches for Be~{\small II} have mainly focused on bright interstellar 
targets like $\zeta$~Per, $\zeta$~Oph, and $\delta$~Sco (see H\'ebrard et al. 
1997), beryllium will be more heavily depleted in these directions if it 
follows a trend similar to that found for boron. Absorption from Be~{\small II} 
should be sought in more diffuse directions where less depletion is expected, 
though even in favorable conditions the equivalent widths will likely be less 
than $\sim0.1$ m\AA. Still, with sufficiently long exposure times, a detection 
should be possible considering the capabilities of modern large aperture 
telescopes.

The near future promises considerable advancements in the field of 
cosmochemistry. For instance, actual measurements of the flux of low-energy 
cosmic rays, needed for accurate predictions of light element synthesis via 
cosmic-ray spallation, are on the horizon. The \emph{Voyager 1} and \emph{2} 
spacecraft, which recently crossed the solar wind termination shock, may be 
able to measure the flux of cosmic rays at energies below 1 GeV nucleon$^{-1}$ 
when they reach the heliopause in the coming decades. These measurements could 
help answer exciting new questions about whether anomalous cosmic rays 
accelerated in astrospheres throughout the Galaxy contribute significantly to 
the GCR flux at low energies, an outcome with direct implications for light 
element nucleosynthesis. The comprehensive survey of interstellar boron 
presented here, when combined with future studies of light element abundances 
and isotopic ratios, an improved treatment of interstellar depletion, and a 
direct determination of the flux of low-energy cosmic rays, will ultimately 
lead to a deeper understanding of the mechanisms responsible for the production 
of the light elements.

\acknowledgments
We are grateful to Chris Howk for aiding in the extraction of STIS edge orders 
and for many helpful discussions and suggestions on this work. We also thank 
Dan Welty for providing optical spectra for the line of sight to X~Per. The 
manual re-calibration of STIS data would not have been accomplished without the 
assistance provided by the individuals at the STIS help desk. This research 
made use of the SIMBAD database operated at CDS, France. D. L. L. thanks the 
Robert A. Welch Foundation for support through grant F-634. The archival work 
presented here was supported by the Space Telescope Science Institute through 
grant HST-AR-11247.01-A. Observations were obtained from the Multimission 
Archive at the Space Telescope Science Institute (MAST). STScI is operated by 
the Association of Universities for Research in Astronomy, Inc., under NASA 
contract NAS5-26555.

\clearpage

\begin{deluxetable}{lcccccccc}
\tablecolumns{9}
\tablewidth{0.8\textwidth}
\tablecaption{Stellar Data for the Boron Sample}
\tablehead{\colhead{} & \colhead{} & \colhead{} & \colhead{} & \colhead{$V$} & 
\colhead{$E(B-V)$\tablenotemark{a}} & \colhead{$l$} & \colhead{$b$} & 
\colhead{$d$\tablenotemark{a}} \\
\colhead{Star} & \colhead{Name} & \colhead{Association} & \colhead{Sp. Type} & 
\colhead{(mag)} & \colhead{(mag)} & \colhead{(deg)} & \colhead{(deg)} & 
\colhead{(pc)} }
\startdata
\multicolumn{9}{c}{STIS Sample} \\
\hline
HD 1383 &\ldots    &\ldots &B1 II  	&7.63   &0.51	&119.02	&$-$0.89 &2860 
\\
HD 13268 &\ldots   &Per OB1 &O8 Vnn 	&8.18   &0.44	&133.96	&$-$4.99 &2290 
\\
HD 13745 &V354 Per &Per OB1 &O9.7 II     &7.90	&0.46	&134.58	&$-$4.96 &2290 
\\
HD 22951 &40 Per   &Per OB2 &B0.5 V 	&4.98   &0.24	&158.92	&$-$16.70 
&283$^{+94}_{-57}$ \\
HD 23180 &$o$ Per  &Per OB2 &B1 III 	&3.86   &0.32	&160.36	&$-$17.74 
&452$^{+277}_{-125}$ \\
HD 24398 &$\zeta$ Per &Per OB2 &B1 Iab	&2.88	&0.30	&162.29	&$-$16.69 
&301$^{+88}_{-56}$ \\
HD 24534 &X Per   &Per OB2 &O9.5pe	&6.10	&0.59	&163.08	&$-$17.14 
&400 \\
\enddata
\end{deluxetable}

\setcounter{table}{0}
\begin{deluxetable}{lcccccccc}
\tablecolumns{9}
\tablewidth{0.8\textwidth}
\tablecaption{(Continued)}
\tablehead{\colhead{} & \colhead{} & \colhead{} & \colhead{} & \colhead{$V$} & 
\colhead{$E(B-V)$\tablenotemark{a}} & \colhead{$l$} & \colhead{$b$} & 
\colhead{$d$\tablenotemark{a}} \\
\colhead{Star} & \colhead{Name} & \colhead{Association} & \colhead{Sp. Type} & 
\colhead{(mag)} & \colhead{(mag)} & \colhead{(deg)} & \colhead{(deg)} & 
\colhead{(pc)} }
\startdata
\multicolumn{9}{c}{GHRS Sample} \\
\hline
HD 36861 &$\lambda$ Ori &Ori OB1 &O8 III &3.39 &0.12 &195.05 &$-$12.00 
&324$^{+109}_{-65}$ \\
HD 38771 &$\kappa$ Ori &Ori OB1 &B0 Iab: &2.05 &0.04 &214.51 &$-$18.50 
&221$^{+45}_{-32}$ \\
HD 143275 &$\delta$ Sco &Sco OB2 &B0.2 IVe &2.29 &0.16 &350.10 &+22.49 
&123$^{+15}_{-12}$ \\
HD 148479 &$\alpha$ Sco B &Sco OB2 &B2.5 V &7 &\ldots &351.95 &+15.07 &160 \\
HD 149757 &$\zeta$ Oph &Sco OB2 &O9 V &2.58 &0.32 &6.28 & +23.59 
&140$^{+16}_{-13}$ \\
\enddata
\tablenotetext{a}{Values of $E(B-V)$ and $d$ are mainly from the references 
that provided H~{\scriptsize I} or H$_2$ column densities (see \S{}~4.1). 
Distances with error bars are from $\geq$ 3-$\sigma$ \emph{Hipparcos} results. 
Other exceptions are noted.}
\tablenotetext{b}{From Cartledge et al. 2004.}
\tablenotetext{c}{Derived by means of spectroscopic parallax using $M_V$ from 
Schmidt-Kaler 1982.}
\tablenotetext{d}{Derived using intrinsic colors $(B-V)_0$ from Schmidt-Kaler 
1982.}
\tablenotetext{e}{The star HD~203338 has a composite spectrum with a hot 
component classified as a B spectral type (see Ginestet et al. 1999).}
\tablenotetext{f}{Corresponds to the distance adopted in Humphreys 1978 for Cep 
OB2.}
\tablenotetext{g}{From Pan et al. 2004.}
\tablenotetext{h}{From Jenkins \& Tripp 2001.}
\end{deluxetable}


\begin{deluxetable}{lcccccc}
\tablecolumns{7}
\tablewidth{0.6\textwidth}
\tablecaption{STIS Datasets}
\tablehead{\colhead{} & \colhead{} & \colhead{Exp. Time} & \colhead{Cen.~Wave} 
& \colhead{} & \colhead{Slit} & \colhead{} \\
\colhead{Star} & \colhead{Dataset} & \colhead{(s)} & \colhead{(\AA)} & 
\colhead{Grating} & \colhead{(arcsec)} & \colhead{$R$} }
\startdata
HD 1383	  &o5c07c	&1440	&1271	&E140H  &0.2$\times$0.2	&82,000 \\
	  &o6lz01	&1200	&1425	&E140M  &0.2$\times$0.2	&38,000 \\
HD 13268  &o63506	&1440	&1425	&E140M  &0.2$\times$0.2	&38,000 \\
HD 13745  &o6lz05	&1200	&1425	&E140M  &0.2$\times$0.2	&38,000 \\
HD 22951  &o64805--09	&36133	&1416	&E140H  &0.2$\times$0.05 &120,000 \\
HD 23180  &o64801--04	&27141	&1416	&E140H  &0.2$\times$0.05 &120,000 \\
HD 24398  &o64810--11	&15100	&1416	&E140H  &0.2$\times$0.05 &120,000 \\
HD 24534  &o66p01	&10779	&1343	&E140H  &0.2$\times$0.09 &108,000 \\
          &o64812--13	&10728	&1416	&E140H  &0.1$\times$0.03 &143,000 \\
HD 36841  &o63516	&1440	&1425	&E140M  &0.2$\times$0.2	&38,000 \\
HD 43818  &o5c07i	&1440	&1271	&E140H  &0.2$\times$0.2	&82,000 \\
HD 52266  &o5c027	&360	&1271	&E140H  &0.2$\times$0.2	&82,000 \\
HD 69106  &o5lh03       &1285	&1271	&E140H  &0.1$\times$0.03 &143,000 \\
HD 79186  &o5c092	&360	&1271	&E140H  &0.2$\times$0.2	&82,000 \\
HD 88115  &o54305	&3900	&1271	&E140H  &0.1$\times$0.03 &143,000 \\
	  &o6lz30	&600	&1425	&E140M  &0.2$\times$0.2	&38,000 \\
HD 92554  &o6lz36	&1200	&1425	&E140M  &0.2$\times$0.2	&38,000 \\
HD 93205  &o4qx01	&780	&1416	&E140H  &0.2$\times$0.09 &108,000 \\
HD 93222  &o4qx02	&1140	&1416	&E140H  &0.2$\times$0.09 &108,000 \\
HD 93237  &o6lj0g	&810	&1271	&E140H  &0.1$\times$0.03 &143,000 \\
	  &o6lj0g	&570	&1453	&E140H  &0.1$\times$0.03 &143,000 \\
HD 99872  &o6lj0i	&1890	&1271	&E140H  &0.1$\times$0.03 &143,000 \\
	  &o6lj0i	&1260	&1453	&E140H  &0.1$\times$0.03 &143,000 \\
HD 99890  &o6lz45	&600	&1425	&E140M  &0.2$\times$0.2	&38,000 \\
\enddata
\end{deluxetable}

\setcounter{table}{1}
\begin{deluxetable}{lcccccc}
\tablecolumns{7}
\tablewidth{0.6\textwidth}
\tablecaption{(Continued)}
\tablehead{\colhead{} & \colhead{} & \colhead{Exp. Time} & \colhead{Cen.~Wave} 
& \colhead{} & \colhead{Slit} & \colhead{} \\
\colhead{Star} & \colhead{Dataset} & \colhead{(s)} & \colhead{(\AA)} & 
\colhead{Grating} & \colhead{(arcsec)} & \colhead{$R$} }
\startdata
HD 104705 &o57r01	&2400   &1271 	&E140H  &0.2$\times$0.09 &108,000 \\
	  &o57r01	&3320   &1453 	&E140H  &0.2$\times$0.09 &108,000 \\
	  &o57r01	&2900   &1489 	&E140H  &0.2$\times$0.09 &108,000 \\
HD 108002 &o6lj08	&1560	&1271	&E140H  &0.1$\times$0.03 &143,000 \\
	  &o6lj08	&1590	&1453	&E140H  &0.1$\times$0.03 &143,000 \\
HD 108610 &o6lj09	&1680	&1271	&E140H  &0.1$\times$0.03 &143,000 \\
	  &o6lj09	&1470	&1453	&E140H  &0.1$\times$0.03 &143,000 \\
HD 108639 &o6lj0a	&1860	&1271	&E140H  &0.2$\times$0.09 &108,000 \\
\enddata
\end{deluxetable}


\begin{deluxetable}{lccccc}
\tablecolumns{6}
\tablewidth{0.55\textwidth}
\tablecaption{McDonald Observations}
\tablehead{\colhead{} & \colhead{$\alpha$(J2000)} & \colhead{$\delta$(J2000)} & 
\colhead{} & \colhead{Exp. Time} & \colhead{} \\
\colhead{Star} & \colhead{(hr)} & \colhead{(deg)} & \colhead{Setting} & 
\colhead{(s)} & \colhead{S/N} }
\startdata
HD 1383		&0.30     &+61.73	&Ca~{\scriptsize II}/K~{\scriptsize I}	
&7200/5400    &55/190 \\
HD 13268\tablenotemark{a} &2.19    &+56.16	&\ldots/K~{\scriptsize I}	
&\ldots/10800 &\ldots/200 \\
HD 13745\tablenotemark{a} &2.26    &+56.00	&\ldots/K~{\scriptsize I}	
&\ldots/7200  &\ldots/235 \\
HD 22951	&3.71     &+33.96	&Ca~{\scriptsize II}/K~{\scriptsize I}	
&1200/1200    &105/250 \\
HD 36841\tablenotemark{a} &5.58    &$-$0.39	&\ldots/K~{\scriptsize I}	
&\ldots/8700  &\ldots/165 \\
HD 43818\tablenotemark{a} &6.32    &+23.47	&\ldots/K~{\scriptsize I}	
&\ldots/3600  &\ldots/245  \\
HD 52266	&7.01     &$-$5.83	&Ca~{\scriptsize II}/K~{\scriptsize I}	
&9000/3600    &80/210 \\
HD 121968	&13.98    &$-$2.92	&Ca~{\scriptsize II}/K~{\scriptsize I}	
&9000/7200    &20/35 \\
\enddata
\tablenotetext{a}{Ca~{\scriptsize II} data for these sight lines are available 
from previous McDonald observations (see Sheffer et al. 2008).}
\tablenotetext{b}{No Ca~{\scriptsize II} data are available for HD~156110 or 
HD~177989 due to inclement weather during the 2008 June observing run.}
\tablenotetext{c}{High-resolution data on K~{\scriptsize I} for HD~198478 
(55~Cyg) are available from Welty \& Hobbs 2001.\\[0.5in]}
\end{deluxetable}


\begin{deluxetable}{lcccc}
\tablecolumns{5}
\tablewidth{0.45\textwidth}
\tablecaption{Atomic Data\tablenotemark{a}}
\tablehead{\colhead{} & \colhead{IP} & \colhead{} & 
\colhead{$\lambda_0$\tablenotemark{b}} & \colhead{} \\
\colhead{Species} & \colhead{(eV)} & \colhead{Transition} & \colhead{(\AA)} & 
\colhead{$f$} }
\startdata
\multicolumn{5}{c}{UV Lines} \\
\hline
O~{\scriptsize I} & 13.62\phn & 2p $^3$P$-$3s $^5$S$_0$ & 1355.598 & 0.00000116 \\
Cu~{\scriptsize II} & 20.29\phn & 3d $^1$S$-$4p $^1$P$_0$ & 1358.773 & 
0.263\phn\phn\phn\phn\phn \\
B~{\scriptsize II} & 25.15\phn & 2s $^1$S$-$2p $^1$P$_0$ & 1362.463 & 
0.996\phn\phn\phn\phn\phn \\
Ga~{\scriptsize II} & 20.52\phn & 4s $^1$S$-$4p $^1$P$_0$ & 1414.402 & 
1.77\phn\phn\phn\phn\phn\phn \\
\hline
\multicolumn{5}{c}{Visible Lines} \\
\hline
Ca~{\scriptsize II} & 11.87\phn & 4s $^2$S$-$4p $^2$P$_0$ & 3933.661 & 
0.6267\phn\phn\phn\phn \\
K~{\scriptsize I} & \phn4.341 & 4s $^2$S$-$4p $^2$P$_0$ & 7698.965 & 
0.3327\phn\phn\phn\phn \\
\enddata
\tablenotetext{a}{Ionization potentials, wavelengths, and oscillator strengths 
are from Morton 2003.}
\tablenotetext{b}{Vacuum wavelengths are quoted for UV lines. For visible 
lines, wavelengths in air are given.}
\end{deluxetable}


\begin{deluxetable}{lccccccccccc}
\tablecolumns{12}
\tablewidth{0.85\textwidth}
\tablecaption{Component Structure for UV Lines\tablenotemark{a}}
\tablehead{\colhead{} & \multicolumn{3}{c}{O~{\scriptsize I} $\lambda$1355} & 
\colhead{} & \multicolumn{3}{c}{Cu~{\scriptsize II} $\lambda$1358} & 
\colhead{} & \multicolumn{3}{c}{Ga~{\scriptsize II} $\lambda$1414} \\
\cline{2-4} \cline{6-8} \cline{10-12} \\
\colhead{} & \colhead{$v_{\mathrm{LSR}}$} & \colhead{$b$} & \colhead{} & 
\colhead{} & \colhead{$v_{\mathrm{LSR}}$} & \colhead{$b$} & \colhead{} & 
\colhead{} & \colhead{$v_{\mathrm{LSR}}$} & \colhead{$b$} & \colhead{} \\
\colhead{Star} & \colhead{(km s$^{-1}$)} & \colhead{(km s$^{-1}$)} & 
\colhead{$N$/$N_{\mathrm{tot}}$} & \colhead{} & \colhead{(km s$^{-1}$)} & 
\colhead{(km s$^{-1}$)} & \colhead{$N$/$N_{\mathrm{tot}}$} & \colhead{} & 
\colhead{(km s$^{-1}$)} & \colhead{(km s$^{-1}$)} & 
\colhead{$N$/$N_{\mathrm{tot}}$} }
\startdata
HD 1383   &$-$54.6 &3.2  &0.09	 &&$-$55.3 &3.2 &0.13	&&$-$54.9 &3.2  &0.12 \\
          &$-$44.0 &1.6	 &0.08	 &&$-$44.7 &2.9	&0.02	&&$-$43.6 &1.6	&0.08 \\
          &$-$34.9 &1.7	 &0.13	 &&$-$35.6 &1.7	&0.11	&&$-$33.9 &2.6	&0.14 \\
          &\ldots&\ldots&\ldots  &&$-$25.6 &1.6	&0.01	&&$-$24.7 &1.7	&0.14 \\
          &$-$19.4 &1.7	 &0.03	 &&$-$20.1 &3.2	&0.17	&&$-$19.7 &2.6	&0.03 \\
          &$-$12.1 &1.6	 &0.17	 &&$-$12.8 &3.1	&0.04	&&$-$13.9 &1.7	&0.11 \\
          &$-$1.8  &1.6	 &0.38	 &&$-$2.5  &2.4	&0.43	&&$-$2.1  &2.6	&0.30 \\
          &+5.1	   &3.2	 &0.12	 &&+4.4	   &1.6	&0.09	&&+4.8	  &1.7	&0.08 \\
\enddata
\tablenotetext{a}{This table is available in its entirety in a machine-readable 
form in the online journal. A portion is shown here for guidance regarding its 
form and content.}
\end{deluxetable}

\begin{deluxetable}{lccccccc}
\tablecolumns{8}
\tablewidth{0.6\textwidth}
\tablecaption{Component Structure for Visible Lines\tablenotemark{a}}
\tablehead{\colhead{} & \multicolumn{3}{c}{Ca~{\scriptsize II} $\lambda$3933} 
& \colhead{} & \multicolumn{3}{c}{K~{\scriptsize I} $\lambda$7698} \\
\cline{2-4} \cline{6-8} \\
\colhead{} & \colhead{$v_{\mathrm{LSR}}$} & \colhead{$b$} & \colhead{} & 
\colhead{} & \colhead{$v_{\mathrm{LSR}}$} & \colhead{$b$} & \colhead{} \\
\colhead{Star} & \colhead{(km s$^{-1}$)} & \colhead{(km s$^{-1}$)} & 
\colhead{$N$/$N_{\mathrm{tot}}$} & \colhead{} & \colhead{(km s$^{-1}$)} & 
\colhead{(km s$^{-1}$)} & \colhead{$N$/$N_{\mathrm{tot}}$} }
\startdata
HD 1383   &$-$54.4 &2.6 &0.05 &&$-$54.9 &2.3 &0.05  \\
          &$-$48.5 &2.1 &0.03 &&\ldots &\ldots &\ldots  \\
          &$-$43.2 &2.9 &0.05 &&$-$43.4 &0.6 &0.05  \\
          &$-$38.2 &2.1 &0.11 &&$-$39.3 &0.9 &0.36  \\
          &$-$35.0 &2.3 &0.11 &&$-$35.2 &1.9 &0.09  \\
          &$-$31.0 &1.6 &0.05 &&$-$31.2 &0.7 &0.04  \\
          &$-$26.3 &2.5 &0.02 &&\ldots &\ldots &\ldots  \\
          &$-$17.9 &2.8 &0.08 &&\ldots &\ldots &\ldots  \\
          &$-$12.3 &2.2 &0.06 &&$-$11.0 &0.5 &0.01  \\
          &$-$7.2  &2.6 &0.09 &&\ldots &\ldots &\ldots  \\
          &$-$3.1  &2.3 &0.07 &&$-$4.3  &1.6 &0.10  \\
          &+1.0    &2.5 &0.07 &&+1.0    &1.6 &0.13  \\
          &+2.4    &3.0 &0.07 &&+4.0    &0.4 &0.04  \\
          &+5.8    &1.9 &0.11 &&+6.2    &0.8 &0.13  \\
          &+22.2   &3.2 &0.02 &&\ldots &\ldots &\ldots  \\
          &+27.0   &1.8 &0.01 &&\ldots &\ldots &\ldots  \\
\enddata
\tablenotetext{a}{This table is available in its entirety in a machine-readable 
form in the online journal. A portion is shown here for guidance regarding its 
form and content.}
\end{deluxetable}

\clearpage
\setlength{\tabcolsep}{0.035in}
\begin{deluxetable}{lccccccccccccc}
\tablecolumns{14}
\tablewidth{\textwidth}
\tablecaption{Results of Profile Synthesis}
\tablehead{ \colhead{} & 
\multicolumn{2}{c}{log $N$(O~{\scriptsize I})\tablenotemark{a}} & \colhead{} & 
\multicolumn{2}{c}{log $N$(Cu~{\scriptsize II})\tablenotemark{a}} & \colhead{} 
& \multicolumn{2}{c}{log $N$(Ga~{\scriptsize II})\tablenotemark{a}} & 
\colhead{} & \multicolumn{4}{c}{log $N$(B~{\scriptsize II})\tablenotemark{b}} \\
\cline{2-3} \cline{5-6} \cline{8-9} \cline{11-14} \\
\colhead{Star} & \colhead{Direct} & \colhead{Template} & \colhead{} & 
\colhead{Direct} & \colhead{Template} & \colhead{} & \colhead{Direct} & 
\colhead{Template} & \colhead{} & \colhead{O} & \colhead{Cu} & \colhead{Ga} & 
\colhead{Visible} }
\startdata
HD 1383\tablenotemark{c} & 18.13$^{+0.04}_{-0.04}$ & 18.12$^{+0.05}_{-0.06}$ && 
12.79$^{+0.05}_{-0.06}$ & 12.80$^{+0.07}_{-0.08}$ && 12.07$^{+0.05}_{-0.05}$ & 
12.07$^{+0.06}_{-0.07}$ && 11.93$^{+0.06}_{-0.07}$ & 11.89$^{+0.07}_{-0.08}$ & 
11.92$^{+0.06}_{-0.07}$ & 11.92$^{+0.08}_{-0.10}$ \\
 & 18.16$^{+0.03}_{-0.03}$ & 18.16$^{+0.03}_{-0.04}$ && 12.80$^{+0.05}_{-0.05}$ & 
12.81$^{+0.06}_{-0.07}$ && \ldots & \ldots && \ldots & \ldots & \ldots & \ldots \\
HD 13268\tablenotemark{d} & 18.04$^{+0.05}_{-0.06}$ & 18.03$^{+0.08}_{-0.09}$ && 
12.67$^{+0.06}_{-0.07}$ & 12.66$^{+0.08}_{-0.10}$ && 11.58$^{+0.10}_{-0.13}$ & 
11.58$^{+0.14}_{-0.20}$ && 11.41$^{+0.13}_{-0.18}$ & 11.40$^{+0.13}_{-0.20}$ & 
11.47$^{+0.11}_{-0.16}$ & 11.41$^{+0.17}_{-0.29}$ \\
HD 13745\tablenotemark{d} & 17.97$^{+0.06}_{-0.06}$ & 17.96$^{+0.08}_{-0.10}$ && 
12.61$^{+0.05}_{-0.06}$ & 12.61$^{+0.08}_{-0.09}$ && 11.76$^{+0.06}_{-0.07}$ & 
11.76$^{+0.08}_{-0.10}$ && 11.58$^{+0.09}_{-0.11}$ & 11.56$^{+0.10}_{-0.13}$ & 
11.59$^{+0.09}_{-0.11}$ & 11.53$^{+0.14}_{-0.22}$ \\
HD 22951 & 17.75$^{+0.01}_{-0.01}$ & 17.76$^{+0.01}_{-0.01}$ && 
12.32$^{+0.01}_{-0.02}$ & 12.33$^{+0.02}_{-0.02}$ && 11.20$^{+0.02}_{-0.02}$ & 
11.20$^{+0.02}_{-0.03}$ && 10.69$^{+0.11}_{-0.14}$ & 10.72$^{+0.12}_{-0.17}$ & 
10.66$^{+0.10}_{-0.13}$ & 10.73$^{+0.11}_{-0.15}$ \\
HD 23180 & 17.75$^{+0.01}_{-0.01}$ & 17.77$^{+0.01}_{-0.01}$ && 
12.27$^{+0.01}_{-0.01}$ & 12.27$^{+0.02}_{-0.02}$ && 11.10$^{+0.04}_{-0.04}$ & 
11.10$^{+0.04}_{-0.04}$ && 10.82$^{+0.08}_{-0.10}$ & 10.83$^{+0.08}_{-0.10}$ & 
10.83$^{+0.10}_{-0.13}$ & 10.83$^{+0.08}_{-0.10}$ \\
HD 24398 & 17.78$^{+0.01}_{-0.01}$ & 17.78$^{+0.01}_{-0.01}$ && 
12.28$^{+0.01}_{-0.01}$ & 12.31$^{+0.02}_{-0.02}$ && 11.23$^{+0.02}_{-0.02}$ & 
11.23$^{+0.02}_{-0.02}$ && 10.66$^{+0.12}_{-0.17}$ & 10.65$^{+0.10}_{-0.13}$ & 
10.69$^{+0.12}_{-0.17}$ & 10.63$^{+0.12}_{-0.16}$ \\
\enddata
\end{deluxetable}

\setcounter{table}{6}
\setlength{\tabcolsep}{0.035in}
\begin{deluxetable}{lccccccccccccc}
\tablecolumns{14}
\tablewidth{\textwidth}
\tablecaption{(Continued)}
\tablehead{ \colhead{} & 
\multicolumn{2}{c}{log $N$(O~{\scriptsize I})\tablenotemark{a}} & \colhead{} & 
\multicolumn{2}{c}{log $N$(Cu~{\scriptsize II})\tablenotemark{a}} & \colhead{} 
& \multicolumn{2}{c}{log $N$(Ga~{\scriptsize II})\tablenotemark{a}} & 
\colhead{} & \multicolumn{4}{c}{log $N$(B~{\scriptsize II})\tablenotemark{b}} \\
\cline{2-3} \cline{5-6} \cline{8-9} \cline{11-14} \\
\colhead{Star} & \colhead{Direct} & \colhead{Template} & \colhead{} & 
\colhead{Direct} & \colhead{Template} & \colhead{} & \colhead{Direct} & 
\colhead{Template} & \colhead{} & \colhead{O} & \colhead{Cu} & \colhead{Ga} & 
\colhead{Visible} }
\startdata
HD 185418 & 18.05$^{+0.03}_{-0.03}$ & 18.07$^{+0.03}_{-0.03}$ && 
12.65$^{+0.04}_{-0.04}$ & 12.66$^{+0.04}_{-0.04}$ && \ldots & \ldots && 
11.29$^{+0.18}_{-0.31}$ & 11.29$^{+0.18}_{-0.32}$ & \ldots & 11.25$^{+0.17}_{-0.27}$ 
\\
HD 187311 & 17.34$^{+0.09}_{-0.12}$ & \ldots && 11.97$^{+0.10}_{-0.13}$ & \ldots && 
$<$ 11.21 & \ldots && 11.34$^{+0.09}_{-0.12}$ & 11.36$^{+0.09}_{-0.12}$ & \ldots & 
\ldots \\
HD 192035 & 17.98$^{+0.05}_{-0.06}$ & 17.98$^{+0.06}_{-0.08}$ && 
12.69$^{+0.04}_{-0.04}$ & 12.66$^{+0.06}_{-0.07}$ && 11.58$^{+0.08}_{-0.09}$  & 
11.53$^{+0.10}_{-0.14}$ && 11.27$^{+0.14}_{-0.22}$ & 11.32$^{+0.12}_{-0.17}$ & 
11.30$^{+0.14}_{-0.20}$ & 11.28$^{+0.18}_{-0.33}$ \\
HD 195965 & 17.78$^{+0.05}_{-0.05}$ & 17.79$^{+0.05}_{-0.06}$ && 
12.42$^{+0.05}_{-0.06}$ & 12.44$^{+0.06}_{-0.08}$ && 11.39$^{+0.05}_{-0.05}$  & 
11.39$^{+0.04}_{-0.05}$ && 11.29$^{+0.20}_{-0.38}$ & 11.28$^{+0.17}_{-0.29}$ & 
11.21$^{+0.20}_{-0.37}$ & 11.22$^{+0.20}_{-0.39}$ \\
HD 198478 & 18.04$^{+0.02}_{-0.02}$ & 18.03$^{+0.03}_{-0.03}$ && 
12.74$^{+0.03}_{-0.03}$ & 12.77$^{+0.03}_{-0.04}$ && \ldots & \ldots && 
11.34$^{+0.14}_{-0.20}$ & 11.34$^{+0.18}_{-0.32}$ & \ldots & 11.33$^{+0.22}_{-0.49}$ 
\\
HD 203338 & 17.83$^{+0.07}_{-0.08}$ & \ldots && 12.50$^{+0.07}_{-0.08}$ & \ldots && 
11.37$^{+0.11}_{-0.14}$ & \ldots && 11.40$^{+0.14}_{-0.21}$ & 11.36$^{+0.16}_{-0.25}$ 
& 11.44$^{+0.12}_{-0.16}$ & \ldots \\
HD 203374\tablenotemark{c} & 18.01$^{+0.04}_{-0.04}$ & 18.01$^{+0.07}_{-0.08}$ && 
12.73$^{+0.03}_{-0.04}$ & 12.73$^{+0.05}_{-0.06}$ && 11.51$^{+0.07}_{-0.09}$  & 
11.53$^{+0.11}_{-0.16}$ && 11.40$^{+0.08}_{-0.11}$ & 11.44$^{+0.10}_{-0.12}$ & 
11.39$^{+0.09}_{-0.12}$ & 11.41$^{+0.15}_{-0.22}$ \\
 & 17.99$^{+0.02}_{-0.02}$ & 17.99$^{+0.02}_{-0.02}$ && 12.70$^{+0.02}_{-0.02}$ & 
12.70$^{+0.03}_{-0.03}$ && \ldots & \ldots && \ldots & \ldots & \ldots & \ldots \\
\enddata
\tablenotetext{a}{Column density in cm$^{-2}$ obtained either by direct profile 
synthesis or by applying a fixed velocity template based on visible data.}
\tablenotetext{b}{B~{\scriptsize II} column density in cm$^{-2}$ obtained by 
fitting the observed spectrum with a fixed profile template derived from either 
the O~{\scriptsize I}, Cu~{\scriptsize II}, or Ga~{\scriptsize II} line, or 
from visible data.}
\tablenotetext{c}{Second line lists results from high-resolution (E140H) data.}
\tablenotetext{d}{Fits to the B~{\scriptsize II} profiles toward HD~13268 and 
HD~13745 exclude components at large negative velocities (see text).}
\end{deluxetable}

\begin{deluxetable}{lcccc}
\tablecolumns{5}
\tablewidth{0.5\textwidth}
\tablecaption{Results for Distinct Cloud Complexes}
\tablehead{\colhead{} & \colhead{$v_{\mathrm{LSR}}$(O~{\scriptsize I})} & 
\colhead{} & \colhead{} & \colhead{} \\
\colhead{Star} & \colhead{(km s$^{-1}$)} & 
\colhead{log $N$(O~{\scriptsize I})} & 
\colhead{log $N$(B~{\scriptsize II})} & \colhead{log (B/O)} }
\startdata
\multicolumn{5}{c}{Perseus Arm} \\
\hline
HD 1383    & $-$43.4 & 17.60$^{+0.08}_{-0.10}$   & $<$ 11.41               & $<$ 
$-$6.19               \\
 & $-$23.2\tablenotemark{a} & $<$ 17.09        & 11.33$^{+0.08}_{-0.10}$   & $>$ 
$-$5.76               \\
	   &  $-$3.8 & 17.98$^{+0.04}_{-0.05}$   & 11.66$^{+0.09}_{-0.11}$   & 
$-$6.32$^{+0.09}_{-0.12}$   \\
HD 13268   & $-$35.8 & 17.44$^{+0.11}_{-0.15}$   & $<$ 11.32		 & $<$ 
$-$6.12	             \\
	   &  $-$8.3 & 17.92$^{+0.06}_{-0.06}$   & 11.42$^{+0.14}_{-0.20}$   & 
$-$6.49$^{+0.14}_{-0.22}$   \\
HD 13745   & $-$42.2 & 17.23$^{+0.16}_{-0.26}$   & $<$ 11.30		 & $<$ 
$-$5.93	             \\
	   &  $-$9.1 & 17.88$^{+0.06}_{-0.06}$   & 11.56$^{+0.10}_{-0.14}$   & 
$-$6.32$^{+0.12}_{-0.16}$   \\
\enddata
\tablenotetext{a}{Velocity of the B~{\scriptsize II} component, since the 
feature is not significantly detected in O~{\scriptsize I}.} \\
\end{deluxetable}


\begin{deluxetable}{lccccccc}
\tablecolumns{8}
\tablewidth{0.65\textwidth}
\tablecaption{Comparison with Previous Studies: O~{\scriptsize I} and Cu~{\scriptsize II}}
\tablehead{ \colhead{} & \multicolumn{3}{c}{log $N$(O~{\scriptsize I})} & \colhead{} 
& \multicolumn{3}{c}{log $N$(Cu~{\scriptsize II})} \\
\cline{2-4} \cline{6-8} \\
\colhead{Star} & \colhead{This Work} & \colhead{Previous Result} & 
\colhead{Ref.} & \colhead{} & \colhead{This Work} & \colhead{Previous Result} & 
\colhead{Ref.} }
\startdata
HD 1383  & 18.13$^{+0.04}_{-0.04}$ & $18.15\pm0.06$ &1 && 12.79$^{+0.05}_{-0.06}$ & 
$12.73\pm0.08$\tablenotemark{a} &2 \\
HD 13268 & 18.04$^{+0.05}_{-0.06}$ & $18.13\pm0.06$ &1 && 12.67$^{+0.06}_{-0.07}$ & 
$12.67\pm0.05$\tablenotemark{a} &2 \\
HD 23180 & 17.75$^{+0.01}_{-0.01}$ & 17.93$^{+0.09}_{-0.11}$\tablenotemark{a} &3 && 
12.27$^{+0.01}_{-0.01}$ & \ldots & \\
HD 24398 & 17.78$^{+0.01}_{-0.01}$ & 17.71$^{+0.05}_{-0.06}$\tablenotemark{a} &3 && 
12.28$^{+0.01}_{-0.01}$ & \ldots & \\
HD 24534 & 17.886$^{+0.003}_{-0.004}$ & 17.83$^{+0.04}_{-0.03}$ &4 && 
12.36$^{+0.01}_{-0.01}$ & \ldots & \\
 & & $17.87\pm0.02$ &5 && & & \\
 & & $18.12\pm0.06$\tablenotemark{a} &6 && & & \\
HD 36841 & 17.64$^{+0.07}_{-0.08}$ & $17.63\pm0.08$ &1 && 12.00$^{+0.10}_{-0.13}$ & 
$12.05\pm0.10$\tablenotemark{a} &2 \\
\enddata
\tablenotetext{a}{Revised using $f$-values for O~{\scriptsize I} $\lambda$1355 and 
Cu~{\scriptsize II} $\lambda$1358 from Morton 2003.}
\tablerefs{(1) Cartledge et al. 2004; (2) Cartledge et al. 2006; (3) Meyer et 
al. 1998; (4) Jensen et al. 2005; (5) Knauth et al. 2003a; (6) Snow et al. 
1998; (7) Andr\'e et al. 2003; (8) Howk et al. 2000; (9) Cartledge et al. 2001; 
(10) Sonnentrucker et al. 2003; (11) Hoopes et al. 2003; (12) Federman et al. 
2005.}
\end{deluxetable}

\begin{deluxetable}{lccccc}
\tablecolumns{6}
\tablewidth{0.5\textwidth}
\tablecaption{Comparison with Previous Studies: B~{\scriptsize II}}
\tablehead{ \colhead{} & \multicolumn{2}{c}{This Work} & \colhead{} & 
\multicolumn{2}{c}{Howk et al. 2000} \\
\cline{2-3} \cline{5-6} \\
\colhead{Star} & \colhead{$W_{\lambda}$(B~{\scriptsize II})\tablenotemark{a}} & 
\colhead{log $N$(B~{\scriptsize II})} & \colhead{} & 
\colhead{$W_{\lambda}$(B~{\scriptsize II})\tablenotemark{a}} & 
\colhead{log $N$(B~{\scriptsize II})} }
\startdata
HD 104705 & $5.8\pm1.0$ & 11.56$^{+0.07}_{-0.08}$ && $5.3\pm1.2$ & 
11.53$^{+0.09}_{-0.11}$ \\
HD 121968 & $1.3\pm0.5$ & 10.89$^{+0.15}_{-0.23}$ && $1.8\pm0.6$ & 
11.06$^{+0.12}_{-0.18}$ \\
HD 177989 & $3.4\pm0.6$ & 11.33$^{+0.07}_{-0.08}$ && $4.8\pm0.4$ & 
11.48$^{+0.04}_{-0.04}$ \\
HD 218915 & $3.9\pm1.4$ & 11.38$^{+0.13}_{-0.19}$ && $2.3\pm0.6$ & 
11.16$^{+0.10}_{-0.13}$ \\
HDE 303308 & $5.5\pm2.1$ & 11.54$^{+0.14}_{-0.21}$ && $8.0\pm1.6$ & 
11.71$^{+0.08}_{-0.10}$ \\
\enddata
\tablenotetext{a}{Fitted (this work) or integrated (Howk et al. 2000) 
equivalent width in m\AA.}
\end{deluxetable}


\begin{deluxetable}{lccccccc}
\tablecolumns{8}
\tablewidth{0.7\textwidth}
\tablecaption{Atomic and Molecular Hydrogen Column Densities}
\tablehead{\colhead{Star} & \colhead{log $N$(H~{\scriptsize I})} & \colhead{Ref.} & 
\colhead{log $N$(H$_2$)} & \colhead{Ref.} & \colhead{log $N_{\mathrm{tot}}$(H)} & 
\colhead{log $f$(H$_2$)} & \colhead{log $\langle n_{\mathrm{H}} \rangle$} }
\startdata
\multicolumn{8}{c}{STIS Sample} \\
\hline
HD 1383   & $21.36\pm0.10$ & 1 & $20.45\pm0.07$ & 2 & 21.46$^{+0.08}_{-0.10}$ & 
$-$0.70 & $-$0.49 \\
HD 13268  & $21.34\pm0.18$ & 1 & $20.51\pm0.08$ & 3 & 21.45$^{+0.14}_{-0.22}$ & 
$-$0.64 & $-$0.40 \\
HD 13745  & $21.25\pm0.10$ & 1 & $20.67\pm0.08$ & 3 & 21.43$^{+0.07}_{-0.09}$ & 
$-$0.46 & $-$0.42 \\
HD 22951  & $21.04\pm0.11$ & 4 & $20.46\pm0.10$ & 5 & 21.22$^{+0.08}_{-0.10}$ & 
$-$0.46 & +0.28 \\
HD 23180  & $20.84\pm0.07$ & 6 & $20.61\pm0.08$ & 5 & 21.18$^{+0.06}_{-0.06}$ & 
$-$0.27 & +0.04 \\
HD 24398  & $20.80\pm0.04$ & 6 & $20.67\pm0.10$ & 5 & 21.20$^{+0.06}_{-0.07}$ & 
$-$0.22 & +0.23 \\
HD 24534  & $20.73\pm0.06$ & 1 & $20.92\pm0.04$ & 7 & 21.34$^{+0.03}_{-0.04}$ & 
$-$0.12 & +0.25 \\
\enddata
\end{deluxetable}

\setcounter{table}{10}
\begin{deluxetable}{lccccccc}
\tablecolumns{8}
\tablewidth{0.7\textwidth}
\tablecaption{(Continued)}
\tablehead{\colhead{Star} & \colhead{log $N$(H~{\scriptsize I})} & \colhead{Ref.} & 
\colhead{log $N$(H$_2$)} & \colhead{Ref.} & \colhead{log $N_{\mathrm{tot}}$(H)} & 
\colhead{log $f$(H$_2$)} & \colhead{log $\langle n_{\mathrm{H}} \rangle$} }
\startdata
HD 185418 & $21.15\pm0.07$ & 10 & $20.71\pm0.10$ & 10 & 21.38$^{+0.06}_{-0.07}$ & 
$-$0.38 & $-$0.08 \\
HD 192035 & $21.09\pm0.19$ & 1 & $20.68\pm0.08$ & 3 & 21.34$^{+0.12}_{-0.17}$ & 
$-$0.36 & $-$0.49 \\
HD 195965 & $20.90\pm0.09$ & 1 & $20.34\pm0.08$ & 3 & 21.09$^{+0.07}_{-0.08}$ & 
$-$0.45 & $-$0.12 \\
HD 198478 & $21.32\pm0.15$ & 2 & $20.87\pm0.15$ & 2 & 21.55$^{+0.11}_{-0.15}$ & 
$-$0.38 & +0.22 \\
HD 203374 & $21.20\pm0.08$ & 9 & $20.70\pm0.01$ & 13 & 21.41$^{+0.05}_{-0.06}$ & 
$-$0.41 & +0.13 \\
HD 207308 & $21.20\pm0.10$ & 11 & $20.86\pm0.01$ & 13 & 21.48$^{+0.06}_{-0.06}$ & 
$-$0.32 & +0.04 \\
HD 207538 & $21.34\pm0.12$ & 1 & $20.91\pm0.05$ & 13 & 21.58$^{+0.08}_{-0.09}$ & 
$-$0.37 & +0.17 \\
HD 208947 & $21.10\pm0.06$ & 9 & $20.09\pm0.21$ & 8 & 21.18$^{+0.06}_{-0.08}$ & 
$-$0.78 & $-$0.01 \\
\enddata
\tablenotetext{a}{Derived from the column density of Kr~{\scriptsize I}, assuming an 
interstellar krypton abundance of log~(Kr/H)$_{\mathrm{ISM}}$~=~$-9.02\pm0.02$ 
(Cartledge et al. 2008).}
\tablerefs{(1) DS94; (2) Cartledge et al. 2004; (3) Sheffer et al. 2008; (4) 
BSD78; (5) Savage et al. 1977; (6) Weighted mean of BSD78 and DS94; (7) 
Rachford et al. 2002; (8) Derived from the CH/H$_2$ relation in Sheffer et al. 
2008; (9) Cartledge et al. 2008; (10) Andr\'e et al. 2003; (11) Jensen \& Snow 
2007a; (12) Burgh et al. 2007; (13) Pan et al. 2005.}
\end{deluxetable}


\begin{deluxetable}{lccccccc}
\tablecolumns{8}
\tablewidth{0.85\textwidth}
\tablecaption{Elemental Abundances of Boron, Oxygen, Copper, and Gallium}
\tablehead{\colhead{Star} & \colhead{log $N$(O~{\scriptsize I})} & 
\colhead{log $N$(B~{\scriptsize II})} & \colhead{log (B/O)} & \colhead{log (B/H)} & 
\colhead{log (O/H)} & \colhead{log (Cu/H)} & \colhead{log (Ga/H)} }
\startdata
\multicolumn{8}{c}{STIS Sample} \\
\hline
HD 1383 &  18.13$^{+0.04}_{-0.04}$ & 11.92$^{+0.07}_{-0.08}$ & 
$-$6.21$^{+0.08}_{-0.09}$ & $-$9.54$^{+0.10}_{-0.14}$ & $-$3.33$^{+0.09}_{-0.11}$ & 
$-$8.67$^{+0.10}_{-0.12}$ & $-$9.39$^{+0.09}_{-0.12}$ \\
HD 13268 & 18.04$^{+0.05}_{-0.06}$ & 11.55$^{+0.13}_{-0.18}$ & 
$-$6.49$^{+0.13}_{-0.20}$ & $-$9.90$^{+0.18}_{-0.32}$ & $-$3.41$^{+0.15}_{-0.24}$ & 
$-$8.78$^{+0.15}_{-0.24}$ & $-$9.87$^{+0.17}_{-0.28}$ \\
HD 13745 & 17.97$^{+0.06}_{-0.06}$ & 11.65$^{+0.10}_{-0.14}$ & 
$-$6.32$^{+0.11}_{-0.16}$ & $-$9.78$^{+0.12}_{-0.17}$ & $-$3.46$^{+0.09}_{-0.11}$ & 
$-$8.82$^{+0.09}_{-0.11}$ & $-$9.67$^{+0.09}_{-0.12}$ \\
HD 22951 & 17.75$^{+0.01}_{-0.01}$ & 10.70$^{+0.11}_{-0.15}$ & 
$-$7.05$^{+0.11}_{-0.15}$ & $-$10.52$^{+0.13}_{-0.19}$ & $-$3.48$^{+0.08}_{-0.11}$ & 
$-$8.90$^{+0.08}_{-0.11}$ & $-$10.02$^{+0.09}_{-0.11}$ \\
HD 23180 & 17.75$^{+0.01}_{-0.01}$ & 10.83$^{+0.09}_{-0.11}$ & 
$-$6.92$^{+0.09}_{-0.11}$ & $-$10.35$^{+0.10}_{-0.13}$ & $-$3.43$^{+0.06}_{-0.06}$ & 
$-$8.90$^{+0.06}_{-0.07}$ & $-$10.08$^{+0.06}_{-0.08}$ \\
HD 24398 & 17.78$^{+0.01}_{-0.01}$ & 10.66$^{+0.12}_{-0.16}$ & 
$-$7.12$^{+0.12}_{-0.16}$ & $-$10.54$^{+0.13}_{-0.18}$ & $-$3.41$^{+0.06}_{-0.07}$ & 
$-$8.92$^{+0.06}_{-0.08}$ & $-$9.97$^{+0.06}_{-0.08}$ \\
HD 24534 & 17.886$^{+0.003}_{-0.004}$ & 10.82$^{+0.04}_{-0.05}$ & 
$-$7.06$^{+0.04}_{-0.05}$ & $-$10.52$^{+0.05}_{-0.06}$ & $-$3.46$^{+0.03}_{-0.04}$ & 
$-$8.98$^{+0.03}_{-0.04}$ & $-$10.13$^{+0.04}_{-0.04}$ \\
\enddata
\end{deluxetable}

\setcounter{table}{11}
\begin{deluxetable}{lccccccc}
\tablecolumns{8}
\tablewidth{0.85\textwidth}
\tablecaption{(Continued)}
\tablehead{\colhead{Star} & \colhead{log $N$(O~{\scriptsize I})} & 
\colhead{log $N$(B~{\scriptsize II})} & \colhead{log (B/O)} & \colhead{log (B/H)} & 
\colhead{log (O/H)} & \colhead{log (Cu/H)} & \colhead{log (Ga/H)} }
\startdata
HD 104705 & 17.88$^{+0.03}_{-0.03}$ & 11.56$^{+0.07}_{-0.08}$ & 
$-$6.32$^{+0.07}_{-0.09}$ & $-$9.63$^{+0.09}_{-0.11}$ & $-$3.30$^{+0.07}_{-0.08}$ & 
$-$8.53$^{+0.06}_{-0.08}$ & $-$9.60$^{+0.06}_{-0.07}$ \\
HD 108002 & 17.81$^{+0.05}_{-0.05}$ & $<$ 11.40 & $<-$6.41 & \ldots & \ldots & 
\ldots & \ldots \\
HD 108639 & 18.03$^{+0.02}_{-0.03}$ & 11.55$^{+0.08}_{-0.11}$ & 
$-$6.48$^{+0.09}_{-0.11}$ & $-$9.84$^{+0.11}_{-0.16}$ & $-$3.36$^{+0.08}_{-0.11}$ & 
$-$8.58$^{+0.08}_{-0.11}$ & $-$9.60$^{+0.08}_{-0.11}$ \\
HD 110434 & 17.57$^{+0.04}_{-0.05}$ & 11.14$^{+0.12}_{-0.18}$ & 
$-$6.43$^{+0.13}_{-0.19}$ & \ldots & \ldots & \ldots & \ldots \\
HD 111934 & 17.95$^{+0.03}_{-0.04}$ & $<$ 11.69 & $<-$6.26 & \ldots & \ldots & 
\ldots & \ldots \\
HD 114441 & 17.93$^{+0.06}_{-0.06}$ & 11.55$^{+0.11}_{-0.16}$ & 
$-$6.38$^{+0.12}_{-0.18}$ & \ldots & \ldots & \ldots & \ldots \\
HD 114886 & 17.97$^{+0.03}_{-0.03}$ & 11.57$^{+0.09}_{-0.12}$ & 
$-$6.41$^{+0.10}_{-0.12}$ & $-$9.85$^{+0.10}_{-0.14}$ & $-$3.44$^{+0.06}_{-0.07}$ & 
$-$8.89$^{+0.07}_{-0.08}$ & $-$9.68$^{+0.06}_{-0.07}$ \\
HD 116781 & 18.00$^{+0.03}_{-0.04}$ & 11.56$^{+0.13}_{-0.18}$ & 
$-$6.44$^{+0.13}_{-0.18}$ & $-$9.69$^{+0.15}_{-0.22}$ & $-$3.25$^{+0.09}_{-0.12}$ & 
$-$8.63$^{+0.10}_{-0.12}$ & \ldots \\
\enddata
\tablenotetext{a}{From Meyer et al. 1998.}
\tablenotetext{b}{From Jura et al. 1996.}
\tablenotetext{c}{From Lambert et al. 1998.}
\tablenotetext{d}{From Howk et al. 2000.}
\tablenotetext{e}{Derived using column densities of Cu~{\scriptsize II} and 
Ga~{\scriptsize II} from Cardelli et al. 1991.}
\end{deluxetable}


\begin{deluxetable}{lcccccccccc}
\tablecolumns{11}
\tablewidth{0.75\textwidth}
\tablecaption{Mean Elemental Abundances and Depletions}
\tablehead{\colhead{} & \colhead{} & \multicolumn{3}{c}{Abundance} && 
\multicolumn{2}{c}{Depletion\tablenotemark{a}} && 
\multicolumn{2}{c}{Sample Size} \\
\cline{3-5} \cline{7-8} \cline{10-11} \\
\colhead{Element} & \colhead{$T_{\mathrm{cond}}$ (K)\tablenotemark{b}} & 
\colhead{Solar\tablenotemark{b}} & \colhead{Warm} & \colhead{Cold} && 
\colhead{Warm} & \colhead{Cold} && \colhead{Warm} & \colhead{Cold} }
\startdata
Oxygen & 180 & $-3.31\pm0.05$ & $-3.36\pm0.13$ & $-3.51\pm0.08$ && 
$-$0.05 & $-$0.20 && 15 & 19 \\
Boron & 908 & $-9.22\pm0.04$ & $-9.63\pm0.09$ & $-10.42\pm0.18$ && 
$-$0.41 & $-$1.20 && 6 & 7 \\
Gallium & 968 & $-8.90\pm0.06$ & $-9.54\pm0.15$ & $-10.02\pm0.14$ && 
$-$0.64 & $-$1.12 && 6 & 10 \\
Copper & 1037 & $-7.74\pm0.06$ & $-8.54\pm0.13$ & $-8.94\pm0.12$ && 
$-$0.80 & $-$1.20 && 9 & 17 \\
\enddata
\tablenotetext{a}{For element X, the logarithmic depletion is defined as 
[X/H]~$\equiv$~log~(X/H)~$-$~log~(X/H)$_{\odot}$.}
\tablenotetext{b}{From Lodders 2003.}
\end{deluxetable}

\begin{deluxetable}{lcc}
\tablecolumns{3}
\tablewidth{0.45\textwidth}
\tablecaption{Stellar Disk and Interstellar Boron Abundances}
\tablehead{\colhead{} & \colhead{log (B/H)} & \colhead{Reference} }
\startdata
Warm ISM & $-9.63\pm0.09$ & This Work \\
F and G Stars & $-9.49\pm0.20$ & Cunha et al. 2000 \\
B Stars & $-9.46\pm0.17$ & Venn et al. 2002 \\
Solar Photosphere & $-9.30^{+0.21}_{-0.12}$ & Cunha \& Smith 1999 \\
Solar System (Meteorites) & $-9.22\pm0.04$ & Lodders 2003 \\
\enddata
\end{deluxetable}


\begin{deluxetable}{lcccccccc}
\tablecolumns{9}
\tablewidth{0.7\textwidth}
\tablecaption{Light Element Abundance Ratios}
\tablehead{\colhead{Star} & \colhead{log (Li/H)} & \colhead{Ref.} & 
\colhead{log (Be/H)} & \colhead{Ref.} & \colhead{log (B/H)} & \colhead{Ref.} & 
\colhead{Li/B\tablenotemark{a}} & \colhead{B/Be\tablenotemark{a}} }
\startdata
$o$ Per & $-$9.56$^{+0.16}_{-0.27}$ & 1 & \ldots & & $-$10.35$^{+0.10}_{-0.13}$ & 2 
& $6.2\pm3.2$ & \ldots \\
$\zeta$ Per & $-$9.48$^{+0.15}_{-0.22}$ & 3 & $<-$12.18 & 4 & 
$-$10.54$^{+0.13}_{-0.18}$ & 2 & $11.4\pm5.7$ & $>$ 40 \\
X Per & $-$10.10$^{+0.14}_{-0.22}$ & 1 & \ldots & & $-$10.52$^{+0.05}_{-0.06}$ & 2 
& $2.6\pm1.1$ & \ldots \\
$\delta$ Sco & \ldots & & $<-$10.98 & 5 & $-$10.16$^{+0.10}_{-0.14}$ & 6 & \ldots 
& $>$ 7 \\
$\zeta$ Oph & $-$9.41$^{+0.14}_{-0.20}$ & 1 & $<-$11.13 & 7 & 
$-$10.56$^{+0.13}_{-0.18}$ & 6 & $14.3\pm7.2$ & $>$ 4 \\
\hline
Solar System & $-8.72\pm0.06$ & 8 & $-10.59\pm0.08$ & 8 & $-9.22\pm0.04$ 
& 8 & $3.2\pm0.6$ & $23\pm5$ \\
\enddata
\tablenotetext{a}{For comparison, standard GCR spallation predicts ratios of 
Li/B = 0.66 and B/Be = 15 (MAR).}
\tablerefs{(1) Knauth et al. 2003b; (2) This Work; (3) Knauth et al. 2000 
(updated in Knauth et al. 2003b; assumes 40\% error); (4) H\'ebrard et al. 
1997; (5) Boesgaard 1985; (6) Lambert et al. 1998; (7) Baade \& Crane 1991; (8) 
Lodders 2003.} \\
\end{deluxetable}


\begin{thebibliography}{}
\bibitem[Abdo et al. (2010)]{a10} Abdo, A. A., et al. 2010, \apj, 712, 459
\bibitem[Acciari et al. (2009)]{a09} Acciari, V. A., et al. 2009, \apjl, 698, 
L133
\bibitem[Aloisi et al. (2007)]{a07} Aloisi, A., Bohlin, R., \& Kim Quijano, J. 
2007, STIS ISR 2007-01, ``New On-Orbit Sensitivity Calibration for All STIS 
Echelle Modes''
\bibitem[Andersson et al. (2002)]{awc02} Andersson, B.-G., Wannier, P. G., \& 
Crawford, I. A. 2002, \mnras, 334, 327
\bibitem[Andr\'e et al. (2003)]{a03} Andr\'e, M. K., et al. 2003, \apj, 591, 
1000
\bibitem[Baade \& Crane (1991)]{bc91} Baade, D., \& Crane, P. 1991, ESO 
Messenger No. 61, 49
\bibitem[Boesgaard (1985)]{b85} Boesgaard, A. M. 1985, \pasp, 97, 37
\bibitem[Boesgaard et al. (1999)]{b99} Boesgaard, A. M., Deliyannis, C. P., 
King, J. R., Ryan, S. G., Vogt, S. S., \& Beers, T. C. 1999, \aj, 117, 1549
\bibitem[Boesgaard \& Heacox (1978)]{bh78} Boesgaard, A. M., \& Heacox, W. D. 
1978, \apj, 226, 888
\bibitem[Boesgaard et al. (2004)]{b04} Boesgaard, A. M., McGrath, E. J., 
Lambert, D. L., \& Cunha, K. 2004, \apj, 606, 306
\bibitem[Bohlin et al. (1978)]{bsd78} Bohlin, R. C., Savage, B. D., \& Drake, 
J. F. 1978, \apj, 224, 132 (BSD78)
\bibitem[Brown et al. (2009)]{b09} Brown, M. S., Federman, S. R., Irving, R. 
E., Cheng, S., \& Curtis, L. J. 2009, \apj, 702, 880
\bibitem[Burgh et al. (2007)]{bfm07} Burgh, E. B., France, , K., \& McCandliss, 
S. R. 2007, \apj, 658, 446
\bibitem[Cardelli et al. (1991)]{cse91} Cardelli, J. A., Savage, B. D., \& 
Ebbets, D. C. 1991, \apjl, 383, L23
\bibitem[Cartledge et al. (2004)]{c04} Cartledge, S. I. B., Lauroesch, J. T., 
Meyer, D. M., \& Sofia, U. J. 2004, \apj, 613, 1037
\bibitem[Cartledge et al. (2006)]{c06} Cartledge, S. I. B., Lauroesch, J. T., 
Meyer, D. M., \& Sofia, U. J. 2006, \apj, 641, 327
\bibitem[Cartledge et al. (2008)]{c08} Cartledge, S. I. B., Lauroesch, J. T., 
Meyer, D. M., Sofia, U. J., \& Clayton, G. C. 2008, \apj, 687, 1043
\bibitem[Cartledge et al. (2001)]{c01} Cartledge, S. I. B., Meyer, D. M., 
Lauroesch, J. T., \& Sofia, U. J. 2001, \apj, 562, 394
\bibitem[Cass\'e et al. (1995)]{clv95} Cass\'e, M., Lehoucq, R., \& 
Vangioni-Flam, E. 1995, \nat, 373, 318
\bibitem[Crinklaw et al. (1994)]{cfj94} Crinklaw, G., Federman, S. R., \& 
Joseph, C. L. 1994, \apj, 424, 748
\bibitem[Cunha \& Smith (1999)]{cs99} Cunha, K., \& Smith, V. V. 1999, \apj, 
512, 1006
\bibitem[Cunha et al. (2000)]{c00} Cunha, K., Smith, V. V., Boesgaard, A. M., 
\& Lambert, D. L. 2000, \apj, 530, 939
\bibitem[Diplas \& Savage (1994)]{ds94} Diplas, A., \& Savage, B. D. 1994, 
\apjs, 93, 211 (DS94)
\bibitem[Duncan et al. (1992)]{dll92} Duncan, D. K., Lambert, D. L., \& Lemke, 
M. 1992, \apj, 401, 584
\bibitem[Duncan et al. (1997)]{d97} Duncan, D. K., Primas, F., Rebull, L. M., 
Boesgaard, A. M., Deliyannis, C. P., Hobbs, L. M., King, J. R., \& Ryan, S. G. 
1997, \apj, 488, 338
\bibitem[Federman et al. (1996a)]{f96a} Federman, S. R., Lambert, D. L., 
Cardelli, J. A., \& Sheffer, Y. 1996a, \nat, 381, 764
\bibitem[Federman et al. (2003)]{f03} Federman, S. R., Lambert, D. L., Sheffer, 
Y., Cardelli, J. A., Andersson, B.-G., van Dishoeck, E. F., \& Zsarg\'o, J. 
2003, \apj, 591, 986
\bibitem[Federman et al. (1993)]{f93} Federman, S. R., Sheffer, Y., Lambert, D. 
L., \& Gilliland, R. L. 1993, \apjl, 413, L51
\bibitem[Federman et al. (2005)]{f05} Federman, S. R., Sheffer, Y., Lambert, D. 
L., \& Smith, V. V. 2005, \apj, 619, 884
\bibitem[Federman et al. (1994)]{f94} Federman, S. R., Strom, C. J., Lambert, 
D. L., Cardelli, J. A., Smith, V. V., \& Joseph, C. L. 1994, \apj, 424, 772
\bibitem[Federman et al. (1996b)]{fwl96b} Federman, S. R., Weber, J., \& 
Lambert, D. L. 1996b, \apj, 463, 181
\bibitem[Fields et al. (2000)]{f00} Fields, B. D., Olive, K. A., Vangioni-Flam, 
E., \& Cass\'e, M. 2000, \apj, 540, 930
\bibitem[Garc\'ia L\'opez et al. (1998)]{gl98} Garc\'ia L\'opez, R. J., 
Lambert, D. L., Edvardsson, B., Gustafsson, B., Kiselman, D., \& Rebolo, R. 
1998, \apj, 500, 241
\bibitem[Gilmore et al. (1992)]{g92} Gilmore, G., Gustafsson, B., Edvardsson, 
B., \& Nissen, P. E. 1992, \nat, 357, 379
\bibitem[Ginestet et al. (1999)]{gcj99} Ginestet, N., Carquillat, J. M., \& 
Jaschek, C. 1999, \aaps, 134, 473
\bibitem[Hartquist et al. (1978)]{hdd78} Hartquist, T. W., Doyle, H. T., \& 
Dalgarno, A. 1978, \aap, 68, 65
\bibitem[H\'ebrard et al. (1997)]{h97} H\'ebrard, G., Lemoine, M., Ferlet, R., 
\& Vidal-Madjar, A. 1997, \aap, 324, 1145
\bibitem[Hobbs et al. (1993)]{h93} Hobbs, L. M., Welty, D. E., Morton, D. C., 
Spitzer, L., \& York, D. G. 1993, \apj, 411, 750
\bibitem[Hoopes et al. (2003)]{h03} Hoopes, C. G., Sembach, K. R., H\'ebrard, 
G., Moos, H. W., \& Knauth, D. C. 2003, \apj, 586, 1094
\bibitem[Howk (2010)]{h10} Howk, J. C. 2010, in Proc. IAU Symp. 268, Light 
Elements in the Universe, ed. C. Charbonnel, M. Tosi, F. Primas, \& C. 
Chiappini (Cambridge: Cambridge Univ. Press), 335
\bibitem[Howk et al. (2000)]{hss00} Howk, J. C., Sembach, K. R., \& Savage, B. 
D. 2000, \apj, 543, 278
\bibitem[Humphreys (1978)]{h78} Humphreys, R. M. 1978, \apjs, 38, 309
\bibitem[Indriolo et al. (2007)]{i07} Indriolo, N., Geballe, T. R., Oka, T., \& 
McCall, B. J. 2007, \apj, 671, 1736
\bibitem[Indriolo et al. (2009)]{ifm09} Indriolo, N., Fields, B. D., \& McCall, 
B. J. 2009, \apj, 694, 257
\bibitem[Jenkins et al. (1986)]{jss86} Jenkins, E. B., Savage, B. D., \& 
Spitzer, L. 1986, \apj, 301, 355
\bibitem[Jenkins \& Tripp (2001)]{jt01} Jenkins, E. B., \& Tripp, T. M. 2001, 
\apjs, 137, 297
\bibitem[Jensen et al. (2005)]{jrs05} Jensen, A. G., Rachford, B. L., \& Snow, 
T. P. 2005, \apj, 619, 891
\bibitem[Jensen \& Snow (2007a)]{js07a} Jensen, A. G., \& Snow, T. P. 2007a, 
\apj, 669, 378
\bibitem[Jensen \& Snow (2007b)]{js07b} Jensen, A. G., \& Snow, T. P. 2007b, 
\apj, 669, 401
\bibitem[Jura et al. (1996)]{j96} Jura, M., Meyer, D. M., Hawkins, I., \& 
Cardelli, J. A. 1996, \apj, 456, 598
\bibitem[Kim Quijano et al. (2003)]{kq03} Kim Quijano, J., et al. 2003, ``STIS 
Instrument Handbook'', Version 7.0, (Baltimore: STScI)
\bibitem[Kiselman \& Carlsson (1996)]{kc96} Kiselman, D., \& Carlsson, M. 1996, 
\aap, 311, 680
\bibitem[Knauth et al. (2003a)]{k03a} Knauth, D. C., Andersson, B.-G., 
McCandliss, S. R., \& Moos, H. W. 2003a, \apjl, 596, L51
\bibitem[Knauth et al. (2003b)]{kfl03b} Knauth, D. C., Federman, S. R., \& 
Lambert, D. L. 2003b, \apj, 586, 268
\bibitem[Knauth et al. (2000)]{k00} Knauth, D. C., Federman, S. R., Lambert, D. 
L., \& Crane, P. 2000, \nat, 405, 656
\bibitem[Lambert et al. (1998)]{l98} Lambert, D. L., Sheffer, Y., Federman, S. 
R., Cardelli, J. A., Sofia, U. J., \& Knauth, D. C. 1998, \apj, 494, 614
\bibitem[Lee et al. (2008)]{l08} Lee, J.-J., Koo, B.-C., Yun, M. S., 
Stanimirovi\'c, S., Heiles, C., \& Heyer, M. 2008, \aj, 135, 796
\bibitem[Lodders (2003)]{l03} Lodders, K. 2003, \apj, 591, 1220
\bibitem[Majaess et al. (2009)]{mtl09} Majaess, D. J., Turner, D. G., \& Lane, 
D. J. 2009, \mnras, 398, 263
\bibitem[McCall et al. (2003)]{mc03} McCall, B. J., et al. 2003, \nat, 422, 500
\bibitem[Meneguzzi et al. (1971)]{mar71} Meneguzzi, M., Audouze, J., \& Reeves, 
H. 1971, \aap, 15, 337 (MAR)
\bibitem[Meneguzzi \& Reeves (1975)]{mr75} Meneguzzi, M., \& Reeves, H. 1975, 
\aap, 40, 99
\bibitem[Meneguzzi \& York (1980)]{my80} Meneguzzi, M., \& York, D. G. 1980, 
\apjl, 235, L111
\bibitem[Meyer et al. (1998)]{mjc98} Meyer, D. M., Jura, M., \& Cardelli, J. A. 
1998, \apj, 493, 222
\bibitem[Morton (2003)]{m03} Morton, D. C. 2003, \apjs, 149, 205
\bibitem[Pan et al. (2004)]{p04} Pan, K., Federman, S. R., Cunha, K., Smith, V. 
V., \& Welty, D. E. 2004, \apjs, 151, 313
\bibitem[Pan et al. (2005)]{p05} Pan, K., Federman, S. R., Sheffer, Y., \& 
Andersson, B.-G. 2005, \apj, 633, 986
\bibitem[Parizot (2000)]{p00} Parizot, E. 2000, \aap, 362, 786
\bibitem[Parizot \& Drury (1999)]{pd99} Parizot, E., \& Drury, L. 1999, \aap, 
349, 673
\bibitem[Parizot \& Drury (2000)]{pd00} Parizot, E., \& Drury, L. 2000, \aap, 
356, L66
\bibitem[Parizot \& Reeves (2001)]{pr01} Parizot, E., \& Reeves, H. 2001, in 
Proc. 27th Int. Cosmic Ray Conf. (OG part 1), ed. M. Simon, E. Lorenz, \& M. 
Pohl (Katlenburg-Lindau, Germany: Copernicus Gesellschaft), 1938
\bibitem[Patel et al. (1998)]{p98} Patel, N. A., Goldsmith, P. F., Heyer, M. 
H., Snell, R. L., \& Pratap, P. 1998, \apj, 507, 241
\bibitem[Perryman et al. (1997)]{p97} Perryman, M. A. C., et al. 1997, \aap, 
323, L49
\bibitem[Prantzos (2007)]{p07} Prantzos, N. 2007, \ssr, 130, 27
\bibitem[Primas et al. (1999)]{pri99} Primas, F., Duncan, D. K., Peterson, R. 
C., \& Thorburn, J. A. 1999, \aap, 343, 545
\bibitem[Proffitt et al. (1999)]{pro99} Proffitt, C. R., J\"onsson, P., 
Litz\'en, U., Pickering, J. C., \& Wahlgren, G. M. 1999, \apj, 516, 342
\bibitem[Proffitt \& Quigley (2001)]{pq01} Proffitt, C. R., \& Quigley, M. F. 
2001, \apj, 548, 429
\bibitem[Przybilla et al. (2008)]{pnb08} Przybilla, N., Nieva, M.-F., \& 
Butler, K. 2008, \apjl, 688, L103
\bibitem[Rachford et al. (2002)]{r02} Rachford, B. L., et al. 2002, \apj, 577, 
221
\bibitem[Ramaty et al. (1996)]{rkl96} Ramaty, R., Kozlovsky, B., \& 
Lingenfelter, R. E. 1996, \apj, 456, 525
\bibitem[Ramaty et al. (1997)]{r97} Ramaty, R., Kozlovsky, B., Lingenfelter, R. 
E., \& Reeves, H. 1997, \apj, 488, 730
\bibitem[Ramaty et al. (2000)]{r00} Ramaty, R., Scully, S. T., Lingenfelter, R. 
E., \& Kozlovsky, B. 2000, \apj, 534, 747
\bibitem[Reeves (1994)]{r94} Reeves, H. 1994, Rev. Mod. Phys., 66, 193
\bibitem[Reeves et al. (1970)]{rfh70} Reeves, H., Fowler, W. A., \& Hoyle, F. 
1970, \nat, 226, 727
\bibitem[Rich \& Boesgaard (2009)]{rb09} Rich, J. A., \& Boesgaard, A. M. 2009, 
\apj, 701, 1519
\bibitem[Ritchey (2009)]{r09} Ritchey, A. M. 2009, Ph.D. Thesis, University of 
Toledo
\bibitem[Ritchey et al. (2010)]{r10} Ritchey, A. M., Federman, S. R., \& 
Lambert, D. L. 2010, \apj, submitted
\bibitem[Savage \& Bohlin (1979)]{sb79} Savage, B. D., \& Bohlin, R. C. 1979, 
\apj, 229, 136
\bibitem[Savage et al. (1977)]{s77} Savage, B. D., Bohlin, R. C., Drake, J. F., 
\& Budich, W. 1977, \apj, 216, 291
\bibitem[Savage \& Sembach (1991)]{ss91} Savage, B. D., \& Sembach, K. R. 1991, 
\apj, 379, 245
\bibitem[Scherer et al. (2008)]{sch08} Scherer, K., Fichtner, H., Ferreira, S. 
E. S., B\"usching, I., \& Potgieter, M. S. 2008, \apjl, 680, L105
\bibitem[Schmidt-Kaler (1982)]{sk82} Schmidt-Kaler, T. 1982, ``Physical 
Parameters of the Stars'', in Landolt-B\"ornstein Numerical Data and Functional 
Relationships in Science and Technology, New Series, Group IV, Vol. 2b (Berlin: 
Springer-Verlag)
\bibitem[Sheffer et al. (2008)]{sh08} Sheffer, Y., Rogers, M., Federman, S. R., 
Abel, N. P., Gredel, R., Lambert, D. L., \& Shaw, G. 2008, \apj, 687, 1075
\bibitem[Sheffer et al. (2007)]{sh07} Sheffer, Y., Rogers, M., Federman, S. R., 
Lambert, D. L., \& Gredel, R. 2007, \apj, 667, 1002
\bibitem[Smith et al. (2001)]{sck01} Smith, V. V., Cunha, K., \& King, J. R. 
2001, \aj, 122, 370
\bibitem[Snell et al. (2005)]{s05} Snell, R. L., Hollenbach, D., Howe, J. E., 
Neufeld, D. A., Kaufman, M. J., Melnick, G. J., Bergin, E. A., \& Wang, Z. 
2005, \apj, 620, 758
\bibitem[Snow et al. (2007)]{sdj07} Snow, T. P., Destree, J. D., \& Jensen, A. 
G. 2007, \apj, 655, 285
\bibitem[Snow et al. (1998)]{s98} Snow, T. P., Hanson, M. M., Black, J. H., van 
Dishoeck, E. F., Crutcher, R. M., \& Lutz, B. L. 1998, \apjl, 496, L113
\bibitem[Snow \& Witt (1996)]{sw96} Snow, T. P., \& Witt, A. N. 1996, \apjl, 
468, L65
\bibitem[Snow \& York (1981)]{sy81} Snow, T. P., \& York, D. G. 1981, \apjl, 
247, L39
\bibitem[Sofia \& Meyer (2001)]{sm01} Sofia, U. J., \& Meyer, D. M. 2001, 
\apjl, 554, L221
\bibitem[Sonnentrucker et al. (2003)]{son03} Sonnentrucker, P., Friedman, S. 
D., Welty, D. E., York, D. G., \& Snow, T. P. 2003, \apj, 596, 350
\bibitem[Sonnentrucker et al. (2007)]{son07} Sonnentrucker, P., Welty, D. E., 
Thorburn, J. A., \& York, D. G. 2007, \apjs, 168, 58
\bibitem[Spitzer (1985)]{s85} Spitzer, L. 1985, \apjl, 290, L21
\bibitem[Spitzer \& Fitzpatrick (1992)]{sf92} Spitzer, L., \& Fitzpatrick, E. 
L. 1992, \apjl, 391, L41
\bibitem[Tan et al. (2009)]{tsz09} Tan, K. F., Shi, J. R., \& Zhao, G. 2009, 
\mnras, 392, 205
\bibitem[Tan et al. (2010)]{tsz10} Tan, K., Shi, J., \& Zhao, G. 2010, \apj, 
713, 458
\bibitem[Tavani et al. (2010)]{t10} Tavani, M., et al. 2010, \apjl, 710, L151
\bibitem[Timmes et al. (1995)]{tww95} Timmes, F. X., Woosley, S. E., \& Weaver, 
T. A. 1995, \apjs, 98, 617
\bibitem[Tull et al. (1995)]{t95} Tull, R. G., Macqueen, P. J., Sneden, C., \& 
Lambert, D. L. 1995, \pasp, 107, 251
\bibitem[Vall\'ee (2005)]{v05} Vall\'ee, J. P. 2005, \aj, 130, 569
\bibitem[Vangioni-Flam et al. (2000)]{vca00} Vangioni-Flam, E., Cass\'e, M., \& 
Audouze, J. 2000, \physrep, 333-334, 365
\bibitem[Vangioni-Flam et al. (1996)]{v96} Vangioni-Flam, E., Cass\'e, M., 
Fields, B. D., \& Olive, K. A. 1996, \apj, 468, 199
\bibitem[Venn et al. (2002)]{v02} Venn, K. A., Brooks, A. M., Lambert, D. L., 
Lemke, M., Langer, N., Lennon, D. J., \& Keenan, F. P. 2002, \apj, 565, 571
\bibitem[Walborn et al. (2002)]{w02} Walborn, N. R., Danks, A. C., Vieira, G., 
\& Landsman, W. B. 2002, \apjs, 140, 407
\bibitem[Walborn et al. (2007)]{w07} Walborn, N. R., Smith, N., Howarth, I. D., 
Vieira Kober, G., Gull, T. R., \& Morse, J. A. 2007, \pasp, 119, 156
\bibitem[Walborn et al. (1998)]{w98} Walborn, N. R., et al. 1998, \apjl, 492, 
L169
\bibitem[Welty \& Hobbs (2001)]{wh01} Welty, D. E., \& Hobbs, L. M. 2001, 
\apjs, 133, 345
\bibitem[Welty et al. (1999)]{w99} Welty, D. E., Hobbs, L. M., Lauroesch, J. 
T., Morton, D. C., Spitzer, L., \& York, D. G. 1999, \apjs, 124, 465
\bibitem[Welty et al. (1996)]{wmh96} Welty, D. E., Morton, D. C., \& Hobbs, L. 
M. 1996, \apjs, 106, 533
\bibitem[Woosley et al. (1990)]{w90} Woosley, S. E., Hartmann, D. H., Hoffman, 
R. D., \& Haxton, W. C. 1990, \apj, 356, 272
\bibitem[Woosley \& Weaver (1995)]{ww95} Woosley, S. E., \& Weaver, T. A. 1995, 
\apjs, 101, 181 (WW95)
\bibitem[Yoshida et al. (2005)]{y05} Yoshida, T., Kajino, T., \& Hartmann, D. 
H. 2005, \prl, 94, 231101
\bibitem[Yoshida et al. (2006)]{y06} Yoshida, T., Kajino, T., Yokomakura, H., 
Kimura, K., Takamura, A., \& Hartmann, D. H. 2006, \apj, 649, 319
\bibitem[Yoshida et al. (2008)]{y08} Yoshida, T., Suzuki, T., Chiba, S., 
Kajino, T., Yokomakura, H., Kimura, K., Takamura, A., \& Hartmann, D. H. 2008, 
\apj, 686, 448
\end{thebibliography}
\end{document}